\documentclass[twocolumn,twocolappendix]{openjournal}
\usepackage{natbib}
\usepackage{graphicx,amsmath,amssymb,amstext}
\usepackage{amsbsy,amsfonts,amsthm,color}
\usepackage[colorlinks,linkcolor=blue,citecolor=blue,urlcolor=blue ]{hyperref}
\usepackage[utf8]{inputenc}
\usepackage{float}
\usepackage[caption=false]{subfig}
\usepackage{upgreek}
\usepackage{lipsum}

\usepackage[usenames,dvipsnames]{xcolor}
\usepackage[normalem]{ulem} 
\usepackage{array}
\usepackage{booktabs} 
\usepackage{multirow}
\usepackage{orcidlink}

\newcommand{\synthesizer}{\mbox{\tt Synthesizer}}

\newcommand{\jwst}{\mbox{\emph{JWST}}}
\newcommand{\hst}{\mbox{\emph{HST}}}

\newcommand{\scsam}{\mbox{\sc sc-sam}}
\newcommand{\jaguar}{\mbox{\sc jaguar}}
\newcommand{\sage}{\mbox{\sc sage}}
\newcommand{\spritz}{\mbox{\sc spritz}}
\newcommand{\eagle}{\mbox{\sc eagle}}
\newcommand{\dream}{\mbox{\sc dream}}
\newcommand{\galform}{\mbox{\sc galform}}

\newcommand{\flags}{\mbox{\sc flags}}

\newcommand{\crest}{\mbox{\tt CREST}}

\newcommand{\sep}{\mbox{\tt SEP}}
\newcommand{\photutils}{\mbox{\tt Photutils}}
\newcommand{\profound}{\mbox{\tt ProFound}}

\newcommand{\ngdeep}{\mbox{\sc ngdeep}}
\newcommand{\ceers}{\mbox{\sc ceers}}
\newcommand{\jades}{\mbox{\sc jades}}
\newcommand{\origins}{\mbox{\sc origins}}
\newcommand{\sapphires}{\mbox{\sc sapphires}}
\newcommand{\primer}{\mbox{\sc primer}}
\newcommand{\panoramic}{\mbox{\sc panoramic}}
\newcommand{\cosmosweb}{\mbox{\sc cosmos-web}}

\newcommand{\cosmos}{\mbox{\sc cosmos}}
\newcommand{\uds}{\mbox{\sc uds}}
\newcommand{\gds}{\mbox{\sc goods-s}}

\newcommand{\egs}{\mbox{\sc egs}}
\newcommand{\hudf}{\mbox{\sc hudf}}

\newcommand{\bpass}{\mbox{\tt BPASS}}
\newcommand{\bc}{\mbox{\tt BC03}}
\newcommand{\fsps}{\mbox{\tt FSPS}}
\newcommand{\ma}{\mbox{\tt M13}}

\newcolumntype{C}[1]{>{\centering\arraybackslash}p{#1}}

\newcommand\blfootnote[1]{%
  \begingroup
  \renewcommand\thefootnote{}\footnote{#1}%
  \addtocounter{footnote}{-1}%
  \endgroup
}

\renewcommand{\farcs}{.\!\!^{\prime\prime}}

\begin{document}

\title{First Light and Assembly of GalaxieS (FLAGS) I: \\ The JWST/NIRCam Number Counts and IGL as Constraints on Galaxy Formation Models \vspace{-1.75em}}

%\author{Jack C. Turner\textsuperscript{1$\star$\,\orcidlink{0000-0001-6247-041X}}}
%\author{Stephen M. Wilkins\textsuperscript{1, 2\,\orcidlink{0000-0003-3903-6935}}}
%\author{Aswin P. Vijayan\textsuperscript{1\,\orcidlink{0000-0002-1905-4194}}}

\author{Jack C. Turner$^{1\star \, \orcidlink{0000-0001-6247-041X}}$}
\author{Stephen M. Wilkins$^{1,2 \, \orcidlink{0000-0003-3903-6935}}$}
\author{Aswin P. Vijayan$^{1 \, \orcidlink{0000-0002-1905-4194}}$}

\blfootnote{$^{\star}$\mbox{Corresponding author, email: \href{mailto:jt458@sussex.ac.uk}{jt458@sussex.ac.uk}}}

% List of institutions
\affiliation{$^1$Astronomy Centre, University of Sussex, Falmer, Brighton BN1 9QH, UK}
\affiliation{$^{2}$Institute of Space Sciences and Astronomy, University of Malta, Msida MSD 2080, Malta}
\begin{abstract}

\jwst\ observations have been used in conjunction with SED fitting to infer the physical properties of galaxies throughout cosmic time, revealing tensions with the predictions of theoretical models. However, the biases associated with this process are poorly understood, which limits its true constraining power. We introduce the First Light and Assembly of GalaxieS (\flags) series, which will leverage forward modelling to confront models with more reliable direct observables. We describe the consistent processing of NIRCam imaging spanning $>1 \ \mathrm{deg}^{2}$ across $30$ independent fields, which can be used to measure the galaxy number counts from $0.9-4.4 \ \mu\mathrm{m}$. The integrated galaxy light (IGL) is constrained with a certainty of $\sim2.5\%$ at the longest wavelengths, producing novel constraints of $6.92^{+0.17}_{-0.17}$ and $3.46^{+0.09}_{-0.08} \ \mathrm{nW\,m^{-2}\,sr^{-1}}$ at 2.77 and $4.10 \ \mathrm{\mu m}$ respectively. We compare these measurements with predictions from galaxy evolution models and find that the IGL is an unreliable measure of model performance. Comparing against the number counts directly reveals \scsam\ as the best-performing model ($\chi^{2}_{\nu}=18.6$), with its superior performance relative to \sage\ attributed to efficient SNe feedback in low-mass halos. We investigate the impact of systematic photometry and forward modelling uncertainties, confirming that the number counts can be a reliable means of evaluating model predictions and performing simulation-based astrophysical parameter inference in the future.
\end{abstract}

\maketitle

%%%%%%%%%%%%%%%%%%%%%%%%%%%%%%%%%%%%%%%%%%%%%%%%%%

%%%%%%%%%%%%%%%%% BODY OF PAPER %%%%%%%%%%%%%%%%%%

\section{Introduction}\label{sec:intro}

Early James Webb Space Telescope (\jwst) observations produced a number of results that appeared at odds with our understanding of both cosmology and galaxy evolution. These came primarily from photometric datasets collected by the Near Infrared Camera \citep[NIRCam;][]{Rieke_2023_NC} and processed with SED fitting to reveal several galaxy candidates at $z>14$ \citep{Austin_2023, Donnan_2023}. Many models failed to reproduce the implied surface densities, suggesting that the mere existence of these galaxies was incompatible with galaxy formation theory \citep{Finkelstein_2022}. To be observable at such a distance, these galaxies would have to be highly luminous and constitute excessive stellar masses \citep{Naidu_2022, Adams_2023, Labbe_2023}. This presents a challenge to $\Lambda$CDM cosmology, from which such objects are unexpected \citep{Boylan-Kolchin_2023, Lovell_2023}.

These frontier candidates were later followed up with the Near Infrared Spectrograph (NIRSpec), revealing previously indistinguishable emission lines and providing tighter redshift constraints than photometry alone. While some photometric redshifts were validated, a large fraction were found to be systematically overestimated. This included a galaxy initially believed to be at $z\sim16$, but subsequently revised to $z<5$, as spectroscopy revealed a combination of nebular emission lines and dust reddening that mimics the photometric colours of distant objects \citep{Arrabal-Haro_2023, Zavala_2023}. This new data resolved some tensions, and the reliability of photometric redshifts began to be called into question \citep{Clausen_2025}.

New model discrepancies continue to arise, now driven by physical properties inferred from the fitting of both photometry and spectra. High-redshift active galactic nuclei (AGN) have been identified in abundance, with enhanced growth mechanisms and massive seeds invoked to explain their extreme mass ratios \citep{Greene_2024, Maiolino_2024}. A population of `little red dots', characterised by small angular sizes and V-shaped SEDs, has been found throughout cosmic time. Despite an extensive literature, theoretical models are yet to converge on a description of their origin \citep{Cenci_2025, Rusakov_2026, Herrero_Carrin_2026}. The inferred number density of massive quiescent galaxies is far in excess of model predictions beyond $z=5$, and accounting for systematic uncertainties cannot resolve a discrepancy of $>2 \ \rm{dex}$ at $z\sim7$ \citep{Weibel_2025, Turner_2026}.  

These physical properties are subject to several biases that complicate model constraints and the interpretation of the underlying physics. Black hole masses inferred using local virial scaling relations can be positively biased, and may not hold in high-redshift environments \citep{Vestergaard_2006, Yue_2010}. AGN components are often omitted from SED fitting, which can cause stellar masses and star formation rates (SFRs) to be overestimated by $0.4$ and $0.6 \ \mathrm{dex}$ respectively \citep{Ciesla_2015}. Outshining, whereby bright, young O and B-type stars obscure the presence of older stellar populations, can cause stellar masses to be underestimated by $>1 \ \mathrm{dex}$ \citep{Narayanan_2024, Harvey_2025_b}. It is not straightforward to quantify how these biases may affect the selection of massive quiescent galaxies from their specific star formation rates. \citet{Harvey_2025} showed that additional $0.1-0.5 \ \mathrm{dex}$ shifts in stellar mass can arise from differing star formation history (SFH) parametrisations and physical property priors. While these caveats do not invalidate \jwst\ results in their entirety, they suggest that the quoted uncertainties may be underestimated and that derived model constraints may be unreliable. Unfortunately, there is currently no cure-all equivalent to spectroscopic redshifts for these properties, so they cannot be validated as easily. This motivates the search for more reliable alternatives.

Inference of physical properties has been necessary to facilitate comparisons to simulations, which generally do not predict the observable properties of galaxies ab initio. Those that do demand a greater computational cost, which restricts the mass and volume scales that can be probed \citep{Somerville_2015_rev}. However, with the advent of modern forward modelling tools, the prediction of observable properties can be treated as a post-processing step \citep{Camps_2015, Wilkins_2016, Vijayan_2021, Fortuni_2023, Lovell_2025_syn, Roper_2025}. These tools predict the integrated rest-frame spectra of galaxies, which can then be mapped onto spectrograph wavelength grids or convolved with filter transmission curves to produce photometry. As the model redshifts are known exactly, it is trivial to shift these properties into the observer frame under an assumed cosmology, and construct lightcones that mimic their continuous distributions in real catalogues \citep{Yung_2022, Drakos_2022}. This raises the question: can valuable model constraints be derived from direct observables alone?

If so, the direct-observable approach offers several advantages over SED fitting. Hydrodynamical models predict the mass, age and metallicity of every star particle. These can be used to self-consistently assign a single stellar population (SSPs) spectrum, which is insensitive to degeneracies between these properties and dust \citep{Worthey_1994, Walcher_2011}. The predicted observables naturally encode the full star formation and chemical enrichment history, where SED fitting is limited to parametric or poorly constrained non-parametric SFHs, which often assume no metallicity evolution \citep{Carnall_2019, Leja_2019}. Any impact of outshining is included consistently, and the contribution of AGN emission can be determined from the mass and accretion rate of the black hole particles. As the relative positions of particles are known, dust attenuation can be implemented using radiative transfer, or line-of-sight models that account for the physical properties of intervening gas \citep{Vijayan_2021}, rather than assuming a uniform screen. Semi-analytic models (SAMs) can also be forward modelled, albeit without the benefits of particle-wise treatments. As all of these physical properties are known exactly up to the resolution limit, forward modelling does not inherently require parameter space exploration. It is therefore orders of magnitude faster than Bayesian SED fitting \citep{Lovell_2025_syn}, allowing thorough quantification of the remaining uncertainties introduced by the assumed SPS model, IMF, and attenuation curve.

\citet{Manzoni_2025} demonstrated that the shape of the galaxy number counts, distributed as a function of apparent magnitude \citep{Hubble_1926, Koo_1992, Marr_2023}, arises from the inverse K-correction of the underlying luminosity functions. This confirms that the counts can be used to probe the same physics of star formation, supernova and AGN feedback, and dust \citep{Benson_2003, Bower_2006, Croton_2006, Gruppioni_2013}, without knowing the redshifts or physical properties of the constituent galaxies. While disentangling these effects may be more difficult, valuable constraints can be derived by comparing with model predictions, where the implemented physics is known. Potential observational biases are limited to those introduced by image processing and source extraction, which affect all catalogue-level studies of galaxies, and can be more feasibly quantified in the absence of SED fitting. The counts can be further simplified by integrating over all magnitudes, yielding the integrated galaxy light (IGL), a single wavelength-dependent quantity that describes the total energy density of observable galaxies and, in the absence of a diffuse component, the observable Universe \citep{Partridge_1967, Dwek_2013}.

There is a wealth of archival number count data with which to compare and validate new results. Large ground-based surveys determine the number of rare bright galaxies to high significance \citep{Bellstedt_2020, Davies_2021}, whereas deep pointed observations extend constraints to faint magnitudes \citep{Koekemoer_2013, Rafelski_2015, Windhorst_2023}. These measurements are often combined to estimate the counts and IGL over a range of magnitudes, but this requires ill-defined colour corrections to homogenise the photometry, making the results uncertain and model-dependent. Harnessing the unprecedented sensitivity of \jwst, it is possible to bypass these corrections and constrain a broad range of magnitudes with NIRCam alone. There are few direct comparisons of these measurements with the predictions of galaxy formation models. These models struggle to produce the observed trends at mid-infrared wavelengths probed by \jwst/MIRI \citep{Wu_2023, Stone_2024, Harish_2025}, but \citet{Windhorst_2023} demonstrated qualitative agreement between the NIRCam counts and a single SAM. Ground-based measurements of the IGL at comparable wavelengths are at odds with the predictions of \textsc{shark} \citep{Lagos_2019, Koushan_2021}, but the same model produces the radio IGL reasonably well \citep{Tompkins_2023}. There is clearly scope for the number counts and IGL to constrain these models.

We introduce the First Light and Assembly of GalaxieS (\flags) series, which will harness observable properties to derive robust constraints on theoretical models of galaxy formation and evolution. In this first instalment, we present the consistent post-processing of $>1 \ \mathrm{deg^{2}}$ of \jwst/NIRCam imaging across $>30$ independent sight-lines, as well as collating and generating model photometry (\S\ref{sec:data}). We use this data to produce galaxy catalogues and compute the number counts in eight NIRCam filters spanning $0.9-4.4 \ \mu\mathrm{m}$ (\S\ref{sec:counts}). After calculating the corresponding IGL (\S\ref{subsec:results_obs}), we compare both to model predictions to evaluate their performance and interpret the underlying physics (\S\ref{subsec:results_models}). We quantify the effect of varying the source extraction parameters governing the photometry and leverage a hydrodynamical model to assess the impact of forward modelling uncertainties (\S\ref{subsec:systematics}). Magnitudes are quoted in the AB system throughout \citep{Oke_1983} and refer to apparent values unless stated otherwise.
\section{Data}\label{sec:data}

\subsection{NIRCam Imaging}\label{subsec:imaging}

The first \flags\ dataset is based on NIRCam imaging from seven key surveys. The calibrated science images were produced by the individual survey teams and are thus optimised for the specific strategies and tuned to remove field-specific artefacts and persistence. These reductions are expected to be of high quality and contain minimal spurious sources, which is particularly important when measuring the galaxy counts. However, differing analysis pipelines may introduce systematic variations between fields \citep{Adams_2024}, which we aim to minimise by performing post-calibration steps consistently.

We do this using Consistent Reduction Extraction and SED fitting Tools (\crest\footnote{\url{https://github.com/jackcturner/crest}}), a Python library assembled to address this concern arising from the varied approaches adopted in the literature. \crest\ collates and builds upon regularly used background subtraction, PSF measurement, and source extraction tools. It does this within an object-oriented framework, in which different tools associated with each step are enclosed in their own individual but consistently interfaced wrappers. This simplifies their use and allows each to be swapped in and out as required to test the associated systematics. In this vein, each tool is now controlled by a configuration file, enabling straightforward tracking and dissemination of parameter choices and ensuring consistency across datasets.

\subsubsection{Surveys} \label{subsubsec:surveys}

\ngdeep\ \citep{Bagley_2024, Leung_2023} is the smallest survey by area considered in this work, consisting of a single NIRCam pointing on the outskirts of the \hudf. This is equivalent to just over $9.8 \ \mathrm{arcmin}^{2}$ of sky coverage, but while small, \ngdeep\ is exceptionally deep. Total exposure times of $70 \ {\rm ks}$ for filters F150W, F200W, F277W and F356W, and twice that for F115W and F444W, will allow the galaxy counts to be constrained at magnitudes fainter than ever before. \jades\ \citep{Eisenstein_2026, Johnson_2026} imaged $210$ and $230 \ \mathrm{arcmin}^{2}$ of the Northern and Southern \mbox{\textsc{goods}} fields respectively. These observations were performed across several filters, from which we select the six listed above, as well as F090W and F410M, which are commonly found in other datasets. The depth varies significantly across both fields, but particularly in \gds, where an area equivalent to a single pointing reaches depths comparable to \ngdeep. We consider this `\origins' field separately from the rest of \gds, which allows us to fully exploit its depth to reduce cosmic variance (CV) at the faintest magnitudes. The CV is reduced further by the inclusion of \sapphires\ \citep{Sun_2025}, which observed two pointings over $17-34 \ \mathrm{ks}$ in the same eight filters. 

The \ceers\ survey \citep{Bagley_2023, Finkelstein_2025} observed $\sim100\ {\rm arcmin^{2}}$ of the \egs\ across ten pointings with exposure times between $3$ and $10 \ \mathrm{ks}$. This achieves a consistent medium depth in all but the shortest of the aforementioned filters. \primer\ (GO-1837) significantly increases the sky coverage in all filters, with $300 \ {\rm arcmin^{2}}$ of medium and $350 \ {\rm arcmin^{2}}$ of shallow imaging across the \cosmos\ and \uds\ fields. Neither of these surveys contributes to the faintest bins, but such areas are vital to obtain robust Poisson statistics around the peak of the magnitude distribution. A similar argument can be made for the inclusion of \cosmosweb\ \citep{Casey_2023, Franco_2026}, which is the largest survey by area considered in this work. A contiguous $\sim 0.6 \ {\rm deg^{2}}$ region of the \cosmos\ field was observed over twenty tiles, from which we select the fourteen that do not overlap with \primer. Such a large area limits the filter coverage to F115W, F150W, F277W and F444W with shallow depths, but significantly reduces the Poisson uncertainties at bright magnitudes. 

While these large surveys greatly reduce Poisson uncertainty, they do little to reduce CV, which is most effectively achieved by considering multiple independent fields. \panoramic\ \citep{Williams_2025} is a pure parallel programme that observed $\sim430 \ {\rm arcmin^{2}}$ of novel \jwst\ area, including 27 associations that are neither overlapping nor adjacent to the legacy fields highlighted above. After omitting three with an overabundance of hot pixels, we increase the number of independent fields considered by a factor $>3$. These include coverage in at least six filters, with variable exposure times. While the other surveys are provided at a consistent $0\farcs03$ pixel resolution, \panoramic\ imaging is made available at $0\farcs02$ and $0\farcs04$ resolution in the short and long-wavelength channels respectively. To maintain consistency between the two channels, we rebin the short-wavelength images in groups of $2\times2$ pixels, summing the science and RMS pixels linearly and in quadrature respectively. These surveys and fields are summarised in Table \ref{tab:surveys}.

\subsubsection{Background subtraction} \label{subsubsec:background}

The background is estimated consistently across all fields, following the tiered source-masking approach outlined by \citet{Bagley_2023}. Large-scale fluctuations are first removed from an image by identifying pixels $5\sigma$ above a rough local background estimate computed without any source masking, and replacing them with the global background value. This image is then median filtered with a ring of inner radius $2\farcs4$ and width $0\farcs12$, with the intention of producing an image with a background that is flat on large scales, allowing a fixed source detection threshold to be used. \crest\ implements tiling and multithreading, allowing this step to be run on the $273$ individual mosaics considered in this work, which constitute $10^{11}$ pixels, in hours rather than days.

The tiered masking process begins by smoothing the filtered image with a Gaussian kernel and detecting sources with a given number of connected pixels $2\sigma$ above the background. The detected sources are masked, and this mask is dilated by a circular footprint before computing a new estimate of the global background. Masked pixels are then replaced with this background estimate before the process repeats. We perform four tiers of masking, iteratively decreasing the Gaussian kernel width, minimum connected pixels and mask dilation radius to remove increasingly faint sources. The masks determined from all filter mosaics within a given field are merged to account for flux that may fall marginally below the detection threshold at some wavelengths. This mask is applied before measuring the final background from the original image, filtered over $5 \times 5$ boxes with area of $0.06 \ {\rm arcsec}^{2}$.

The use of a global background derived from a ring-median-filtered image is highly effective for surveys like \ceers, where the RMS is already reasonably consistent across the entire field. However, if the RMS across large regions of an image varies by more than an order of magnitude, the lower-depth regions can be overmasked if the noise fluctuates above average. This is particularly limiting when using a merged source mask, as only one of the filter mosaics needs to be overmasked for the remainder to be affected, resulting in systematic residual background light. To account for this, \crest\ introduces a tunable scaling of the detection threshold based on the weight of each pixel relative to the median. Through visual inspection, we set this scaling to turn on only when the weight is at least a factor of 30 below the median and limit it to a factor of 2.5. We find this approach to be more effective than simply using the weight image to define the detection threshold, as this limits the detectability of the faintest objects.

\begin{figure}
    \centering
    \includegraphics[width=\columnwidth]{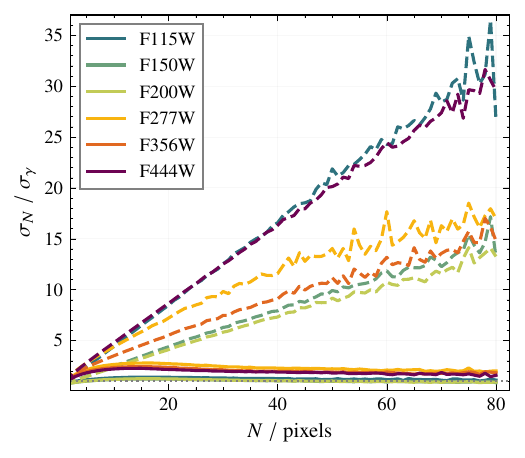}
    \caption{Comparison of the background RMS in \ngdeep, measured from both the original science (dashed) and background-subtracted images (solid). The RMS is measured over different scales $N$ and normalised by the minimum expected from photon counting noise.}
    \label{fig:bkg}
\end{figure}

Figure \ref{fig:bkg} demonstrates the effectiveness of this approach when applied to \ngdeep. Using the merged source mask, we identify blank regions of the image and rebin these in blocks of increasing side length $N$ pixels. At each scale, we compute the standard deviation across all blocks and rescale to a per-pixel value to quantify the variation in the background over that scale $\sigma_{N}$. We plot this RMS as a function of scale, after normalising by the minimum expected from photon-counting statistics alone, approximated as the error map mean $\sigma_{\gamma}$. Inter-pixel correlations introduced by drizzling suppress the RMS on small scales, while the wings of galaxies missed by source masking slightly inflate it on scales of $\sim10$ pixels. The background subtraction performs remarkably well at larger scales, quickly converging towards the ideal RMS, and produces mosaics in which the background is consistently an order of magnitude smoother than the original science images.

\subsubsection{Point-spread functions} \label{subsubsec:psfs}

Point-spread functions (PSFs) are measured empirically from each image using an adapted version of the \texttt{Aperpy} \citep{Weaver_2024} star stacking procedure. Firstly, peaks are identified as regions of the image with flux $10\sigma$ above the global median absolute deviation. Following the methodology of \cite{Skelton_2014} and \cite{Whitaker_2019}, stars are assumed to have a size, defined by the ratio between the fluxes in $0\farcs16$ and $0\farcs32$ diameter apertures, that is approximately constant with respect to apparent magnitude. Stars are separated from other peaks, such as AGN or hot pixels, by fitting a linear function to the size-magnitude plane and selecting sources that fall within a $10\%$ range of the fit. These sources are extracted from the image in square cutouts of width $3\farcs0$, and recentred based on the centre of mass determined from image moments. Cutouts are discarded if recentring requires a shift of $>2$ pixels. Sources are then ensured to be both well detected and unsaturated by requiring a signal-to-noise ratio $>1000$ and an apparent magnitude $15<m_{\mathrm{AB}}<25$. The remaining cutouts undergo two iterations of pixel-wise $3\sigma$-clipping, with clipped pixels masked by a circular footprint of radius $\sim0\farcs85$. Cutouts are discarded if $>75\%$ of the pixels are masked, or if $>50\%$ of the pixels within a $0\farcs24$ radius of the centre are masked, with only the remainder used to create a final stacked PSF. The NIRCam position angle is not consistent across all fields, so these should be considered field-averaged PSFs, but this will have little impact on integrated photometry. The \sapphires\ and \panoramic\ mosaics do not contain a sufficient number of stars from which to measure a high signal-to-noise PSF. In these cases, simulated PSFs are generated at the appropriate pixel scale using \texttt{STPSF} \citep{Perrin_2014}.

\begin{figure}
    \centering
    \includegraphics[width=\columnwidth]{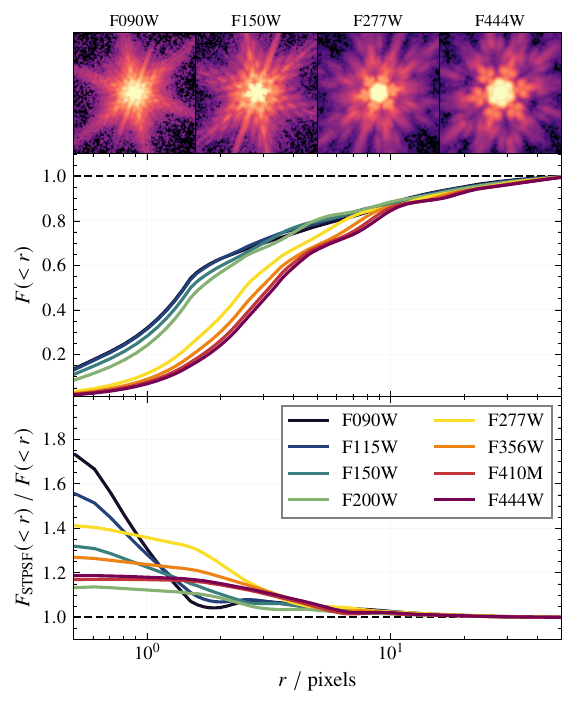}
    \caption{\emph{Top}: Four $101\times101 \ \mathrm{pixel} \ (9.2 \ \mathrm{arcsec}^{2})$ example PSFs measured from the \primer/\cosmos\ field using the empirical star stacking procedure. \emph{Middle}: Curves of growth for all PSFs measured from the same field, with each filter indicated by a different colour. \emph{Bottom}: The ratio between \texttt{STPSF} simulated and empirical PSFs.\vspace{0.2em}}
    \label{fig:psfs}
\end{figure}

The top panel of Figure \ref{fig:psfs} displays four PSFs measured from the \primer/\cosmos\ mosaics, demonstrating the level of detail that can be recovered with this approach. The middle panel shows the normalised curves of growth of all eight PSFs measured from the same field. They appear as expected, with the FWHM decreasing as a function of wavelength from $\sim0 \farcs 06$ in F090W to $\sim0\farcs16$ in F444W, including a sharp increase between the SW and LW channels. These values vary from field to field, but by no more than $3\%$, and the energy enclosed beyond 10 pixels is approximately equal across all filters. The bottom panel shows the ratio between the energy enclosed by the \texttt{STPSF} PSFs and those measured empirically. The simulated PSFs are far narrower, with the difference exceeding $70\%$ within the smallest apertures. However, given the estimated magnitude completeness limit of \sapphires\ and \panoramic\ (\S\ref{subsec:counts}) and the average angular size of galaxies at that magnitude, the minimum aperture radius expected to contribute to the number counts is $\sim5 \ \mathrm{pixels}$. The difference quickly falls below $5\%$ at larger radii, so we do not consider the use of simulated PSFs to have a significant effect on our results.

Filter-dependent PSFs can introduce bias when comparing their number count distributions, as confusion could, in theory, shift the long-wavelength counts towards brighter galaxies as nearby objects appear blended. However, the high resolution of NIRCam means this is unlikely to be a significant issue, and standard deblending procedures can easily overcome it (\S\ref{subsubsec:fiducial}). However, the flux measured in a fixed aperture is inversely proportional to the PSF width, potentially biasing the counts in the opposite direction. This is usually accounted for by PSF homogenisation, but this artificially smooths the background, lowering the global RMS. This greatly increases the sensitivity to noise unless a stacked detection image is used, which is suboptimal for measuring the true wavelength-dependent counts. We therefore choose to analyse the images at their native resolution, adopt apertures of varying size, and apply a PSF-informed aperture correction to scale the measured fluxes to the total.

\subsubsection{Depths} \label{subsubsec:depths}

Average image depths are measured from blank regions identified from the merged source masks generated during background subtraction (\S\ref{subsubsec:background}). Non-overlapping apertures of radius $0\farcs15$ are placed randomly within these regions, with a density of $100 \ {\rm arcmin}^{-2}$ to ensure broad coverage and robust statistics. The flux within each aperture is calculated and scaled to the total based on the fraction of the PSF enclosed. The $5\sigma$ point-source depth is computed from the median absolute deviation of these fluxes and listed in Table \ref{tab:surveys}. It is important to note that the absolute depth value is strongly dependent on the aperture size, so these values should only be used to compare the relative depth of each field or to compare with other works adopting the same approach.

\renewcommand{\arraystretch}{1.2}
\begin{table*}
    \centering
    \begin{tabular}{l c c c c c c c c c}
        \toprule
        \multirow{2}{*}{\textbf{Survey (Field)}} & \multirow{2}{*}{\textbf{Area [arcmin$^2$]}} & \multicolumn{8}{c}{\textbf{5$\sigma$ Depth [AB mag]}} \\ 
        \cmidrule(lr){3-10}
        & & \textbf{F090W} & \textbf{F115W} & \textbf{F150W} & \textbf{F200W} & \textbf{F277W} & \textbf{F356W} & \textbf{F410M} & \textbf{F444W} \\ 
        \midrule
        CEERS (EGS) & 84.8 & -- & 28.7 & 28.7 & 28.9 & 29.0 & 29.0 & 28.3 & 28.6 \\ 
        COSMOS-Web (COSMOS) & 1286.2 & -- & 27.0 & 27.3 & -- & 28.0 & -- & -- & 27.7 \\ 
        JADES (GOODS-N) & 180.5 & 29.1 & 28.6 & 28.7 & 29.3 & 29.4 & 29.2 & 28.8 & 28.9 \\  
        JADES (GOODS-S) & 178.7 & 29.3 & 29.4 & 29.4 & 29.4 & 29.8 & 29.8 & 29.3 & 29.5 \\
        JADES (ORIGINS) & 12.6 & 29.7 & 29.8 & 30.1 & 30.0 & 30.3 & 30.3 & 29.8 & 30.1 \\
        NGDEEP (HUDF) & 9.6 & -- & 30.1 & 30.0 & 30.0 & 30.2 & 30.1 & -- & 30.0 \\
        PANORAMIC* & 207.7 & -- & 27.5 & 27.7 & 27.8 & 28.3 & 28.3 & 28.4 & 28.0 \\ 
        PRIMER (COSMOS) & 140.0 & 27.8 & 27.8 & 28.1 & 28.2 & 28.6 & 28.6 & 28.0 & 28.3 \\ 
        PRIMER (UDS) & 247.3 & 27.6 & 27.8 & 27.9 & 28.1 & 28.4 & 28.4 & 27.7 & 28.0 \\ 
        SAPPHIRES & 14.6 & 29.1 & 28.9 & 29.1 & 29.3 & 29.6 & 29.5 & 28.9 & 29.4 \\ 

        \bottomrule
    \end{tabular}
    \caption{Properties of the surveys considered in this work and the legacy fields that they cover. Due to small differences in coverage across filters, the quoted value of the unmasked science area is the maximum available for that field. The $5\sigma$ point-source depths are averaged over each field and measured in circular apertures of radius $0\farcs15$. For \panoramic, we quote the sum of the maximum area and average depth across associations.}
    \label{tab:surveys}
\end{table*}

\subsection{Models}\label{subsec:modeldata}

\begin{table*}
\centering
\begin{tabular}{ccccccc}
\toprule
\textbf{Model} & \textbf{Type} & \textbf{Area/Volume} & $M_{\mathrm{min}} \ / \ \mathrm{M_{\odot}}$ & \textbf{SPS} & \textbf{Dust Attenuation} & \textbf{Nebular Emission} \\ \midrule
DREaM & SHAM & $1 \ \mathrm{deg}^{2}$ & $10^{5}$ & \texttt{FSPS} & \citet{Charlot_2000} & \citet{Byler_2017} \\
EAGLE & Hydrodynamical & $(100 \ \mathrm{cMpc})^{3}$ & $10^{7}$ & \texttt{BPASS} & \citet{Vijayan_2021} & \citet{Wilkins_2020} \\
GALFORM & SAM & $(800 \ \mathrm{cMpc})^{3}$ & $10^{5}$ & \texttt{M05} & \citet{Silva_1998} & None\\ 
JAGUAR & SEM & $100 \ \mathrm{arcmin}^{2}$ & $10^{6}$ & \texttt{BC03} & \citet{Charlot_2000} & \citet{Gutkin_2016} \\
SAGE & SAM & $1000 \ \mathrm{arcmin}^{2}$ & $10^{6}$ & \texttt{BC03} & \citet{Devriendt_1999} & None \\
SCSAM & SAM & $1000 \ \mathrm{arcmin}^{2}$ & $10^{7}$ & \texttt{BC03} & \citet{Somerville_2012} & \citet{Hirschmann_2019}\\
SPRITZ & SEM & $46 \ \mathrm{arcmin}^{2}$ & $10^{4}$ & \texttt{BC03} & \citet{Charlot_2000} & \citet{Bonato_2019} \\ \bottomrule
\end{tabular}
\caption{Properties of the models considered in this work. The first four columns indicate the model name, type, volume or sky area over which predictions are made and the minimum stellar mass. The remaining columns describe the assumptions made during forward modelling, and/or when inferring the physical properties of empirical SEDs. All bar \galform\ use a \citet{Chabrier_2003} IMF, which uses different IMFs for each star-forming phase \citep{Lacey_2016}.}
\label{tab:model_info}
\end{table*}

We compare these observations to the predictions of three broad classes of models: semi-empirical (SEMs) including subhalo abundance matching (SHAM), semi-analytic (SAMs) and hydrodynamical, which we summarise below. Details of physics implemented by each model can be found in the associated papers, but aspects particularly relevant to this work will be highlighted throughout as appropriate. See \citet{Somerville_2015_rev} for a review of these classes.

Semi-empirical models are constructed by sampling from observed relations that describe the number density of galaxies per unit volume, as a function of properties such as luminosity or stellar mass. Galaxies can be distributed throughout cosmic time by considering the redshift evolution of these functions, thereby allowing the construction of lightcone catalogues of arbitrary volume. Additional physical properties can be derived from a series of connected scaling relations and observables predicted by matching to observed template SEDs, or by assigning a spectrum from an SPS model. In theory, these models have a grounding in observational truth, but this relies on SED fitting inferences. However, these models still add value, as we can demonstrate the discriminating power of the direct observable approach by comparing their predictions. This work considers ten realisations of the \jaguar\ lightcones equivalent to $\sim100 \ \mathrm{arcmin}^{2}$ \citep{Williams_2018} and the $1 \ \mathrm{deg}^{2}$ \dream\ SHAM lightcone \citep{Drakos_2022}. We also include two $\sim46 \ \mathrm{arcmin}^{2}$ realisations of the \spritz\ model, which we refer to as \spritz$_{1}$ and \spritz$_{4}$ to denote the different exponents used to extrapolate far-infrared luminosity functions to high redshift $\propto(1 + z)^{k_{\Phi}}$, where $k_{\Phi} = -1,-4$ \citep{Bisigello_2021}.

Semi-analytic models apply analytical prescriptions of the physics governing galaxy evolution to the outputs of dark matter only (DMO) simulations. By comparing their predictions to observations, we can directly constrain our theoretical understanding of processes such as gas cooling and star formation, and their interplay with feedback mechanisms. The stellar mass and metallicity naturally evolve through time, producing star formation and chemical enrichment histories that facilitate self-consistent forward modelling. \citet{Cowley_2018} made predictions for the \jwst\ galaxy counts from a $(800 \ \mathrm{cMpc})^{3}$ box, using both the fiducial \galform\ model \citep{Lacey_2016} and a second where supernova feedback is weaker at high redshift \citep[\galform$_{\mathrm{EFB}}$;][]{Hou_2016}. These models are confronted with the observations, alongside a $1000 \ \mathrm{arcmin}^{2}$ \scsam\ lightcone \citep{Somerville_2015_scsam, Yung_2022} built upon the $250 \ \rm{Mpc/h}$ Bolshoi-Planck \mbox{MultiDark} DMO simulation \citep{Klypin_2016}. We use the Theoretical Astrophysical Observatory \citep{Bernyk_2016} to construct a \sage\ \citep{Croton_2006, Croton_2016} lightcone with identical area, based on the same DMO simulation. We make consistent choices for the SPS model and the initial mass function (IMF) used to generate intrinsic photometry, and both SAMs use a slab dust model. 

Each of these SEMs, SAMs, and the \dream\ SHAM model implements a different forward modelling approach. This may influence the conclusions we draw in the observer frame regardless of the underlying physics. In an ideal scenario, the forward modelling would be performed consistently and implement variations to quantify these effects, but such datasets are not yet available. However, the different forward modelling components are summarised in Table \ref{tab:model_info}, in an attempt to illustrate how comparable the predictions are.

Finally, we compare to the predictions of the \eagle\ cosmological hydrodynamical simulation \citep{Schaye_2014, Crain_2015}. For this work, we generate new photometry predictions from a $(100 \ \mathrm{cMpc})^{3}$ box, with gas and dark matter particle resolution of $1.81\times10^{6}$ and $9.70 \times10^{6} \ \mathrm{M_{\odot}}$ respectively. We do this using \synthesizer\ \citep{Lovell_2025_syn, Roper_2025}, accounting for the physical properties and relative positions of star and gas particles. Galaxies must exceed a stellar mass $M_{\ast}>10^{7} \ \mathrm{M_{\odot}}$, to ensure that their predicted properties are reliable. The observables are measured within a $30 \ \rm{pkpc}$ aperture, which captures the majority of the total flux while minimising the influence of distant but gravitationally bound particles that would not be included in real observations. The fiducial approach assigns spectra to star particles using the \texttt{BPASS-v2.2.1} \citep{Stanway_2018} binary SPS models, assuming a \citet{Chabrier_2003} IMF ranging from $0.1-300 \ \rm{M_{\odot}}$. Photoionisation modelling follows the approach of \citet{Wilkins_2020} and \citet{Vijayan_2026}, who employed \texttt{Cloudy} \citep{Ferland_2017} assuming an escape fraction $f_{\rm{esc}} = 0$. Dust attenuation is modelled using a line of sight (LOS) model, whereby stars are only attenuated by gas particles that intervene in the line of sight to the observer, by an amount dependent on their dust column density \citep{Vijayan_2019, Vijayan_2021}. This assumes a power-law attenuation curve where $\tau\propto\lambda^{-1}$ and additional birth cloud attenuation proportional to stellar metallicity is applied to stars younger than  $10 \ \rm{Myr}$ \citep{Charlot_2000}. After modelling the effects of IGM absorption \citep{Inoue_2014}, observer-frame photometry is measured at each redshift snapshot by convolving the integrated spectra with the NIRCam transmission curves.

As discussed previously, different forward modelling approaches can influence observer-frame predictions regardless of the underlying physics. To quantify the potential impact of these inconsistencies, we generate versions of the \eagle\ photometry that vary key components in turn, and compare their predictions in Section \ref{subsec:systematics}. This includes the stellar models of \bc\ \citep{Bruzual_2003}, \fsps\ \citep{Conroy_2010} and \ma\ \citep{Maraston_2011}, building upon the original \texttt{M05} models \citep{Maraston_2005}. These allow the impact of isochrone, spectra and binary star assumptions to be investigated. The fiducial photometry assumes a power-law dust attenuation curve, which is compared to the local Milky Way (MW) and Small and Large Magellanic Cloud (SMC, LMC) measurements \citep{Gordon_2003} and that of \citet{Calzetti_2000}, as well as an unattenuated model. Finally, we include a set of photometry without additional \texttt{Cloudy} modelling to investigate whether nebular emission can significantly affect an integrated quantity such as the galaxy counts.

\section{Galaxy Counts}\label{sec:counts}

\subsection{Fiducial Catalogues}\label{subsubsec:fiducial}

\renewcommand{\arraystretch}{1.2}
\begin{table}
\centering
\begin{tabular}{cc}
\toprule
\textbf{Parameter} & \textbf{Fiducial Value}\\

\midrule

BACK\_VALUE & 0.0\\
BACKPHOTO\_TYPE & LOCAL\\
BACKPHOTO\_THICK & 18\\

\midrule

WEIGHT\_TYPE & MAP\_WEIGHT\\
DETECT\_THRESH & 1.5\\
DETECT\_MINAREA & 6\\
FILTER\_NAME & gauss\_3.0\_5x5\\
DEBLEND\_MINCONT & 0.005\\
DEBLEND\_NTHRESH & 32\\
CLEAN & Y\\
CLEAN\_PARAM & 5.0\\

\midrule

PHOT\_AUTOPARAMS & 2.5, 1.5\\
MASK\_TYPE & BLANK\\

\bottomrule
\end{tabular}

\caption{The \texttt{Source Extractor} parameter values used to generate the fiducial individual band catalogues. Pixel-based values are quoted as used for the $0\farcs03$ imaging, but these are scaled appropriately before application to \panoramic.}
\label{tab:fiducial}

\end{table}

The catalogue creation approach is motivated by the science goal of accurately measuring the wavelength-dependent number counts. This rules out the use of a stacked detection image, which inherently biases against the detection of objects that emit in only a subset of the selected filters. For this reason, we perform source extraction on each background-subtracted image individually. A significant limitation of this approach is that the contribution of random noise fluctuations cannot be suppressed, and segmentation must be performed accordingly.

We use \texttt{Source~Extractor~v2.28.0} \citep[SE; ][]{Bertin_1996}  to identify sources and measure their photometric fluxes. Table \ref{tab:fiducial} lists the parameter values used to generate fiducial catalogues from the $0\farcs03 \ / \ \mathrm{pixel}$ imaging, which are scaled when applied to the coarser resolution of \panoramic. These parameters are selected based on rigorous comparison of the generated segmentation maps across multiple bands. The science images are first convolved with a $5\times5$ pixel Gaussian filter with FWHM of 3 pixels, thereby enhancing the detection of faint sources on this scale. To guard against the identification of spurious noise fluctuations, we adopt a conservative source detection requirement of six connected pixels that are $1.5\sigma$ brighter than the local background. This threshold is defined by a weight map that excludes Poisson noise, which SE scales to an absolute RMS using the internally measured background map. This is favoured over using the RMS maps generated by the reduction pipelines, which introduced unrealistic fluctuations in the faint-end counts.

Photometry is measured in Kron apertures \citep{Kron_1980}, which can vary in size and shape depending on the properties of each object in each filter. The local background is estimated within an annulus of width $0\farcs54$ around each source, and subtracted to account for any missed by dilating the tiered source masks. These Kron apertures have been shown to capture $>90\%$ of the total flux when scaled by a factor $2.5$, regardless of the source morphology \citep{Graham_2005}. An aperture correction is applied to account for the filter-dependent PSFs and capture closer to the total flux. For each scaled Kron aperture, we construct a circular aperture that encompasses the same sky area, and the Kron flux is scaled by the fraction of the corresponding PSF that is enclosed within. These total fluxes are finally corrected for galactic extinction using the \cite{Schlafly_2011} dust maps by matching the NIRCam filter to the \hst\ or UKIRT band that is closest in wavelength. The surveys considered in this work are small on full sky scales, so the same correction measured at the image centre is applied to all galaxies. This correction constitutes no more than $0.1 \ \mathrm{mag}$ at the NIR wavelengths considered.

There is no particular feature implemented in SE that gives it a science advantage over the other commonly used source extraction tools, such as \sep\ \citep{Barbary_2016}, \photutils\ \citep{Bradley_2025}, or \profound\ \citep{Robotham_2018}. However, SE is the fastest, allowing us to investigate the impact of these parameter assumptions and make robust estimates of the catalogue completeness (\S\ref{subsec:counts}).

These catalogues contain many spurious sources, despite the cautious approach to their extraction. Hot pixels, which can exceed the minimum pixel threshold due to dithering and detection filtering, can inflate the counts at all but the brightest magnitudes. These are difficult to mask manually due to their abundance and naturally small extent, but all real astrophysical sources should have a FWHM that exceeds that of the PSF. We therefore remove hot pixels by requiring that the FWHM of each source be at least $90\%$ as broad as the PSF. This $10\%$ tolerance accounts for uncertainties in the FWHMs measured by SE, and guards against the removal of small PSF-shaped sources, which are likely to be quasars.

\begin{figure}
    \centering
    \includegraphics[width=\columnwidth]{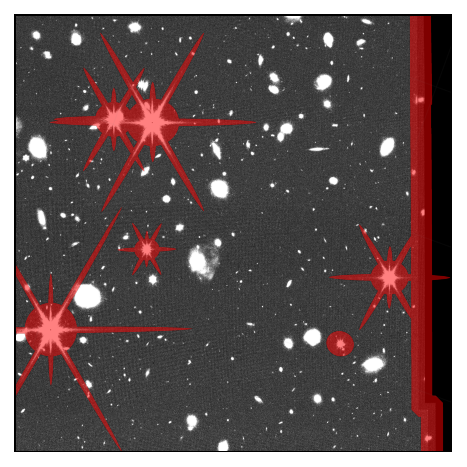}
    \caption{A $3000\times3000$ pixel square cutout of a S/N image constructed from four \cosmosweb\ images, equivalent to $2.25 \ \mathrm{arcmin}^{2}$. The automated edge mask and semi-automated star masks are shown as a red overlay.}
    \label{fig:mask}
\end{figure}

Image edges introduce several issues, as the higher noise level not only inflates the faint-end counts, but sources can also be cut off, leading to underestimated fluxes. These effects are mitigated by generating edge masks, which cover any area within $1\farcs5$ of a boundary and are merged across filters. Stars and their large diffraction spikes can inflate the counts across the magnitude distribution following deblending, so both are removed by a semi-automated procedure. We start by constructing a single `S/N' image for each field, where each pixel value corresponds to the maximum S/N measured across all filters. The positions and classifications of objects in the GAIA DR3 catalogue \citep{GAIA_2016, GAIA_2023} are then used to identify sources more likely to be stars than quasars or galaxies. The S/N images are then visually inspected using SAOImageDS9, and elliptical regions covering the maximum extent of each star are overlaid. Obvious misclassifications of galaxies, as revealed by extended light inconsistent with the PSF, are ignored. Additional contaminants, such as persistence, snowball residuals and dragon's breath produced by very bright stars, are also masked at this stage. The star and edge masks are applied after catalogue creation by removing sources whose measured centre falls on a masked pixel. The total sky area of each image is calculated precisely by summing unmasked pixels, and the maximum of each survey is given in Table \ref{tab:surveys}. This totals $>0.65 \ \mathrm{deg}^{2}$, equivalent to a factor $>30$ increase over the area considered by previous NIRCam studies of the galaxy counts \citep{Windhorst_2023}, resulting in lower Poisson and CV uncertainties. An example S/N image and its corresponding mask are shown in Figure \ref{fig:mask}, demonstrating the significant area preserved by using multiple elliptical regions rather than a single circle.

\subsection{Observed Galaxy Counts}\label{subsec:counts}

\begin{figure}
    \centering
    \includegraphics[width=\columnwidth]{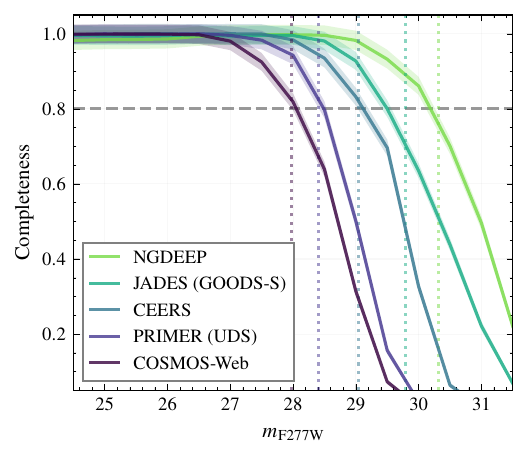}
    \caption{The magnitude-dependent completeness of five fields spanning a representative range of depths, as measured by injecting synthetic point sources. Shaded regions span the $1\sigma$ uncertainty range, and dotted lines show the $5\sigma$ point source depth of the imaging. The grey dashed line denotes the $80\%$ completeness limit.}
    \label{fig:completeness}
\end{figure}

These catalogues are now ready for use in measuring the galaxy number counts. We convert the measured fluxes to AB magnitudes and use the nearest-grid-point approach to bin the data into $0.5 \ \mathrm{mag}$ width bins centred on integer and half-integer magnitudes. Due to the limited survey depths, the number of sources in an image that are included in the associated catalogue decreases with apparent magnitude. This also depends on the source extraction approach, as fewer faint sources are identified when, for example, the detection threshold or the minimum connected pixels are increased. We estimate this magnitude-dependent completeness using the source-injection approach of \citet{Stone_2024}. We randomly sample $1500$ fluxes from each $0.5 \ \mathrm{mag}$ width magnitude bin, and scale empirical or simulated PSFs to these total values. We randomly place one of these synthetic sources into an unmasked area of the image for every $10^{6}$ unmasked pixels, as determined by the tiered source masks produced during background subtraction. We repeat this for multiple copies of the image until all $1500$ have been placed, ensuring representative coverage without introducing artificial crowding. We do not apply any additional dilation to the source mask, which allows the effect of source blending to be naturally accounted for. The completeness is measured as the fraction of these sources recovered by SE using the fiducial approach, requiring a $>3\sigma$ detection within $0\farcs12$ of the inserted location and a flux within $50\%$ of the true value. This process is repeated for every magnitude bin, filter and field. Figure \ref{fig:completeness} shows the completeness measured in the F277W images of five surveys with differing depths. A survey only contributes to the counts in a magnitude bin if its completeness is $>80\%$, which minimises the sensitivity of our results to these corrections. This limit generally coincides with the $5\sigma$ point source depth, but can fall $<0.5 \ \mathrm{mag}$ below.

The counts are affected by three main sources of statistical uncertainty. Sample variance is inherent to all counting operations and can be assumed to follow a Poisson distribution. We apply the full asymmetric treatment given by \citet{Garwood_1936}

\begin{equation}
    \sigma_{P,+} = \frac{1}{2} \cdot Q_{\chi^2_{2(n+1)}}\left(p+ \frac{1 - p}{2} \right) - n,
\end{equation}

\begin{equation}
    \sigma_{P,-} = n - \frac{1}{2} \cdot Q_{\chi^2_{2n}}\left( \frac{1 - p}{2} \right),
\end{equation}  where $\sigma_{P,+}$ and $\sigma_{P,-}$ are the upper and lower uncertainties on the count $n$, $Q$ is the percent point function and $p=0.68$ is the confidence limit. Eddington bias, whereby sparsely populated bins are preferentially impacted by scatter, can contribute to the uncertainty at the bright end. We quantify this effect by resampling the flux of each source from a Gaussian centred on the true flux, with a standard deviation equal to the uncertainty, before recomputing the counts. This is repeated over 5000 iterations, and the Eddington bias term $\sigma_{E}$ is defined as the standard deviation of the resampled counts.

The third statistical uncertainty contribution is cosmic variance, arising from field-to-field variation in the large-scale structure of the Universe. Common CV estimators rely on knowledge of the total volume probed by the observations \citep{Driver_2010}, but this cannot be determined without source redshifts, which we aim to avoid. We instead utilise \scsam\ to estimate the CV semi-analytically. \citet{Yung_2022} produced $\sim1000 \ \mathrm{arcmin}^{2}$ lightcones for five fields, each with eight realisations. These realisations use the same physics model, so any differences between them result from variations in the underlying dark matter structure introduced by box tiling or stochastic elements of the model. The CV term is estimated by measuring the F277W galaxy counts from each of these forty realisations and calculating the median count in each bin. In the absence of cosmic variance, one would expect $\sim68\%$ of the individual measurements to agree with the median within $1\sigma$. We define the CV term ($\sigma'_{\mathrm{CV}}$) as the minimum additional percentage error on the counts required to achieve this consistency. The function is smoothed with a spline to allow extrapolation towards bright galaxies that are not produced by the simulation. This process is repeated for multiple square cutouts of differing area, to create a grid of $\sigma'_{\mathrm{CV}}$ as a function of both magnitude and survey area. 

\begin{figure}
    \centering
    \includegraphics[width=\columnwidth]{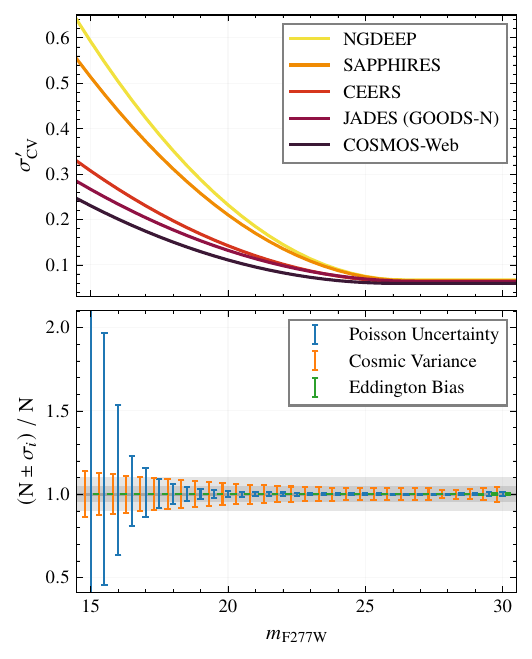}
    \caption{\emph{Top}: The relative cosmic variance as a function of F277W apparent magnitude, as measured using \scsam\ for a subset of fields spanning a representative range of survey areas. \emph{Bottom}: The contribution of Poisson (blue), cosmic variance (orange) and Eddington bias (green) terms to the total uncertainty budget of the F277W galaxy counts. Those within the dark and light grey shaded bands contribute no more than 5 and $10\%$ respectively.}
    \label{fig:count_errors}
\end{figure}

The top panel of Figure \ref{fig:count_errors} shows the relative cosmic variance as a function of apparent F277W magnitude for a sample of fields with representative areas. These evolve as expected, with the small fields more susceptible to CV at the bright end because they do not capture the rare overdense environments in which these galaxies are most likely to form. The CV converges towards a common value $\sigma'_{\mathrm{CV}} \sim 0.05$ at the faint end, which is less sensitive to these extreme density fluctuations. The total relative cosmic variance contributing to a magnitude bin $i$ is 

\begin{equation}
    \sigma_{\mathrm{CV},i}' = \frac{\sqrt{\sum_{j}A_{j}^{2}\sigma'_{\mathrm{CV},j}{}^{2}}}{\sum_{j}A_{j}} \ ,
\end{equation} where $A_{j}$ and $\sigma_{\mathrm{CV}}'$ are the sky area and relative cosmic variance associated with a survey $j$ \citep{Moster_2011}. The bottom panel shows the contribution of each source of uncertainty to the F277W counts in each bin. By considering $>30$ independent fields, the combined CV contributes an uncertainty of only $\sim10\%$ at the bright end, and quickly falls below $5\%$ at intermediate magnitudes. It does not decrease monotonically with magnitude, as the completeness of the shallower fields begins to fall below the $80\%$ threshold required to contribute at $m_{\mathrm{F277W}} > 28$. The Poisson uncertainty is significant at the bright end, but quickly falls to the same level as the Eddington bias, which is negligible at all magnitudes.

\subsection{Model Counts}

The galaxy counts predicted by the semi-empirical and semi-analytical models can be extracted straightforwardly from the lightcone catalogues, after binning on the same magnitude grid and normalising by the equivalent sky area. For \galform, we use the number counts produced by \citet{Cowley_2018} directly. The \eagle\ photometry is output at discrete redshift snapshots, so the galaxy counts cannot be calculated in the same way. Instead, we calculate the counts per unit comoving volume $V$ in each of the 28 snapshots, which span $0<z<15$ and preferentially sample the lowest redshifts, which are expected to make the most significant contribution \citep{Manzoni_2025}. This volume density is still computed as a function of apparent magnitude as observed at $z=0$, so it can be collated into a quantity $n(z)$, which describes the redshift evolution of the volume density of sources in each of the observed magnitude bins. If measured across the entire Universe, the total number of galaxies in a given bin is

\begin{equation}
    \mathcal{N} = 4\pi\int_{0}^{15}{ n(z) \times\frac{dV}{dz} \ dz},
\end{equation} which can be normalised by the total area of the sky to give the number within $1 \ \mathrm{deg}^{2}$. Repeating this for all bins produces the counts as a function of apparent magnitude.

\begin{figure*}
    \centering
    \includegraphics[width=\textwidth]{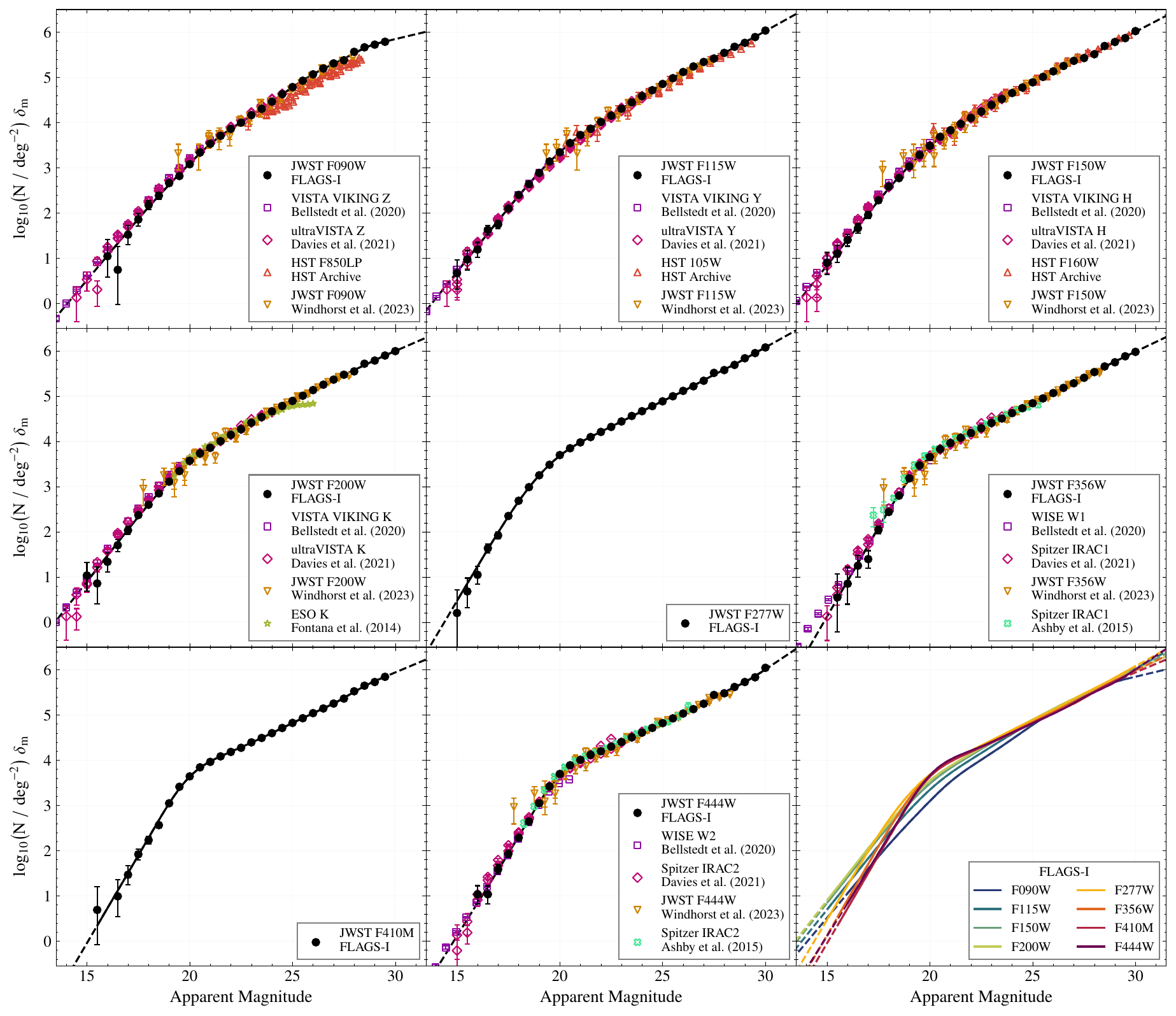}
    \caption{The completeness corrected galaxy number counts measured in each \jwst/NIRCam filter using the \flags\ dataset (black circles). A black line shows the GAM fit, with the dashed region indicating where it has been extrapolated beyond the data. The GAMs measured in each filter are compared in the bottom-right panel. The unfilled coloured points show the counts measured at similar wavelengths by \citet{Fontana_2014}, \citet{Ashby_2015}, \citet{Bellstedt_2020}, \citet{Davies_2021} and \citet{Windhorst_2023}. The HST Archive refers to a compilation of HST measurements, including those of \citet{Windhorst_2011}, \citet{Koekemoer_2013} and \citet{Rafelski_2015}. Archival data is not available at the wavelengths probed by F277W and F410M.}
    \label{fig:obs_counts}
\end{figure*}

\section{Results}\label{sec:results}

\subsection{Observations}\label{subsec:results_obs}

Figure \ref{fig:obs_counts} shows the completeness corrected counts measured in each of the eight NIRCam filters, which span $0.9-4.4 \ \mathrm{\mu m}$. These take the expected form, where the short-wavelength counts evolve following a smooth power law, with a knee at $\sim 20 \mathrm{th}$ magnitude that becomes increasingly well defined at longer wavelengths. \citet{Manzoni_2025} showed that while this knee originates in the underlying luminosity functions, its wavelength dependence in the observer frame is due to the frequency component of the K-correction. The counts continue to grow following a power law beyond the knee, where the inclusion of \ngdeep\ and the \origins\ field provides constraints at 30th magnitude in all but F090W and F410M. This makes the \flags\ compilation the deepest measurement of the galaxy number counts to date, confirming that the densities continue to follow the same faint-end power law up to at least 30th magnitude.

\renewcommand{\arraystretch}{1.65}
\begin{table*}
\centering
\begin{tabular}{cccccc}
\toprule
\textbf{Filter} & \textbf{Wavelength} & \textbf{IGL} & \textbf{eIGL} & \textbf{eIGL Peak} & \textbf{eIGL IQR}\\

 & $\mu \mathrm{m}$ & $\mathrm{nW/m^{-2}/sr^{-1}}$ & $\mathrm{nW/m^{-2}/sr^{-1}}$ & AB-mag & AB-mag \\

\midrule

F090W & 0.90 & $7.87^{+0.26}_{-0.25}$ & $9.03^{+1.48}_{-0.71}$ & $20.81^{+0.16}_{-0.16}$ & $4.87^{+0.76}_{-0.36}$\\

F115W & 1.15 & $9.76^{+0.25}_{-0.22}$ & $10.76^{+0.73}_{-0.47}$ & $20.13^{+0.15}_{-0.16}$ & $4.40^{+0.30}_{-0.18}$\\

F150W & 1.50 & $9.77^{+0.24}_{-0.22}$ & $10.74^{+0.65}_{-0.45}$ & $19.93^{+0.14}_{-0.15}$ & $4.19^{+0.25}_{-0.18}$\\

F200W & 2.00 & $8.17^{+0.22}_{-0.20}$ & $8.93^{+0.56}_{-0.41}$ & $19.80^{+0.11}_{-0.12}$ & $4.03^{+0.26}_{-0.18}$\\

F277W & 2.77 & $6.74^{+0.15}_{-0.16}$ & $6.92^{+0.17}_{-0.17}$ & $19.71^{+0.09}_{-0.09}$ & $3.20^{+0.07}_{-0.06}$\\

F356W & 3.56 & $4.38^{+0.10}_{-0.09}$ & $4.48^{+0.11}_{-0.10}$ & $19.94^{+0.07}_{-0.06}$ & $2.93^{+0.05}_{-0.05}$\\

F410M & 4.1 & $3.40^{+0.08}_{-0.08}$ & $3.46^{+0.09}_{-0.08}$ &  $20.22^{+0.07}_{-0.07}$ & $2.77^{+0.05}_{-0.05}$\\

F444W & 4.44 & $3.33^{+0.07}_{-0.07}$ & $3.47^{+0.09}_{-0.08}$ & $20.27^{+0.06}_{-0.05}$ & $2.71^{+0.05}_{-0.05}$\\

\bottomrule
\end{tabular}

\caption{The IGL measured from the galaxy number counts in each \jwst/NIRCam filter. From left to right, the columns indicate the filter code, pivot wavelength, lower bound IGL, total extrapolated IGL, magnitude at which the energy density distribution peaks, and the interquartile range of the energy density distribution. Median values are quoted alongside the 16th and 84th percentiles for each measured quantity.}
\label{tab:obs_ebl}

\end{table*}

We compare with previous studies conducted at comparable but not identical wavelengths, although no prior observations are sufficiently close to those probed by F277W and F410M. The combination of a large sky area and $>30$ independent fields ensures robust constraints at the bright end, with \cosmosweb\ greatly reducing the Poisson uncertainties in four of the filters. There is broad agreement with large ground-based surveys, and with \emph{Spitzer}/IRAC observations at long wavelengths \citep{Bellstedt_2020, Davies_2021}. The only possible tension appears closer to the knee, with the NIRCam data of \citet{Windhorst_2023} and IRAC1 observations of \citet{Ashby_2015} suggesting a shallower bright-end slope. However, this occurs at the bright limit of this data, which may be biased by cosmic variance or stellar contamination. Given the agreement between our counts and wider-area surveys, we consider them to be accurate in this regime. There is very close agreement faintwards of the knee, with only the archival \hst\ F850LP and F105W data predicting $0.1-0.2 \ \mathrm{dex}$ lower counts. This could be due to systematic differences in the source extraction approaches or to these studies overestimating completeness. Reassuringly, there is consistency with the NIRCam observations of \citet{Windhorst_2023} at all wavelengths.

A linear Generalised Additive Model (GAM) is constructed from the counts in each filter using \texttt{pyGAM} \citep{Serven_2025}, considering only the NIRCam measurements produced by this work. We fit a single spline term consisting of twenty third-order splines, which we weight by the inverse variance, including all three uncertainty terms. To maintain consistency with previous studies, the smoothing parameter is set independently for each filter to ensure 10 effective degrees of freedom \citep{Driver_2016, Koushan_2021}, but we find that values between 8 and 12 have little impact on our results. A GAM is favoured over standard spline fitting, as erratic behaviour outside the bounds of the data can be penalised, and we can enforce that the counts increase monotonically with magnitude. The GAMs are shown in their respective panels of Figure \ref{fig:obs_counts} and compared to one another in the lower right panel. The extrapolations behave reliably and agree closely with the literature beyond the bright limit of the \jwst\ data.

These splines can be used to estimate the IGL, which is given by \begin{equation}
    \mathrm{IGL} = \frac{c}{\lambda} \int_{m_{\mathrm{min}}}^{m_{\mathrm{max}}}{N_{m} \times10^{-0.4m} \ dm},
\end{equation} where $\lambda$ is the effective wavelength of the filter, $N_{m}$ is the number of galaxies per magnitude bin between $m$ and $m+\delta m$, and the factor $10^{-0.4m}$ converts these counts to an equivalent energy density. We obtain a lower limit on the IGL by integrating the GAMs over the magnitude range $\sim15-30$ covered by the NIRCam data. The total (eIGL) is estimated by extrapolating and integrating the GAM over the range $m_{\mathrm{min}}=-100$ to $m_{\mathrm{max}}=100$, which produces a result no different to using $\pm\infty$. We estimate the associated uncertainty by performing 3000 resampling iterations of each count from a Gaussian with a width defined by the combination of Poisson, CV and Eddington terms. A fourth term accounting for uncertainty in the absolute flux calibration of the NIRCam filters is included separately. We assume a conservative wavelength-independent uncertainty of $1\%$\footnote{\href{https://jwst-docs.stsci.edu/jwst-calibration-status/nircam-calibration-status/nircam-imaging-calibration-status}{https://jwst-docs.stsci.edu/jwst-calibration-status/nircam-calibration-status/nircam-imaging-calibration-status}}, which corresponds to a $0.01$ shift in magnitude or $2\%$ of the magnitude bin width. This is included in the iterative resampling by systematically perturbing all bin centres with a value drawn from a Gaussian distribution with mean zero and standard deviation equal to this value. A new GAM is constructed for each of the 3000 iterations, and we report the median IGL properties with uncertainties based on the 16th and 84th percentiles in Table \ref{tab:obs_ebl}. 

\begin{figure*}
    \centering
    \includegraphics[width=\textwidth]{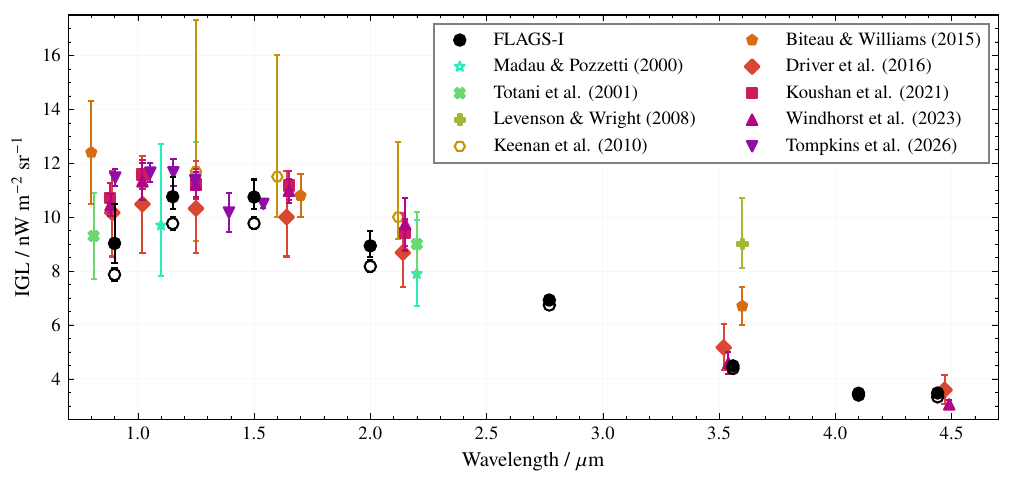}
    \caption{The total extrapolated eIGL (black circles) and lower limits (unfilled circles) measured by integrating the GAM constructed from \jwst/NIRCam galaxy counts over a $-100<m_{\mathrm{AB}}<100$ range. Points and error bars show the median and 16th-84th percentiles respectively, determined from 3000 resampling iterations. Other lower limits from \citet{Madau_2000} and \citet{Keenan_2010} are shown as unfilled points. Additional measurements of the extrapolated IGL from \citet{Totani_2001}, \citet{Levenson_2008}, \citet{Driver_2016}, \citet{Koushan_2021}, \citet{Windhorst_2023} and \citet{Tompkins_2026} are shown as filled points. The VHE measurement of \citet{Biteau_2015} is also shown.}
    \label{fig:obs_ebl}
\end{figure*}

These measurements are also shown in Figure \ref{fig:obs_ebl}, with the lower limits and extrapolated total denoted by unfilled and filled circles respectively. There is little difference between the IGL and eIGL measurements at $\lambda>2.5 \ \mathrm{\mu m}$, which can be attributed to narrower energy density distribution peaks, with fewer galaxies outside the magnitude range directly probed by these observations. This narrowness is quantified by an interquartile range eIGL IQR, which is simply the magnitude range enclosing the central $50\%$ of the energy density distribution. At $\lambda=0.90 \ \mathrm{\mu m}$, the eIGL IQR is $4.9 \ \mathrm{mag}$, but decreases with increasing wavelength to as low as $2.7 \ \mathrm{mag}$. As such, the difference between the lower limit and the total eIGL increases from $4\%$ at $4.44 \ \mathrm{\mu m}$, to $13\%$ at $0.9 \ \mathrm{\mu m}$. \citet{Windhorst_2023} posit that this can arise from early-type galaxies at $1<z<2$ dominating the counts at long wavelengths, which probe the peak of their old metal-enriched stellar emission at $\lambda_{\mathrm{rest}}\sim1.25 \ \mathrm{\mu m}$. This interpretation was corroborated by \citep{Manzoni_2025}, who showed that the contribution of bulge-dominated galaxies to the $m_{\mathrm{F444W}}<23 \ \mathrm{mag}$ \galform\ counts peaks strongly at $z=1.5$. Disk-dominated galaxies span a broader range of redshifts and stellar ages, resulting in a larger, more varied contribution at shorter wavelengths. These FWHMs can therefore be considered a prediction of galaxy evolution models.

Similarly, the peak magnitude of the energy density distribution encodes information about the underlying galaxy population. The wavelength-dependent values are also listed in Table \ref{tab:obs_ebl} and brighten from $20.81 \ \mathrm{mag}$ in F090W to $19.71 \ \mathrm{mag}$ in F277W, before dimming again towards longer wavelengths. This is another potentially valuable model constraint, as the peak will shift brighter or fainter if the photon budget at that wavelength is dominated by a handful of evolved massive galaxies or many low-mass galaxies respectively. The physics of structure formation and gas accretion are therefore encoded and can be probed through model comparisons.

\begin{figure*}
    \centering
    \includegraphics[width=\textwidth]{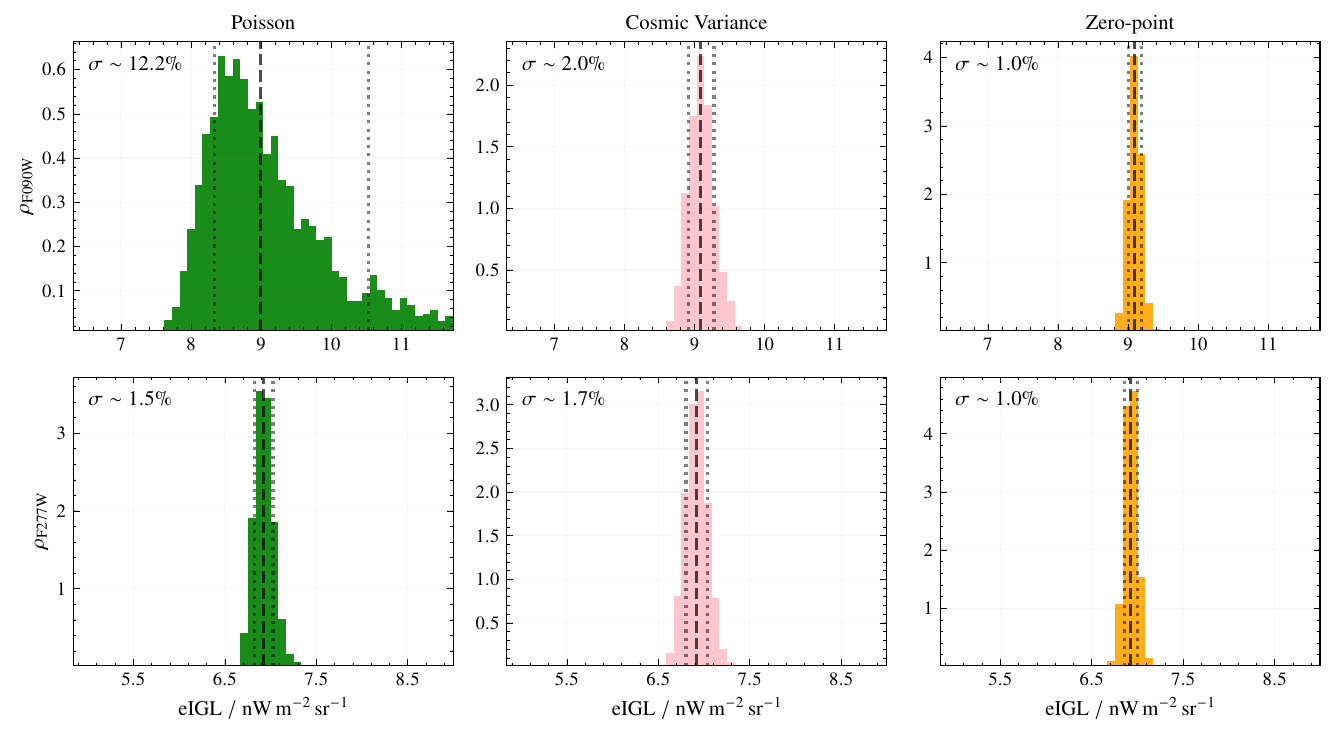}
    \caption{Distribution of eIGL values recovered when resampling the galaxy number counts from their Poisson (green), CV (pink) and zero-point (orange) uncertainty terms individually. Each panel shows the density of sources in 50 bins spanning a $30\%$ uncertainty range, where the dashed and dotted lines indicate the median and 16th-84th percentile range respectively. The corresponding approximate percentage uncertainty is shown in the top left of each panel. The results for F090W and F277W are shown on the top and bottom rows respectively.}
    \label{fig:ebl_errors}
\end{figure*}

The differing reliance on extrapolation introduced by the wavelength-dependent IQR influences the confidence with which the total eIGL can be measured at a given wavelength. The percentage uncertainty exceeds $16\%$ at $0.9 \ \mathrm{\mu m}$, but this quickly drops to $\sim7\%$ in the remaining filters at $<2 \ \mathrm{\mu m}$ and to $\sim2.5\%$ at longer wavelengths. However, the exaggerated uncertainty at $0.9 \ \mathrm{\mu m}$ is primarily driven by increased Poisson and, to a lesser extent, CV uncertainties, as only five of the fields include F090W imaging. This is demonstrated in Figure \ref{fig:ebl_errors}, with each panel showing the distribution of eIGL values recovered when resampling the counts from each uncertainty distribution individually. This is again performed over 3000 iterations, where we find that as suggested by Figure \ref{fig:count_errors}, the Eddington bias has a negligible impact and is thus omitted from the figure. Looking at F090W in the top panel, while CV contributes an uncertainty of only $2\%$, the Poisson term exceeds $12\%$. The bottom row shows the same distributions constructed from the F277W counts, which demonstrates the true benefit of \cosmosweb\ and \panoramic\ to reduce the Poisson term to only $\sim1.5\%$.

Regardless of this limitation in F090W, the longer-wavelength measurements are very competitive in the context of the broader literature, with previous extrapolated IGL studies achieving a total uncertainty of $\sim4\%$ \citep{Koushan_2021, Windhorst_2023}. These new measurements can therefore be used to inform very high energy (VHE) measurements, which harness attenuation of $\gamma$-rays emitted by distant blazars to determine the normalisation of a predefined EBL model \citep{MAGIC-Collaboration_2018, Hess_2013, Desai_2017}. The eIGL uncertainties could be reduced further with precise filter-dependent estimates of the zero-point uncertainties, which will be lower than the conservative value adopted here.

NIRCam probes the peak of the cosmic optical background, which the eIGL measurements demonstrate clearly at $\sim1.3 \ \mathrm{\mu m}$, before declining quickly at longer wavelengths. This general trend is in good agreement with prior data employing a range of methods, sourced from individual studies and the compilation of \citet{Dwek_2013}. We omit direct estimates, which aim to measure the combination of the IGL and the unresolved light which make up the extragalactic background light (EBL). These estimates have been shown to predict an EBL that is more than twice as bright as the IGL, which \citet{Driver_2016} attribute to uncertainties in the removal of Galactic foreground contamination. These measurements also cannot be compared to model predictions without full image-based forward modelling, so they are not relevant to this work. 

Our results are consistent with most of the previous studies within $1\sigma$, including the most recent IGL extrapolations of \citet{Driver_2016}, \citet{Koushan_2021}, \citet{Windhorst_2023} and \citet{Tompkins_2026}. These studies combine data from multiple observatories and apply colour corrections to the magnitudes or binned counts to shift them into a common reference filter. This requires a characteristic spectral shape to be assumed, which introduces a high degree of uncertainty. A range of populations contribute to the counts, each with different age, metallicity and dust distributions, and the effect of nebular emission is highly redshift-dependent \citep{Wilkins_2014}. In addition, if source extraction is not performed consistently across these datasets, there may be systematic shifts between them. It is therefore unlikely that this colour correction can be applied without introducing bias, which is why our results are based solely on the NIRCam data. It is reassuring that these are consistent with the ensemble studies, but it is also worth noting that our median values are systematically below those of the three most recent studies at $\le 2 \ \mathrm{\mu m}$. \citet{Koushan_2021} posit that their higher IGL values relative to \citet{Driver_2016} are a result of the \texttt{ProFound} source extraction tool collecting a larger fraction of the flux from each galaxy. While its curve of growth dilation approach is likely superior to fixed apertures, we expect the aperture correction to account for this in all but the most extended galaxies. \emph{Euclid} and \emph{Roman} surveys may also reach sufficient depths to constrain the bright and faint ends without ancillary data, which can help resolve this without the colour correction caveat \citep{Euclid-Collaboration_2025}.

The most significant discrepancies occur at $3.56 \ \mathrm{\mu m}$, with the \citet{Levenson_2008} eIGL measurement in tension at the $4\sigma$ level. Their approach differs in that they employ profile fitting to extract source photometry. They find that this overestimates the flux of synthetic sources, which they attempt to rectify while fitting a broken power law to the overestimated counts. This makes the results highly model-dependent, and comparing instead to the completeness-corrected aperture fluxes reduces the tension to $\sim2\sigma$, although they find these to be underestimated. The most likely explanation for the tension is an underestimation of the extrapolation uncertainty, as the \emph{Spitzer}/IRAC data they use is only $80\%$ complete at $\sim 19\mathrm{th}$ magnitude. Figure \ref{fig:obs_counts} shows that this is still brighter than the knee, so the gradient of their faint-end power law fit is heavily dependent on completeness corrections, which exceed $60\%$.

\citet{Biteau_2015} employ a novel VHE approach, in which both the normalisation and wavelength dependence of the EBL are measured after making a set of assumptions about its redshift evolution. Unlike direct galaxy-count measurements, VHE constraints probe the total EBL through photon pair-production attenuation of blazar emission \citep{Gould_1997}. They are therefore sensitive to any light that crosses the line of sight, including any diffuse component that is not contained within the resolved galaxies. This measurement may therefore represent something closer to the total EBL, but the diffuse component would need to contribute $\sim50\%$ for the $2\sigma$ tension with our result to be resolved. This seems unlikely, given that the expectation is $<20\%$ \citep{Driver_2016}. While a full exploration of VHE uncertainties is beyond the scope of this work, it should be noted that large systematics arise from assumptions about the intrinsic blazar spectra, intervening magnetic fields, and the redshift evolution of the EBL. Considering this, along with the close agreement with other eIGL measurements at this wavelength and the smooth evolution between adjacent wavelengths, we do not consider the tension with \citet{Biteau_2015} indicative of an issue with our measurement.

These comparisons suggest that our \jwst/NIRCam galaxy number counts and the derived IGL measurements are reliable. As such, they can be used to determine whether these quantities can, in general, distinguish between models and constrain the underlying physics.

\subsection{Model Comparisons}\label{subsec:results_models}

\begin{figure}
    \centering
    \includegraphics[width=\columnwidth]{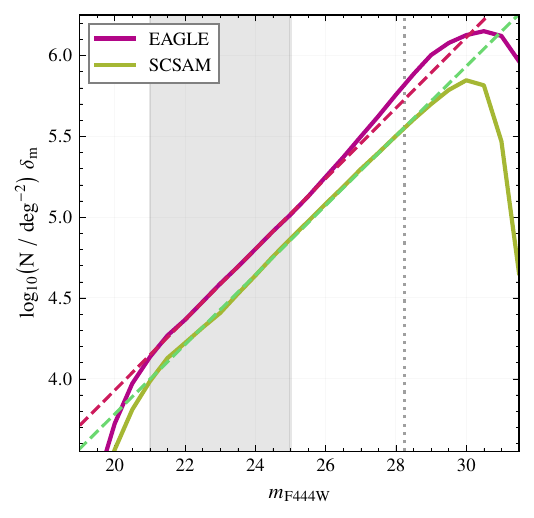}
    \caption{The power-law fits used to determine the faint-end limit of the model comparisons. The model-predicted counts are shown by solid lines, with the fits over the magnitude range $21<m_{\mathrm{AB}}<25$ (shaded region) overlaid as dashed lines. The derived faint-magnitude limit is indicated by the dotted grey line.}
    \label{fig:faint_limit}
\end{figure}

Low-mass galaxies constitute a larger fraction of the faint-end counts as they are intuitively fainter than more massive galaxies at the same redshift. They also constitute a larger fraction of the galaxy population at high redshift, which has not had sufficient time to assemble large stellar masses and will naturally appear fainter than low-redshift analogues due to their distance. The models introduced in Section \ref{subsec:modeldata} have differing mass resolutions, either due to different particle masses in the underlying simulations or to limitations of the observations used to construct them. This restricts the range over which their predictions can be evaluated, and potentially introduces incompleteness bias when comparing models at the faint end. The $M_{\ast}>10^{7} \ \mathrm{M_{\odot}}$ resolution of \scsam\ and \eagle\ is the lowest, so these will be the most affected by incompleteness. To determine the affected magnitudes, we fit a simple power law to the F444W \scsam\ counts from 21st to 25th magnitude, where \citet{Manzoni_2025} showed that $M_{\ast}<10^{7} \ \mathrm{M_{\odot}}$ galaxies make up no more than $0.1\%$ of the \galform\ counts. We then extrapolate this parametrisation to fainter magnitudes and consider the counts to be incomplete when they deviate from the fit by $>1\%$. As shown in Figure \ref{fig:faint_limit}, this occurs at magnitude $28.5$, so the $28\mathrm{th}$ magnitude bin is the faintest used for further comparisons. The \eagle\ gradient increases outside of the fit range, so we favour this \scsam\ result rather than the 30th magnitude limit that it suggests. Ideally, the models would probe comparable volumes, but we impose a limit of $m_{\mathrm{AB}}>18$ at the bright end to limit the effect of cosmic variance.

\begin{figure*}
    \centering
    \includegraphics[width=\textwidth]{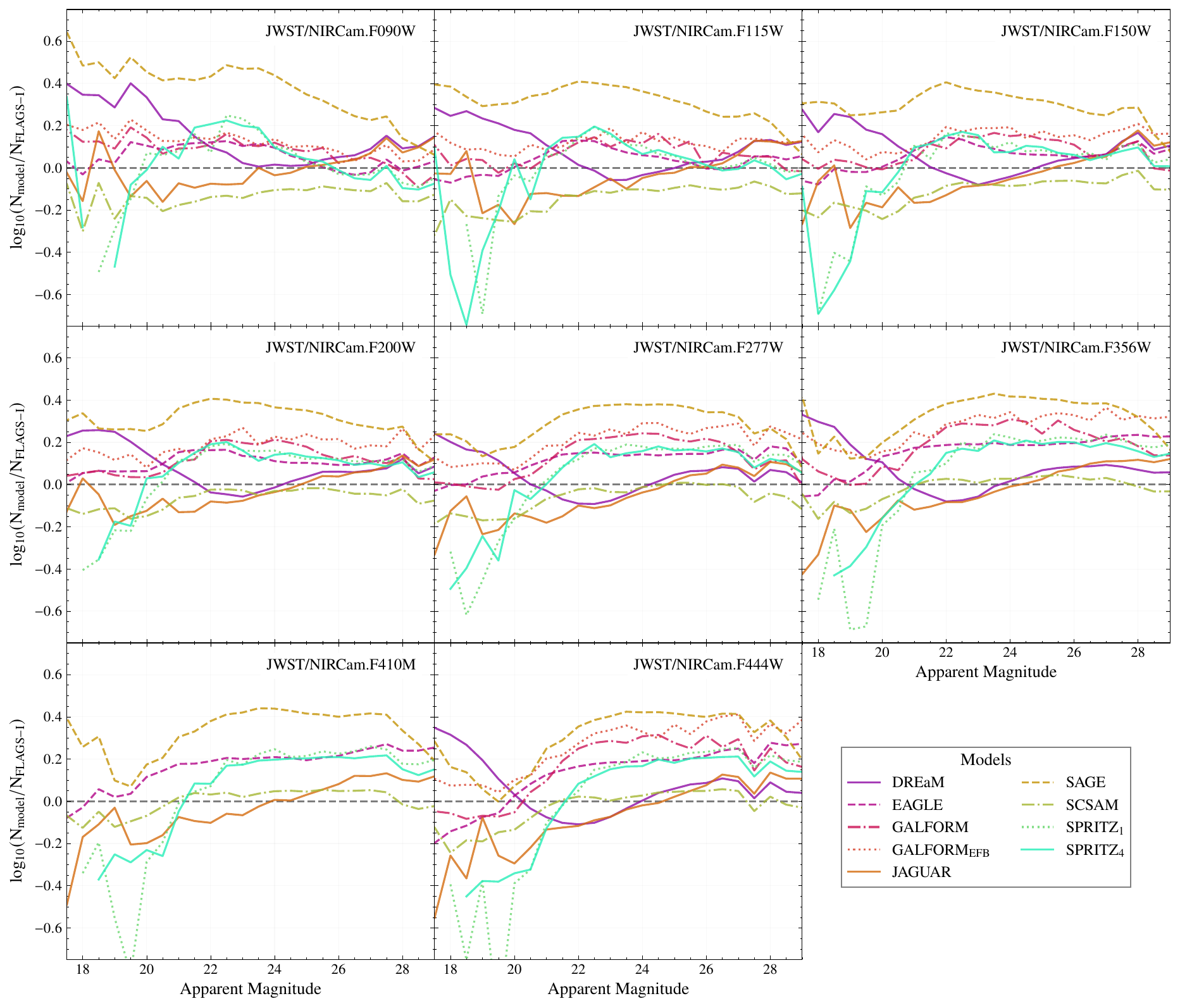}
    \caption{The ratio between the observed counts and those predicted by the \dream, \eagle, \galform, \jaguar, \sage, \scsam\ and \spritz\ models in each of the eight NIRCam filters. The grey-shaded regions indicate the observational uncertainty bounds.}
    \label{fig:model_counts}
\end{figure*}

Figure \ref{fig:model_counts} shows the ratio between model and observed galaxy counts across this magnitude range. There are clear systematic offsets, with models generally overpredicting the number of galaxies across the magnitude distribution, but particularly at the faint end in long-wavelength filters. \sage\ demonstrates the most significant discrepancy, consistently predicting a $0.3-0.4 \ \mathrm{dex}$ excess, which rises as high as $0.6 \ \mathrm{dex}$ at the brightest magnitudes. In contrast, while \scsam\ underpredicts the counts at short wavelengths, it recovers those at $>2 \ \mathrm{\mu m}$ remarkably well. 

As discussed in Section \ref{subsec:modeldata}, the \scsam\ and \sage\ lightcones are based on the same underlying DMO simulation and assume the same SPS model and IMF when generating photometry. While both dust attenuation models are dependent on the gas metallicity and density, \sage\ applies two additional scaling factors to the optical depth. These reduce the attenuation of galaxies by a factor $(1-\omega_{\lambda})^{1/2}(1+z)^{-1/2}$, where $\omega_{\lambda}$ is the albedo. We can therefore expect the $A_{\mathrm{V}}$ of a \sage\ galaxy at $z=1$ to be $30\%$ lower than an analogue in \scsam, even when assuming $\omega_{\lambda}=0$. However, a lack of improvement at long wavelengths, where the effect of dust attenuation is less significant, suggests that this is not the primary driver of the discrepancy. Given the consistency with which the models are applied and forward modelled, the observed differences must manifest from physical processes.

\begin{figure}
    \centering
    \includegraphics[width=\columnwidth]{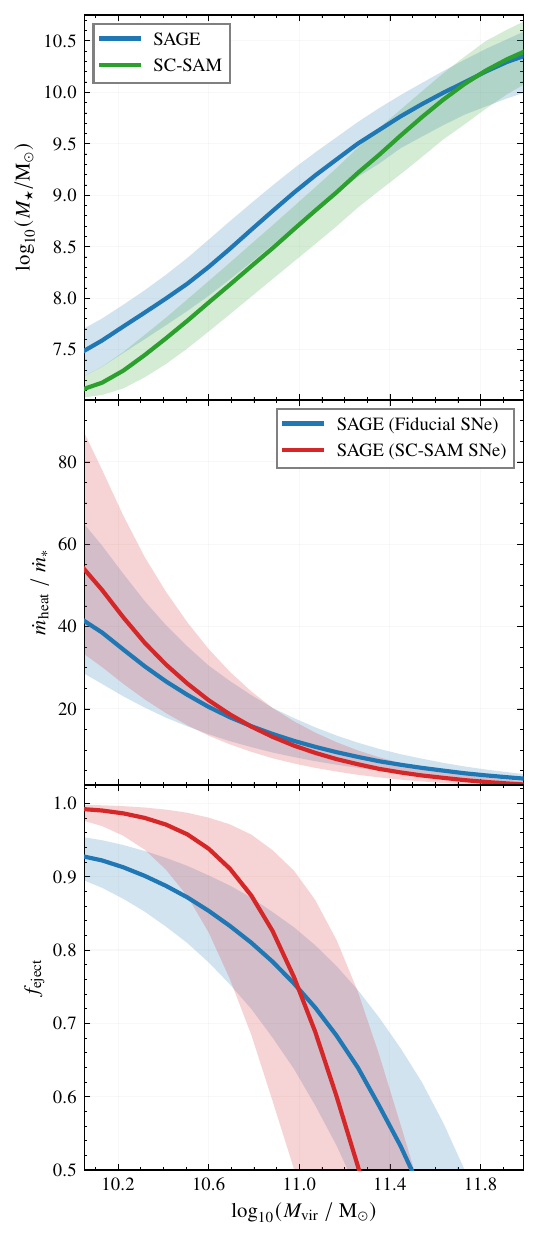}
    \caption{The running median (solid lines) and 16th and 84th percentiles (shaded regions) of quantities as a function of dark matter halo virial mass. \emph{Top}: The stellar-halo mass relation predicted by \sage\ (blue) and \scsam\ (green). \emph{Middle}: The supernova feedback loading factor calculated from \sage\ halo properties using the fiducial (blue) and \scsam\ prescriptions (red). \emph{Bottom}: The fraction of gas heated by SNe feedback that is also expelled into the IGM, again for the fiducial \sage\ and \scsam\ prescriptions.}
    \label{fig:feedback}
\end{figure}

While observationally inferred physical properties should be avoided when evaluating model performance, we can use those predicted by the models to identify areas for improvement. The top panel of Figure \ref{fig:feedback} shows the running median relationship between stellar mass and halo virial mass for central galaxies in \scsam\ and \sage. While consistent at $M_{\mathrm{vir}} \gtrsim 10^{11.6} \ \mathrm{M_{\odot}}$, \sage\ predicts systematically larger stellar-halo mass ratios at lower halo masses, with the discrepancy exceeding $0.4 \ \mathrm{dex}$ at $M_{\mathrm{vir}} \lesssim 10^{10.8} \ \mathrm{M_{\odot}}$. \scsam\ predicts $<0.03\%$ of the galaxies in this regime to have mass ratios $>0.01$, whereas \sage\ predicts $>4\%$, including some with ratio $>0.1$. Despite the associated uncertainties, it should be noted that such galaxies have not been observed in abundance \citep{Zu_2015, Zaritsky_2023}. It appears that galaxies in low-mass \sage\ halos accumulate stellar mass too quickly.

The assumed baryon fraction, cooling prescriptions \citep{White_1991} and photoionisation squelching \citep{Gnedin_2000} can all affect the rate at which stars form. However, these are implemented largely consistently, although squelching is actually marginally less effective in \scsam. \sage\ galaxies form stars when the gas mass exceeds a critical value \citep{Kauffmann_1996}, and then proceed following a Kennicutt-Schmidt-like (KS) relation \citep{Kennicutt_1998}. \scsam\ adopts a more complex `two-slope' model dependent on the abundance of molecular hydrogen, which effectively reduces the fraction of gas available for star formation. The effect of this cannot be evaluated self-consistently using the lightcone catalogues, but \citet{Somerville_2015_scsam} showed that a KS model produced only marginally higher stellar-halo mass ratios in low-mass halos, when compared to molecular hydrogen-based recipes under the same conditions. Merger-triggered star formation is implemented consistently by both models \citep{Somerville_2008}.

\begin{figure*}
    \centering
    \includegraphics[width=\textwidth]{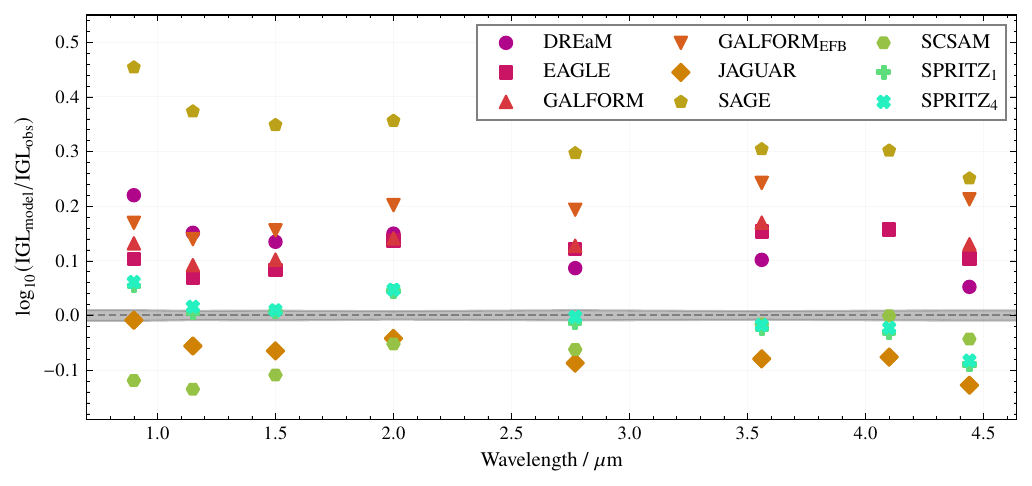}
    \caption{The ratio of the model-predicted and observed IGL calculated by summing the binwise energy densities over the \mbox{$18.0<m_{\mathrm{AB}}<28.5$} magnitude range. The grey shaded region denotes the observational uncertainties.}
    \label{fig:model_ebl}
\end{figure*}

\renewcommand{\arraystretch}{1.2}
\begin{table*}
    \centering
    \begin{tabular}{c c c c c c c c c c c}
        \toprule
        \multirow{2}{*}{\textbf{Model}} & \multicolumn{9}{c}{\textbf{Galaxy Counts $\chi^{2}_{\nu}$ ($18.0<m_{\mathrm{AB}}<28.5$)}} & \multirow{2}{*}{\textbf{Rank}} \\
        \cmidrule(lr){2-10}
        & \textbf{F090W} & \textbf{F115W} & \textbf{F150W} & \textbf{F200W} & \textbf{F277W} & \textbf{F356W} & \textbf{F410M} & \textbf{F444W} & \textbf{Total}\\
        \midrule
        DREaM  & 30.9 & 27.5 & 26.8 & 30.8 & 14.5 & 34.6 & -- & 20.7 & 26.6 & 2 (7)\\
        EAGLE & 12.8 & 11.5 & 22.3 & 75.2 & 71.5 & 238.3 & 169.7 & 125.4 & 90.8 & 4 (5)\\
        GALFORM  & 20.1 & 24.5 & 38.0 & 122.6 & 96.8 & 346.1 & -- & 185.6 & 119.1 & 6 (6)\\
        GALFORM$_{\mathrm{EFB}}$ & 31.1 & 50.4 & 81.4 & 231.4 & 198.0 & 572.1 & -- & 317.6 & 211.7 & 8 (8)\\
        JAGUAR & 13.9 & 28.0 & 32.0 & 31.9 & 27.2 & 47.8 & 31.0 & 35.2 & 30.9 & 3 (3)\\
        SAGE & 334.1 & 288.5 & 275.3 & 573.0 & 321.6 & 825.9 & 557.4 & 410.9 & 448.3 & 9 (9)\\
        SCSAM & 38.5 & 39.3 & 23.6 & 14.2 & 9.1 & 7.5 & 7.7 & 8.7 & 18.6 & 1 (4)\\
        SPRITZ$_{1}$ & 33.0 & 34.5 & 44.6 & 100.2 & 96.2 & 291.7 & 202.4 & 189.9 & 125.1 & 7 (2) \\
        SPRITZ$_{4}$ & 36.0 & 41.8 & 47.0 & 92.0 & 75.8 & 209.4 & 142.1 & 113.9 & 94.1 & 5 (1)\\
        \midrule
        \textbf{Mean} & 61.2 & 60.7 & 65.7 & 141.3 & 101.2 & 285.9 & 185.0 & 156.4 & 129.5\\
        
        \bottomrule
    \end{tabular}
    \caption{The $\chi^{2}_{\nu}$ computed by comparing the observed and model-predicted binwise galaxy counts over the range $18.0<m_{\mathrm{AB}}<28.5$. Columns show either the value computed in the corresponding band only or the total computed across all filters simultaneously. The last row shows the mean $\chi^{2}_{\nu}$ across models. The last column shows the model ranking based on the total $\chi^{2}_{\nu}$, with the ranking based on the combined IGL $\chi^{2}_{\nu}$ shown in brackets.}
    \label{tab:model_counts}
\end{table*}

These considerations make feedback the most likely cause of this overly efficient stellar mass growth. AGN feedback is unlikely to be influential in this halo mass regime, as the black holes are yet to grow sufficiently \citep{Booth_2013}, leaving stellar feedback from supernovae (SNe) as the focus of further analysis. Both models allow this feedback to heat the surrounding gas and to expel a fraction of it into an external reservoir that represents the IGM. The mass of gas heated by SNe feedback is dependent on the dark matter halo virial velocity in \sage\ \citep[see][section 8]{Croton_2016}, and the maximum circular velocity in \scsam\ \citep[see][section 2.7 and table 3 respectively]{Somerville_2008, Yung_2022}, as well as the star formation rate and several free parameters. As these are independent of galaxy properties, we can compare their effects somewhat self-consistently, without rerunning the models. We therefore evaluate both prescriptions using the halo properties of \sage\ central galaxies as they appear in the lightcone. The middle panel of Figure \ref{fig:feedback} shows the median mass of gas heated by SNe feedback per unit stellar mass formed, referred to as the loading factor $\eta=\dot{m}_{\mathrm{heat}}/\dot{m}_{\ast}$. At low halo masses, the \sage\ feedback is up to $\sim50\%$ weaker, thereby allowing stellar mass to be assembled more rapidly. The bottom panel shows the fraction of this heated gas that is also expelled into the IGM component. \scsam\ expels almost $100\%$ at the lowest halo masses, representing their shallow potential wells, and consistently expels a larger fraction than \sage\ until $M_{\mathrm{vir}} \gtrsim 10^{11} \ \mathrm{M_{\odot}}$. This gas must be reaccreted before it can begin cooling, which slows star formation over an extended timescale. Inefficient heating and gas expulsion in low-mass \sage\ halos lead to excessive stellar mass growth, thereby inflating galaxy number counts across all magnitudes. This could be confirmed by rerunning the \sage\ model with the \scsam\ prescription self-consistently implemented.

Turning again to Figure \ref{fig:model_counts}, we can use \galform\ as an example of a self-consistent feedback comparison. The fiducial model consistently produces $0.1-0.2 \ \mathrm{dex}$ fewer galaxies than the evolving feedback model. This brings it into closer agreement with the observations, albeit while still over-predicting the number of galaxies. The fiducial model enforces a power-law connection between the SNe loading factor and the halo circular velocity $V_{\mathrm{c}}$, leading to highly efficient feedback ($\eta>10^{5}$) in low-mass halos. The evolving model not only lowers the normalisation of this power law as a function of redshift, but introduces a turn off at $V_{\mathrm{c}}<50 \ \mathrm{km \ s^{-1}}$ that reduces feedback strength even further, preferentially so in low-mass halos. \citet{Hou_2016} favour this model based on its more accurate prediction of the $z>7$ rest-frame UV luminosity functions, while maintaining agreement with optical and NIR measurements at $z=0$, and it has been adopted by more recent \galform\ studies \citep{Lu_2025}. However, like in \sage, the weaker feedback produces too many bright galaxies at the redshifts probed by the galaxy counts. The direct observable approach favours the more efficient feedback produced by the fiducial model.

\begin{figure*}
    \centering
    \includegraphics[width=\textwidth]{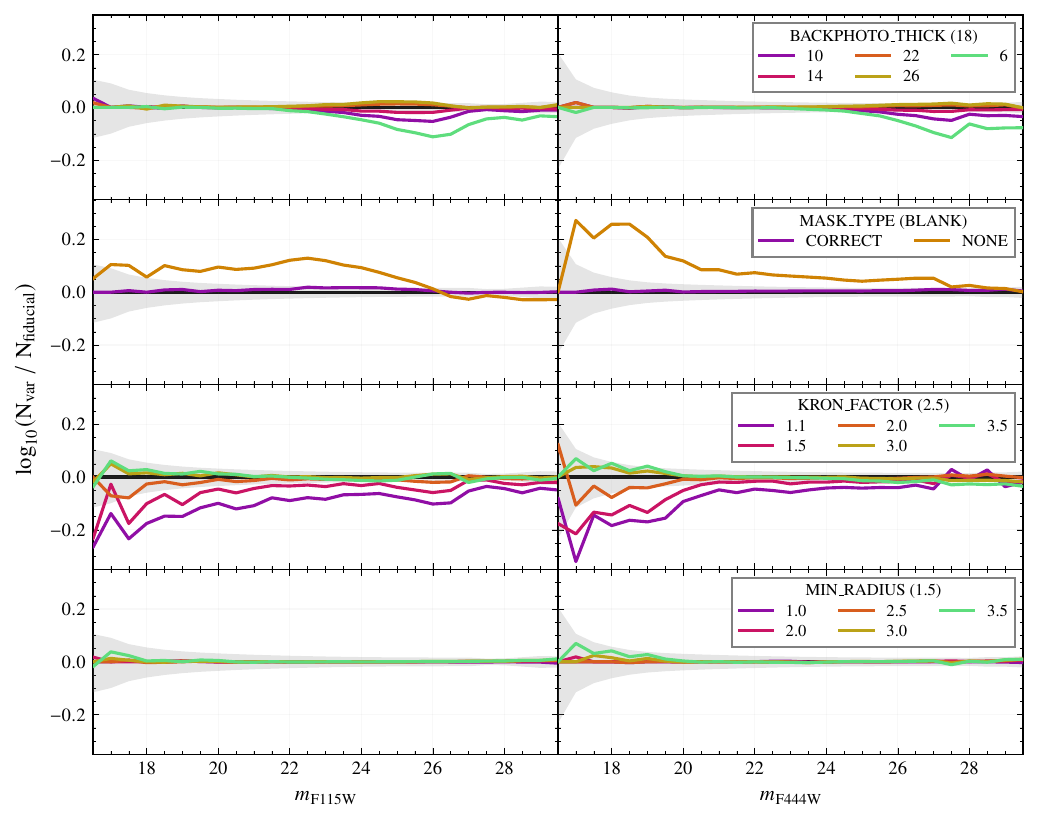}
    \caption{The effect of varying SE parameters governing photometry on the measured galaxy number counts. Each row shows a different parameter, with the left and right columns showing the F115W and F444W counts respectively. The coloured lines show the shift introduced by each parameter variation, and the fiducial value is shown in brackets. Shaded regions indicate the $1\sigma$ observational uncertainties.}
    \label{fig:param_variation}
\end{figure*}

Table \ref{tab:model_counts} shows the total binwise $\chi^{2}_{\nu}$ of each model, as well as those determined from individual filters, computed over the $18.0<m_{\mathrm{AB}}<28.5$ range. Consistent with the previous discussion, \scsam\ is the best performing model overall with $\chi_{\nu}^{2}=18.6$, whereas \sage\ performs most poorly with $\chi_{\nu}^{2}=448.3$. The fiducial \galform\ model performs better than the evolving model in all filters. While semi-empirical models cannot be used to interpret the physical mechanisms governing galaxy evolution, they can help demonstrate this constraining power.  While \dream\ and \jaguar\ perform at comparable levels of $\chi^{2}_{\nu} = 26.6$ and $30.9$ respectively, these can be clearly distinguished from \spritz. Both \spritz\ implementations perform more poorly, but the $k_{\phi}=-4$ model outperforms $k_{\phi}=-1$ at all wavelengths. \eagle\ and the semi-analytical models generally perform best at short wavelengths. This can be attributed to calibrations to the $z=0$ GSMF function, which is primarily inferred observationally from UV-optical data. The declining performance at longer wavelengths could be attributed to increasing uncertainty in stellar modelling \citep{Byrne_2023}, but this seems unlikely given the reversed trend in \scsam. If galaxies do indeed evolve too quickly in the other models, due to some combination of efficient star formation or inefficient feedback, older, more metal-enriched galaxies will constitute a larger fraction of the population. These galaxies will be redder, thereby preferentially inflating the long-wavelength counts. These results highlight that feedback physics remains uncertain, but that the galaxy number counts are capable of differentiating between model implementations. There is therefore potential for application to simulation suites that explore systematic variations of this physics, such as \textsc{camels} \citep{Villaescusa-Navarro_2021}.

We now seek to determine whether the same conclusions can be drawn from the IGL and show the ratio between observed and model-predicted IGL computed over the same $18.0<m_{\mathrm{AB}}<28.5$ magnitude range in Figure \ref{fig:model_ebl}. This is computed by summing the binwise energy densities, rather than integrating a GAM, as the GAM model struggled to fit the \spritz\ counts. Observational uncertainties are therefore determined by resampling the counts within their uncertainties and summing the new energy densities over $3000$ iterations. Unsurprisingly, \sage\ overpredicts the IGL as it does the number counts, and the fiducial \galform\ model predicts a lower energy density than the evolving feedback model. However, there is a significant discrepancy when compared to Figure \ref{fig:model_counts}. \spritz$_{4}$ and \spritz$_{1}$ appear to be the best predictors of the IGL, despite the binwise measurements ranking them 5th and 7th respectively. Unlike \sage\ and \galform, which overpredict the number of galaxies at all magnitudes, \spritz\ predicts too many faint galaxies, but too few bright galaxies. These effectively cancel each other out, yielding an IGL that appears to agree exceptionally well with the observations. This demonstrates that, while the IGL is an interesting quantity in its own right, it is an unreliable indicator of model performance. This should instead be quantified using the galaxy counts directly where available. This is not necessarily a detriment, as comparisons can be made binwise as in this work, rather than relying on smoothing and extrapolation.

\begin{figure*}
    \centering
    \includegraphics[width=\textwidth]{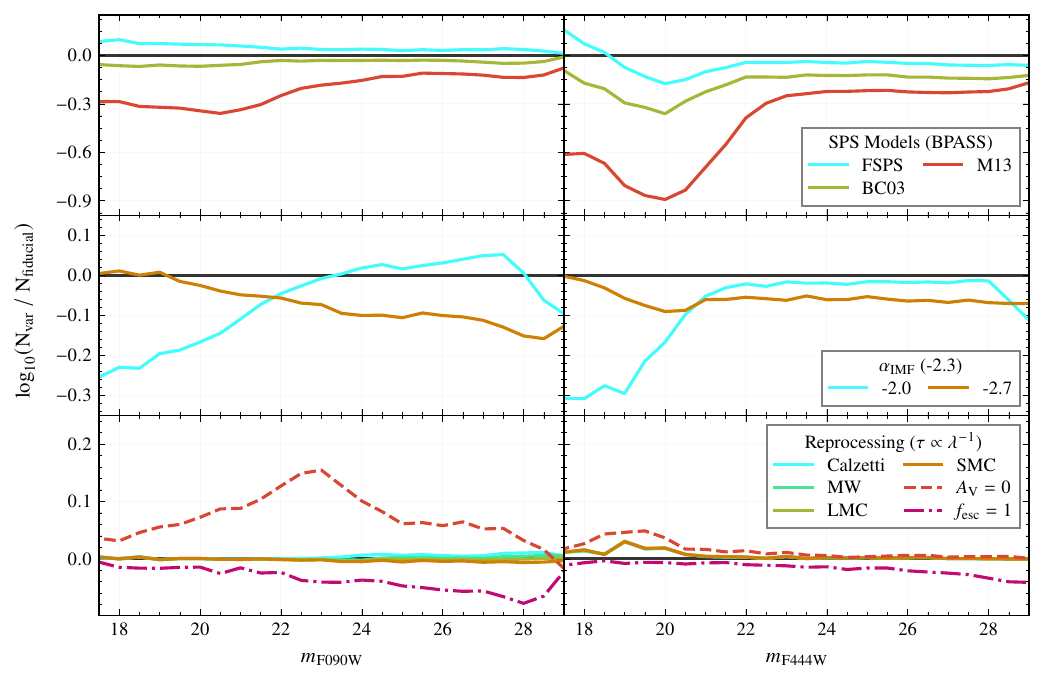}
    \caption{The effect of varying forward modelling assumptions on the galaxy number counts predicted by \eagle\ in F090W (left) and F444W (right). The top row shows the effect of stellar modelling, the middle row the specific effect of the IMF, and the bottom row shows the effect of additional reprocessing. The coloured lines show the shift introduced by each variation, with the fiducial value in brackets.}
    \label{fig:sim_variation}
\end{figure*}

\subsection{Systematics}\label{subsec:systematics}

While these constraints are insensitive to the biases introduced by SED fitting, source extraction assumptions could still introduce systematic shifts in the observed number counts. Unlike the random effect or Poisson noise or the systematic offset of zero-point errors, it is not straightforward to include this in the quoted uncertainties. There is a range of somewhat reasonable options for each SE parameter that could be sampled to estimate the associated uncertainty, but this is not well defined. This is also complicated by the coarser pixel scale of the \panoramic\ imaging, which prevents consistent parameter scaling in some cases. In addition, the fiducial values listed in Table \ref{tab:fiducial} are chosen based on rigorous visual inspection of segmentation maps, so they are believed to most accurately recover the number of galaxies based on this evidence. Randomly sampling from an arbitrary range of parameters may therefore misrepresent the uncertainty. As a compromise, we generate a new set of catalogues by varying the parameters that directly influence photometry, after omitting \panoramic\ to ensure consistent pixel-based values. These parameters cannot be as easily optimised by eye, so it is reasonable to consider them uncertain. 

Figure \ref{fig:param_variation} shows the ratio between the fiducial counts and those determined from each new set of catalogues for both F115W and F444W. Decreasing the radius of the annulus used to correct for local residual background can reduce the faint-end counts by up to $0.1 \ \mathrm{dex}$. However, these annuli are so thin that they are likely biased by the diffuse source flux, thereby overestimating the correction. More reasonable values shift the counts by no more than $0.05 \ \mathrm{dex}$, and we consider this a more accurate representation of the associated uncertainty. When summing the flux within the Kron apertures, pixels assigned to other objects can be included, set to zero, or corrected to cancel out local noise. The latter two options produce no significant change in the counts, so this masking does not contribute any uncertainty. However, it should be noted that if nearby sources are not masked, as was the case in early \texttt{SEP} versions, the fluxes can be systematically overestimated. This preferentially affects longer wavelengths, where sources are more likely to be blended. 

We do not consider the reduced counts recovered when applying a more conservative scaling to the Kron aperture to be an indicator of systematic uncertainty. In fact, the convergence towards consistent counts as the factor is increased demonstrates that our apertures capture close to the total flux and that those at $m_{\mathrm{AB}}>20$ are insensitive to any that is missed. The fluxes of bright extended galaxies may be somewhat underestimated, but any shift is within the uncertainties. Furthermore, a lack of clear wavelength dependence demonstrates that the aperture correction is working as intended. The minimum radius is the smallest pixel radius of the Kron aperture used for photometry. Sources measured to be more compact have their aperture fixed at this size, to minimise the impact of noise. Increasing this floor produces deviations of $\sim0.05-0.1 \ \mathrm{dex}$ in the brightest bins, potentially indicating underestimation of the flux from compact AGN, although this is still within the statistical uncertainty.

Overall, systematic uncertainties in photometry propagate to a shift in counts of $\lesssim0.1 \ \mathrm{dex}$, and we therefore consider them robust and conclude that these systematics cannot resolve the model offsets. It is important to note that this is dependent on the imaging resolution and depth, and that other instruments may be more sensitive.

These uncertainties affect our understanding of the true galaxy number counts, but forward-modelling systematics can impact model predictions. These are slightly different, as they can be considered an element of the wider model, but they nonetheless influence how well we deem the physical mechanisms to represent the real Universe. The effects cannot be quantified for all models without the raw data, so we use \eagle\ as an example in Figure \ref{fig:sim_variation}. This shows the shift from the fiducial counts constructed using the \texttt{BPASS} stellar models, a \citet{Chabrier_2003} IMF with high-mass slope $\alpha_{\mathrm{IMF}}=-2.3$, and a power-law attenuation curve (\S\ref{subsec:modeldata}). The two columns now show the results for F090W and F444W, with the former likely to be the most sensitive to dust.

We adopt \bpass\ as the fiducial SPS model for the \eagle\ photometry given its inclusion of binary stars, which are expected to be commonplace in all galaxies \citep{Sana_2012} and significantly alter their integrated spectra \citep{Stanway_2016}. Nonetheless, differing SPS model components such as isochrones and atmospheres can further alter the predicted emission \citep{Conroy_2009}, so we compare to the counts generated using \bc, \ma\ and \fsps\ in the top panel of Figure \ref{fig:sim_variation}. The latter is flexible, so we choose MIST isochrones \citep{Choi_2016, Dotter_2016} and MILES stellar spectra \citep{Miles_2006}, with the latter also used by \bc. It should be noted that due to the non-linear nature of the number counts, a shift of some magnitude does not imply that the underlying galaxies are fainter by the same degree. Relatively small shifts can greatly alter the bright-end counts in particular, similarly to the Eddington bias effect discussed in Section \ref{subsec:counts}. Even with this in mind, the uncertainties introduced by SPS modelling assumptions are clear, with the predictions deviating by up to $\sim0.3$ and $0.9 \ \mathrm{dex}$ in F090W and F444W respectively. The variations between \bpass, \fsps\ and \bc\ are reasonably small and can be attributed to differences in the adopted model inputs. For example, the brighter galaxies produced by \fsps\ relative to \bc\ are a consequence of MIST isochrones predicting longer stellar main-sequence lifetimes than those on the Padova tracks \citep{Girardi_2000}. For this reason, \eagle\ agrees most closely with the observations when using \bc\ to generate the photometry (see Appendix \ref{app:eagle_sps}). However, this is only an indicator of SPS model performance if the \eagle\ physics is assumed to be representative.

\ma\ systematically predicts fainter galaxies and therefore fewer counts in the magnitude range probed. Relative to \bpass, the J-band luminosities of galaxies at $z=0$ are up to $0.25\ \mathrm{dex}$ fainter when using \ma, and only the faintest below $\sim10^{26}\ \mathrm{erg\,s^{-1}\,Hz^{-1}}$ are relatively brighter by $<0.1 \ \mathrm{dex}$. This effect is significant with respect to all three other models, and is most notable in nearby galaxies hosting the oldest stellar populations. This difference may therefore arise from the treatment of post-main-sequence stars, specifically the use of the fuel consumption theorem rather than isochrones \citep{Renzini_1981}. Uncertainties in our understanding of stellar populations clearly propagate into model constraints, and Vijayan et al. (\emph{in prep}) will present a detailed analysis of these in the wider context of cosmological hydrodynamical simulations.

The \eagle\ photometry is produced using a \citet{Chabrier_2003} IMF to match the \eagle\ subgrid model \citep{Crain_2015}, where it controls stellar evolution including feedback and chemical enrichment timescales. This is therefore the most self-consistent choice, but we demonstrate the effect of possible variations in the middle panel of Figure \ref{fig:sim_variation}. A shallower slope of $\alpha_{\mathrm{IMF}}=-2.0$ assigns a larger fraction of the mass formed in a population to hot and short-lived massive stars. At the higher redshifts and younger ages probed by the faint counts, the galaxy light is dominated by massive O and B types. These emit harder ionising radiation than their lower-mass counterparts, and as there are now more of them, the galaxies become brighter at shorter wavelengths \citep{Stanway_2016}. In contrast, at the older ages probed by the bright-end counts, a larger proportion of the stars have evolved off the main sequence, and the galaxies become fainter. The opposite is true for a steeper IMF of $\alpha_{\mathrm{IMF}}=-2.7$, which produces more low-mass stars with high mass-to-light ratios that spend longer on the main sequence. These variations shift the counts by no more than $0.3 \ \mathrm{dex}$. 
 
The stellar spectra must be reprocessed by nebular gas and dust to produce SEDs that resemble those of real galaxies. The solid lines in the bottom panel of Figure \ref{fig:sim_variation} show the change in counts when assuming different dust attenuation curves. Reassuringly, this has little effect, with the \citet{Calzetti_2000} and local curves all consistent with one another, and differing from the $\tau \propto\lambda^{-1}$ curve by $<0.04 \ \mathrm{dex}$. Removing dust entirely gives an upper limit on how much it can be responsible for overpredicting the counts. Unsurprisingly, dust attenuation is important in accurately producing the shape of the short-wavelength counts, but reduces the normalisation by $\lesssim0.15 \ \mathrm{dex}$. At long wavelengths, the effect is negligible in all but the most evolved dusty galaxies. Setting $f_{\mathrm{esc}}=1$ demonstrates the importance of including nebular emission, which increases the counts by up to $\sim0.1 \ \mathrm{dex}$. The faint-end counts are preferentially affected, as these galaxies are likely to appear early in their evolution and thus host a larger fraction of young stars responsible for this emission.

Taken together, these results show that while forward-modelled galaxy number counts are largely insensitive to the reprocessing of stellar emission, the model used to predict that intrinsic component can have a significant effect. This may influence the interpretation of which models agree most closely with the observations, and therefore which physical mechanisms are most representative of the real Universe. However, no combination of these parameters would simultaneously resolve the tensions in both F090W and F444W. If the constraining power of direct observables is to be maximised, this forward modelling should be performed consistently, and the effect of variations quantified.
\section{Conclusions}\label{sec:conclusion}

We have presented the first instalment of the First Light and Assembly of Galaxies series. \flags\ will harness forward modelling to evaluate galaxy formation models using direct observables, thereby reducing associated bias and systematic uncertainty. The first \flags\ imaging dataset is constructed by consistently background-subtracting \jwst/NIRCam imaging from eight surveys, which span $>1 \ \mathrm{deg^{2}}$ across more than 30 independent sightlines. Consistent source extraction, aperture corrections and masking produce galaxy catalogues with an effective area of $0.65 \ \mathrm{deg^{2}}$, representing a $>1.5 \ \mathrm{dex}$ increase over previous NIRCam galaxy number count studies. 

We use these catalogues to produce new measurements of the galaxy counts in eight filters, without the need for colour corrections, and with reduced sensitivity to Poisson and cosmic variance uncertainties. Source injection simulations yield robust completeness estimates exceeding $80\%$ at 30th magnitude, enabling the counts to be probed deeper than ever before. We reliably extrapolate these counts by fitting GAMs, and derive robust constraints on the total $0.9-4.44 \ \mathrm{\mu m}$ IGL, achieving uncertainties as low as $\sim2.5\%$ at the longest wavelengths. \jwst\ facilitates novel measurements of $\mathrm{eIGL} =6.92^{+0.17}_{-0.17}$ and $3.46^{+0.09}_{-0.08} \ \mathrm{nW\,m^{-2}\,sr^{-1}}$ at 2.77 and $4.10 \ \mathrm{\mu m}$ respectively.

These observables are then used to evaluate the predictions of semi-empirical, semi-analytic and hydrodynamical models to demonstrate their constraining power. This reveals that despite being an interesting quantity, the IGL is an unreliable diagnostic of model performance. Overprediction of the faint counts by \spritz\ is mitigated by underestimating the number of bright galaxies, causing it to erroneously appear as the best model. The number counts themselves provide a far more discriminating test and can be employed without fitting or extrapolation. Models generally overpredict the total number of galaxies, particularly those which are faint at long wavelengths. This is not the case for \scsam, which emerges as the best-performing model at all wavelengths with a combined $\chi^{2}_{\nu}=18.6$. \sage\ performs most poorly ($\chi^{2}_{\nu}=448.3$), and we show through comparison with \scsam\ that this is due to inefficient stellar feedback in low-mass halos. A similar comparison between the fiducial and evolving-feedback \galform\ models suggests that the weaker feedback in the latter degrades its performance. 

We further demonstrate that systematic uncertainties arising from photometry parameters contribute to the counts by no more than $\sim 0.05 \ \mathrm{dex}$, well below the differences between models. This confirms that the direct observable approach is not limited by systematic uncertainties and establishes the galaxy number counts as a robust means of evaluating galaxy formation models. Uncertainties in SPS modelling are more significant, which motivates consistent forward modelling where possible and further model development. This approach is ready for application to simulation suites sampling large parameter spaces such as \textsc{camels}, enabling simulation-based inference of the astrophysical parameters that govern star formation and feedback.

\section*{Acknowledgements}

This work is based on observations made with the NASA/ESA/CSA James Webb Space Telescope. Some data was obtained from the Mikulski Archive for Space Telescopes at the Space Telescope Science Institute, which is operated by the Association of Universities for Research in Astronomy, Inc., under NASA contract NAS 5-03127 for \jwst.

We wish to acknowledge the \jwst\ programmes 1176, 1180, 1181, 1210, 1264, 1283, 1286, 1287, 1345, 1727, 1895, 1963, 2079, 2198, 2514, 2516, 2674, 2750, 3215, 3577, 3990, 4540, 4762, 5398, 5997, 6434, 6511, and 6541, which collected the imaging used throughout this work.

We also wish to acknowledge the following open source software packages used in the analysis: \texttt{Astropy} \citep{Astropy_2022}, \texttt{CMasher} \citep{van-der-Velden_2020}, \texttt{h5py}, \texttt{Matplotlib} \citep{Hunter_2007}, \texttt{Numpy} \citep{Harris_2020} and \texttt{Scipy} \citep{Virtanen_2020}.

This work used the DiRAC@Durham facility managed by the Institute for Computational Cosmology on behalf of the STFC DiRAC HPC Facility (www.dirac.ac.uk). The equipment was funded by BEIS capital funding via STFC capital grants ST/K00042X/1, ST/P002293/1, ST/R002371/1 and ST/S002502/1, Durham University and STFC operations grant ST/R000832/1. DiRAC is part of the National e-Infrastructure. 

Jack C. Turner would like to acknowledge Jonathan Loveday and Vivienne Wild, whose comments greatly improved the manuscript.

JCT is supported by an STFC PhD studentship ST/X508822/1.
APV and SMW acknowledge support from the Sussex Astronomy Centre STFC Consolidated Grant (ST/X001040/1).

We list here the roles and contributions of the authors according to the Contributor Roles Taxonomy (CRediT)\footnote{\url{https://credit.niso.org/}}.
\textbf{JCT}: Conceptualisation, Data curation, Methodology, Investigation, Formal Analysis, Visualisation, Writing - original draft. \textbf{SMW}: Conceptualisation, Methodology, Supervision, Writing - review \& editing. \textbf{APV}: Methodology, Data curation, Writing - review \& editing.

\section*{Data Availability}

The raw imaging of each survey is publicly available on MAST. Calibrated mosaics are publicly available for \ceers, \cosmosweb, \jades, \panoramic\ and \sapphires. The processed imaging produced as part of this work is available upon reasonable request. The galaxy catalogues and associated analysis code will be made available following publication of the manuscript.

%%%%%%%%%%%%%%%%%%%% REFERENCES %%%%%%%%%%%%%%%%%%

% The best way to enter references is to use BibTeX:
\bibliographystyle{mnras}
\bibliography{counts} 

@ARTICLE{Adams_2023,
       author = {{Adams}, N.~J. and {Conselice}, C.~J. and {Ferreira}, L. and {Austin}, D. and {Trussler}, J.~A.~A. and {Juod{\v{z}}balis}, I. and {Wilkins}, S.~M. and {Caruana}, J. and {Dayal}, P. and {Verma}, A. and {Vijayan}, A.~P.},
        title = "{Discovery and properties of ultra-high redshift galaxies (9 < z < 12) in the JWST ERO SMACS 0723 Field}",
      journal = {\mnras},
         year = 2023,
        month = jan,
       volume = {518},
       number = {3},
        pages = {4755-4766},
          doi = {10.1093/mnras/stac3347},
archivePrefix = {arXiv},
       eprint = {2207.11217},
 primaryClass = {astro-ph.GA},
       adsurl = {https://ui.adsabs.harvard.edu/abs/2023MNRAS.518.4755A}
}

@ARTICLE{Adams_2024,
       author = {{Adams}, Nathan J. and {Conselice}, Christopher J. and {Austin}, Duncan and {Harvey}, Thomas and {Ferreira}, Leonardo and {Trussler}, James and {Juod{\v{z}}balis}, Ignas and {Li}, Qiong and {Windhorst}, Rogier and {Cohen}, Seth H. and {Jansen}, Rolf A. and {Summers}, Jake and {Tompkins}, Scott and {Driver}, Simon P. and {Robotham}, Aaron and {D'Silva}, Jordan C.~J. and {Yan}, Haojing and {Coe}, Dan and {Frye}, Brenda and {Grogin}, Norman A. and {Koekemoer}, Anton M. and {Marshall}, Madeline A. and {Pirzkal}, Nor and {Ryan}, Russell E. and {Maksym}, W. Peter and {Rutkowski}, Michael J. and {Willmer}, Christopher N.~A. and {Hammel}, Heidi B. and {Nonino}, Mario and {Bhatawdekar}, Rachana and {Wilkins}, Stephen M. and {Bradley}, Larry D. and {Broadhurst}, Tom and {Cheng}, Cheng and {Dole}, Herv{\'e} and {Hathi}, Nimish P. and {Zitrin}, Adi},
        title = "{EPOCHS. II. The Ultraviolet Luminosity Function from 7.5 < z < 13.5 Using 180 arcmin$^{2}$ of Deep, Blank Fields from the PEARLS Survey and Public JWST Data}",
      journal = {\apj},
         year = 2024,
        month = apr,
       volume = {965},
       number = {2},
          eid = {169},
        pages = {169},
          doi = {10.3847/1538-4357/ad2a7b},
archivePrefix = {arXiv},
       eprint = {2304.13721},
 primaryClass = {astro-ph.GA},
       adsurl = {https://ui.adsabs.harvard.edu/abs/2024ApJ...965..169A}
}

@ARTICLE{Arrabal-Haro_2023,
       author = {{Arrabal Haro}, Pablo and {Dickinson}, Mark and {Finkelstein}, Steven L. and {Kartaltepe}, Jeyhan S. and {Donnan}, Callum T. and {Burgarella}, Denis and {Carnall}, Adam C. and {Cullen}, Fergus and {Dunlop}, James S. and {Fern{\'a}ndez}, Vital and {Fujimoto}, Seiji and {Jung}, Intae and {Krips}, Melanie and {Larson}, Rebecca L. and {Papovich}, Casey and {P{\'e}rez-Gonz{\'a}lez}, Pablo G. and {Amor{\'\i}n}, Ricardo O. and {Bagley}, Micaela B. and {Buat}, V{\'e}ronique and {Casey}, Caitlin M. and {Chworowsky}, Katherine and {Cohen}, Seth H. and {Ferguson}, Henry C. and {Giavalisco}, Mauro and {Huertas-Company}, Marc and {Hutchison}, Taylor A. and {Kocevski}, Dale D. and {Koekemoer}, Anton M. and {Lucas}, Ray A. and {McLeod}, Derek J. and {McLure}, Ross J. and {Pirzkal}, Norbert and {Seill{\'e}}, Lise-Marie and {Trump}, Jonathan R. and {Weiner}, Benjamin J. and {Wilkins}, Stephen M. and {Zavala}, Jorge A.},
        title = "{Confirmation and refutation of very luminous galaxies in the early Universe}",
      journal = {\nat},
         year = 2023,
        month = oct,
       volume = {622},
       number = {7984},
        pages = {707-711},
          doi = {10.1038/s41586-023-06521-7},
archivePrefix = {arXiv},
       eprint = {2303.15431},
 primaryClass = {astro-ph.GA},
       adsurl = {https://ui.adsabs.harvard.edu/abs/2023Natur.622..707A}
}

@ARTICLE{Ashby_2015,
       author = {{Ashby}, M.~L.~N. and {Willner}, S.~P. and {Fazio}, G.~G. and {Dunlop}, J.~S. and {Egami}, E. and {Faber}, S.~M. and {Ferguson}, H.~C. and {Grogin}, N.~A. and {Hora}, J.~L. and {Huang}, J. -S. and {Koekemoer}, A.~M. and {Labb{\'e}}, I. and {Wang}, Z.},
        title = "{S-CANDELS: The Spitzer-Cosmic Assembly Near-Infrared Deep Extragalactic Survey. Survey Design, Photometry, and Deep IRAC Source Counts}",
      journal = {\apjs},
         year = 2015,
        month = jun,
       volume = {218},
       number = {2},
          eid = {33},
        pages = {33},
          doi = {10.1088/0067-0049/218/2/33},
archivePrefix = {arXiv},
       eprint = {1506.01323},
 primaryClass = {astro-ph.GA},
       adsurl = {https://ui.adsabs.harvard.edu/abs/2015ApJS..218...33A}
}

@ARTICLE{Astropy_2022,
       author = {{Astropy Collaboration} and {Price-Whelan}, Adrian M. and {Lim}, Pey Lian and {Earl}, Nicholas and {Starkman}, Nathaniel and {Bradley}, Larry and {Shupe}, David L. and {Patil}, Aarya A. and {Corrales}, Lia and {Brasseur}, C.~E. and {N{"o}the}, Maximilian and {Donath}, Axel and {Tollerud}, Erik and {Morris}, Brett M. and {Ginsburg}, Adam and {Vaher}, Eero and {Weaver}, Benjamin A. and {Tocknell}, James and {Jamieson}, William and {van Kerkwijk}, Marten H. and {Robitaille}, Thomas P. and {Merry}, Bruce and {Bachetti}, Matteo and {G{"u}nther}, H. Moritz and {Aldcroft}, Thomas L. and {Alvarado-Montes}, Jaime A. and {Archibald}, Anne M. and {B{'o}di}, Attila and {Bapat}, Shreyas and {Barentsen}, Geert and {Baz{'a}n}, Juanjo and {Biswas}, Manish and {Boquien}, M{'e}d{'e}ric and {Burke}, D.~J. and {Cara}, Daria and {Cara}, Mihai and {Conroy}, Kyle E. and {Conseil}, Simon and {Craig}, Matthew W. and {Cross}, Robert M. and {Cruz}, Kelle L. and {D'Eugenio}, Francesco and {Dencheva}, Nadia and {Devillepoix}, Hadrien A.~R. and {Dietrich}, J{"o}rg P. and {Eigenbrot}, Arthur Davis and {Erben}, Thomas and {Ferreira}, Leonardo and {Foreman-Mackey}, Daniel and {Fox}, Ryan and {Freij}, Nabil and {Garg}, Suyog and {Geda}, Robel and {Glattly}, Lauren and {Gondhalekar}, Yash and {Gordon}, Karl D. and {Grant}, David and {Greenfield}, Perry and {Groener}, Austen M. and {Guest}, Steve and {Gurovich}, Sebastian and {Handberg}, Rasmus and {Hart}, Akeem and {Hatfield-Dodds}, Zac and {Homeier}, Derek and {Hosseinzadeh}, Griffin and {Jenness}, Tim and {Jones}, Craig K. and {Joseph}, Prajwel and {Kalmbach}, J. Bryce and {Karamehmetoglu}, Emir and {Ka{l}uszy{'n}ski}, Miko{l}aj and {Kelley}, Michael S.~P. and {Kern}, Nicholas and {Kerzendorf}, Wolfgang E. and {Koch}, Eric W. and {Kulumani}, Shankar and {Lee}, Antony and {Ly}, Chun and {Ma}, Zhiyuan and {MacBride}, Conor and {Maljaars}, Jakob M. and {Muna}, Demitri and {Murphy}, N.~A. and {Norman}, Henrik and {O'Steen}, Richard and {Oman}, Kyle A. and {Pacifici}, Camilla and {Pascual}, Sergio and {Pascual-Granado}, J. and {Patil}, Rohit R. and {Perren}, Gabriel I. and {Pickering}, Timothy E. and {Rastogi}, Tanuj and {Roulston}, Benjamin R. and {Ryan}, Daniel F. and {Rykoff}, Eli S. and {Sabater}, Jose and {Sakurikar}, Parikshit and {Salgado}, Jes{'u}s and {Sanghi}, Aniket and {Saunders}, Nicholas and {Savchenko}, Volodymyr and {Schwardt}, Ludwig and {Seifert-Eckert}, Michael and {Shih}, Albert Y. and {Jain}, Anany Shrey and {Shukla}, Gyanendra and {Sick}, Jonathan and {Simpson}, Chris and {Singanamalla}, Sudheesh and {Singer}, Leo P. and {Singhal}, Jaladh and {Sinha}, Manodeep and {Sip{H{o}}cz}, Brigitta M. and {Spitler}, Lee R. and {Stansby}, David and {Streicher}, Ole and {{{S}}umak}, Jani and {Swinbank}, John D. and {Taranu}, Dan S. and {Tewary}, Nikita and {Tremblay}, Grant R. and {Val-Borro}, Miguel de and {Van Kooten}, Samuel J. and {Vasovi{'c}}, Zlatan and {Verma}, Shresth and {de Miranda Cardoso}, Jos{'e} Vin{'i}cius and {Williams}, Peter K.~G. and {Wilson}, Tom J. and {Winkel}, Benjamin and {Wood-Vasey}, W.~M. and {Xue}, Rui and {Yoachim}, Peter and {Zhang}, Chen and {Zonca}, Andrea and {Astropy Project Contributors}},
        title = "{The Astropy Project: Sustaining and Growing a Community-oriented Open-source Project and the Latest Major Release (v5.0) of the Core Package}",
      journal = {\apj},
         year = 2022,
        month = aug,
       volume = {935},
       number = {2},
          eid = {167},
        pages = {167},
          doi = {10.3847/1538-4357/ac7c74},
archivePrefix = {arXiv},
       eprint = {2206.14220},
 primaryClass = {astro-ph.IM},
       adsurl = {https://ui.adsabs.harvard.edu/abs/2022ApJ...935..167A}
}

@ARTICLE{Austin_2023,
       author = {{Austin}, Duncan and {Adams}, Nathan and {Conselice}, Christopher J. and {Harvey}, Thomas and {Ormerod}, Katherine and {Trussler}, James and {Li}, Qiong and {Ferreira}, Leonardo and {Dayal}, Pratika and {Juod{\v{z}}balis}, Ignas},
        title = "{A Large Population of Faint 8 < z < 16 Galaxies Found in the First JWST NIRCam Observations of the NGDEEP Survey}",
      journal = {\apjl},
         year = 2023,
        month = jul,
       volume = {952},
       number = {1},
          eid = {L7},
        pages = {L7},
          doi = {10.3847/2041-8213/ace18d},
archivePrefix = {arXiv},
       eprint = {2302.04270},
 primaryClass = {astro-ph.GA},
       adsurl = {https://ui.adsabs.harvard.edu/abs/2023ApJ...952L...7A}
}

@ARTICLE{Bagley_2023,
       author = {{Bagley}, Micaela B. and {Finkelstein}, Steven L. and {Koekemoer}, Anton M. and {Ferguson}, Henry C. and {Arrabal Haro}, Pablo and {Dickinson}, Mark and {Kartaltepe}, Jeyhan S. and {Papovich}, Casey and {P{\'e}rez-Gonz{\'a}lez}, Pablo G. and {Pirzkal}, Nor and {Somerville}, Rachel S. and {Willmer}, Christopher N.~A. and {Yang}, Guang and {Yung}, L.~Y. Aaron and {Fontana}, Adriano and {Grazian}, Andrea and {Grogin}, Norman A. and {Hirschmann}, Michaela and {Kewley}, Lisa J. and {Kirkpatrick}, Allison and {Kocevski}, Dale D. and {Lotz}, Jennifer M. and {Medrano}, Aubrey and {Morales}, Alexa M. and {Pentericci}, Laura and {Ravindranath}, Swara and {Trump}, Jonathan R. and {Wilkins}, Stephen M. and {Calabr{\`o}}, Antonello and {Cooper}, M.~C. and {Costantin}, Luca and {de la Vega}, Alexander and {Hilbert}, Bryan and {Hutchison}, Taylor A. and {Larson}, Rebecca L. and {Lucas}, Ray A. and {McGrath}, Elizabeth J. and {Ryan}, Russell and {Wang}, Xin and {Wuyts}, Stijn},
        title = "{CEERS Epoch 1 NIRCam Imaging: Reduction Methods and Simulations Enabling Early JWST Science Results}",
      journal = {\apjl},
         year = 2023,
        month = mar,
       volume = {946},
       number = {1},
          eid = {L12},
        pages = {L12},
          doi = {10.3847/2041-8213/acbb08},
archivePrefix = {arXiv},
       eprint = {2211.02495},
 primaryClass = {astro-ph.IM},
       adsurl = {https://ui.adsabs.harvard.edu/abs/2023ApJ...946L..12B}
}

@ARTICLE{Bagley_2024,
       author = {{Bagley}, Micaela B. and {Pirzkal}, Nor and {Finkelstein}, Steven L. and {Papovich}, Casey and {Berg}, Danielle A. and {Lotz}, Jennifer M. and {Leung}, Gene C.~K. and {Ferguson}, Henry C. and {Koekemoer}, Anton M. and {Dickinson}, Mark and {Kartaltepe}, Jeyhan S. and {Kocevski}, Dale D. and {Somerville}, Rachel S. and {Yung}, L.~Y. Aaron and {Backhaus}, Bren E. and {Casey}, Caitlin M. and {Castellano}, Marco and {Ch{\'a}vez Ortiz}, {\'O}scar A. and {Chworowsky}, Katherine and {Cox}, Isabella G. and {Dav{\'e}}, Romeel and {Davis}, Kelcey and {Estrada-Carpenter}, Vicente and {Fontana}, Adriano and {Fujimoto}, Seiji and {Gardner}, Jonathan P. and {Giavalisco}, Mauro and {Grazian}, Andrea and {Grogin}, Norman A. and {Hathi}, Nimish P. and {Hutchison}, Taylor A. and {Jaskot}, Anne E. and {Jung}, Intae and {Kewley}, Lisa J. and {Kirkpatrick}, Allison and {Larson}, Rebecca L. and {Matharu}, Jasleen and {Natarajan}, Priyamvada and {Pentericci}, Laura and {P{\'e}rez-Gonz{\'a}lez}, Pablo G. and {Ravindranath}, Swara and {Rothberg}, Barry and {Ryan}, Russell and {Shen}, Lu and {Simons}, Raymond C. and {Snyder}, Gregory F. and {Trump}, Jonathan R. and {Wilkins}, Stephen M.},
        title = "{The Next Generation Deep Extragalactic Exploratory Public (NGDEEP) Survey}",
      journal = {\apjl},
         year = 2024,
        month = apr,
       volume = {965},
       number = {1},
          eid = {L6},
        pages = {L6},
          doi = {10.3847/2041-8213/ad2f31},
archivePrefix = {arXiv},
       eprint = {2302.05466},
 primaryClass = {astro-ph.GA},
       adsurl = {https://ui.adsabs.harvard.edu/abs/2024ApJ...965L...6B}
}

@article{Barbary_2016, doi = {10.21105/joss.00058}, url = {https://doi.org/10.21105/joss.00058}, year = {2016}, publisher = {The Open Journal}, volume = {1}, number = {6}, pages = {58}, author = {Kyle Barbary}, title = {SEP: Source Extractor as a library}, journal = {Journal of Open Source Software} }

@ARTICLE{Benson_2003,
       author = {{Benson}, A.~J. and {Bower}, R.~G. and {Frenk}, C.~S. and {Lacey}, C.~G. and {Baugh}, C.~M. and {Cole}, S.},
        title = "{What Shapes the Luminosity Function of Galaxies?}",
      journal = {\apj},
         year = 2003,
        month = dec,
       volume = {599},
       number = {1},
        pages = {38-49},
          doi = {10.1086/379160},
archivePrefix = {arXiv},
       eprint = {astro-ph/0302450},
 primaryClass = {astro-ph},
       adsurl = {https://ui.adsabs.harvard.edu/abs/2003ApJ...599...38B}
}

@ARTICLE{Bellstedt_2020,
       author = {{Bellstedt}, Sabine and {Driver}, Simon P. and {Robotham}, Aaron S.~G. and {Davies}, Luke J.~M. and {Bogue}, Kamran R.~J. and {Cook}, Robin H.~W. and {Hashemizadeh}, Abdolhosein and {Koushan}, Soheil and {Taylor}, Edward N. and {Thorne}, Jessica E. and {Turner}, Ryan J. and {Wright}, Angus H.},
        title = "{Galaxy And Mass Assembly (GAMA): assimilation of KiDS into the GAMA database}",
      journal = {\mnras},
         year = 2020,
        month = aug,
       volume = {496},
       number = {3},
        pages = {3235-3256},
          doi = {10.1093/mnras/staa1466},
archivePrefix = {arXiv},
       eprint = {2005.11215},
 primaryClass = {astro-ph.GA},
       adsurl = {https://ui.adsabs.harvard.edu/abs/2020MNRAS.496.3235B}
}

@ARTICLE{Bernyk_2016,
       author = {{Bernyk}, Maksym and {Croton}, Darren J. and {Tonini}, Chiara and {Hodkinson}, Luke and {Hassan}, Amr H. and {Garel}, Thibault and {Duffy}, Alan R. and {Mutch}, Simon J. and {Poole}, Gregory B. and {Hegarty}, Sarah},
        title = "{The Theoretical Astrophysical Observatory: Cloud-based Mock Galaxy Catalogs}",
      journal = {\apjs},
         year = 2016,
        month = mar,
       volume = {223},
       number = {1},
          eid = {9},
        pages = {9},
          doi = {10.3847/0067-0049/223/1/9},
archivePrefix = {arXiv},
       eprint = {1403.5270},
 primaryClass = {astro-ph.GA},
       adsurl = {https://ui.adsabs.harvard.edu/abs/2016ApJS..223....9B}
}

@ARTICLE{Bertin_1996,
       author = {{Bertin}, E. and {Arnouts}, S.},
        title = "{SExtractor: Software for source extraction.}",
      journal = {\aaps},
         year = 1996,
        month = jun,
       volume = {117},
        pages = {393-404},
          doi = {10.1051/aas:1996164},
       adsurl = {https://ui.adsabs.harvard.edu/abs/1996A&AS..117..393B}
}

@ARTICLE{Bisigello_2021,
       author = {{Bisigello}, L. and {Gruppioni}, C. and {Feltre}, A. and {Calura}, F. and {Pozzi}, F. and {Vignali}, C. and {Barchiesi}, L. and {Rodighiero}, G. and {Negrello}, M.},
        title = "{Simulating the infrared sky with a SPRITZ}",
      journal = {\aap},
         year = 2021,
        month = jul,
       volume = {651},
          eid = {A52},
        pages = {A52},
          doi = {10.1051/0004-6361/202039909},
archivePrefix = {arXiv},
       eprint = {2011.07074},
 primaryClass = {astro-ph.GA},
       adsurl = {https://ui.adsabs.harvard.edu/abs/2021A&A...651A..52B}
}

@ARTICLE{Biteau_2015,
       author = {{Biteau}, J. and {Williams}, D.~A.},
        title = "{The Extragalactic Background Light, the Hubble Constant, and Anomalies: Conclusions from 20 Years of TeV Gamma-ray Observations}",
      journal = {\apj},
         year = 2015,
        month = oct,
       volume = {812},
       number = {1},
          eid = {60},
        pages = {60},
          doi = {10.1088/0004-637X/812/1/60},
archivePrefix = {arXiv},
       eprint = {1502.04166},
 primaryClass = {astro-ph.CO},
       adsurl = {https://ui.adsabs.harvard.edu/abs/2015ApJ...812...60B}
}

@ARTICLE{Bonato_2019,
       author = {{Bonato}, Matteo and {De Zotti}, Gianfranco and {Leisawitz}, David and {Negrello}, Mattia and {Massardi}, Marcella and {Baronchelli}, Ivano and {Cai}, Zhen-Yi and {Bradford}, Charles M. and {Pope}, Alexandra and {Murphy}, Eric J. and {Armus}, Lee and {Cooray}, Asantha},
        title = "{Origins Space Telescope: Predictions for far-IR spectroscopic surveys}",
      journal = {\pasa},
         year = 2019,
        month = apr,
       volume = {36},
          eid = {e017},
        pages = {e017},
          doi = {10.1017/pasa.2019.8},
archivePrefix = {arXiv},
       eprint = {1903.00946},
 primaryClass = {astro-ph.GA},
       adsurl = {https://ui.adsabs.harvard.edu/abs/2019PASA...36...17B}
}

@ARTICLE{Booth_2013,
       author = {{Booth}, C.~M. and {Schaye}, Joop},
        title = "{The interaction between feedback from active galactic nuclei and supernovae}",
      journal = {Scientific Reports},
         year = 2013,
        month = may,
       volume = {3},
          eid = {1738},
        pages = {1738},
          doi = {10.1038/srep01738},
archivePrefix = {arXiv},
       eprint = {1203.3802},
 primaryClass = {astro-ph.CO},
       adsurl = {https://ui.adsabs.harvard.edu/abs/2013NatSR...3.1738B}
}

@ARTICLE{Bower_2006,
       author = {{Bower}, R.~G. and {Benson}, A.~J. and {Malbon}, R. and {Helly}, J.~C. and {Frenk}, C.~S. and {Baugh}, C.~M. and {Cole}, S. and {Lacey}, C.~G.},
        title = "{Breaking the hierarchy of galaxy formation}",
      journal = {\mnras},
         year = 2006,
        month = aug,
       volume = {370},
       number = {2},
        pages = {645-655},
          doi = {10.1111/j.1365-2966.2006.10519.x},
archivePrefix = {arXiv},
       eprint = {astro-ph/0511338},
 primaryClass = {astro-ph},
       adsurl = {https://ui.adsabs.harvard.edu/abs/2006MNRAS.370..645B}
}

@ARTICLE{Boylan-Kolchin_2023,
       author = {{Boylan-Kolchin}, Michael},
        title = "{Stress testing {\ensuremath{\Lambda}}CDM with high-redshift galaxy candidates}",
      journal = {Nature Astronomy},
         year = 2023,
        month = jun,
       volume = {7},
        pages = {731-735},
          doi = {10.1038/s41550-023-01937-7},
archivePrefix = {arXiv},
       eprint = {2208.01611},
 primaryClass = {astro-ph.CO},
       adsurl = {https://ui.adsabs.harvard.edu/abs/2023NatAs...7..731B}
}

@software{Bradley_2025,
  author       = {Larry Bradley and
                  Brigitta Sipőcz and
                  Thomas Robitaille and
                  Erik Tollerud and
                  Zé Vinícius and
                  Christoph Deil and
                  Kyle Barbary and
                  Tom J Wilson and
                  Ivo Busko and
                  Axel Donath and
                  Hans Moritz Günther and
                  Mihai Cara and
                  P. L. Lim and
                  Sebastian Meßlinger and
                  Simon Conseil and
                  Michael Droettboom and
                  Azalee Bostroem and
                  E. M. Bray and
                  Lars Andersen Bratholm and
                  Zach Burnett and
                  William Jamieson and
                  Adam Ginsburg and
                  Dan Taranu and
                  Geert Barentsen and
                  Matt Craig and
                  Brett M. Morris and
                  Marshall Perrin and
                  Shivangee Rathi},
  title        = {astropy/photutils: 2.3.0},
  month        = sep,
  year         = 2025,
  publisher    = {Zenodo},
  version      = {2.3.0},
  doi          = {10.5281/zenodo.17129028},
  url          = {https://doi.org/10.5281/zenodo.17129028},
  swhid        = {swh:1:dir:dd51869167d76d722ba87e3f80f9f4199ec08c3f;origin=https://doi.org/10.5281/zenodo.596036;visi
                   t=swh:1:snp:30a5f50b0586911dc674668853d9abc352a2bc
                   22;anchor=swh:1:rel:e97861da904cf010c499a4211cd8a6
                   12373e912a;path=astropy-photutils-2294e35
                  },
}

@ARTICLE{Bruzual_2003,
       author = {{Bruzual}, G. and {Charlot}, S.},
        title = "{Stellar population synthesis at the resolution of 2003}",
      journal = {\mnras},
         year = 2003,
        month = oct,
       volume = {344},
       number = {4},
        pages = {1000-1028},
          doi = {10.1046/j.1365-8711.2003.06897.x},
archivePrefix = {arXiv},
       eprint = {astro-ph/0309134},
 primaryClass = {astro-ph},
       adsurl = {https://ui.adsabs.harvard.edu/abs/2003MNRAS.344.1000B}
}

@ARTICLE{Byler_2017,
       author = {{Byler}, Nell and {Dalcanton}, Julianne J. and {Conroy}, Charlie and {Johnson}, Benjamin D.},
        title = "{Nebular Continuum and Line Emission in Stellar Population Synthesis Models}",
      journal = {\apj},
         year = 2017,
        month = may,
       volume = {840},
       number = {1},
          eid = {44},
        pages = {44},
          doi = {10.3847/1538-4357/aa6c66},
archivePrefix = {arXiv},
       eprint = {1611.08305},
 primaryClass = {astro-ph.GA},
       adsurl = {https://ui.adsabs.harvard.edu/abs/2017ApJ...840...44B}
}

@ARTICLE{Byrne_2023,
       author = {{Byrne}, C.~M. and {Stanway}, E.~R.},
        title = "{On the impact of spectral template uncertainties in synthetic stellar populations}",
      journal = {\mnras},
         year = 2023,
        month = jun,
       volume = {521},
       number = {4},
        pages = {4995-5012},
          doi = {10.1093/mnras/stad832},
archivePrefix = {arXiv},
       eprint = {2303.16920},
 primaryClass = {astro-ph.GA},
       adsurl = {https://ui.adsabs.harvard.edu/abs/2023MNRAS.521.4995B}
}

@ARTICLE{Calzetti_2000,
       author = {{Calzetti}, Daniela and {Armus}, Lee and {Bohlin}, Ralph C. and {Kinney}, Anne L. and {Koornneef}, Jan and {Storchi-Bergmann}, Thaisa},
        title = "{The Dust Content and Opacity of Actively Star-forming Galaxies}",
      journal = {\apj},
         year = 2000,
        month = apr,
       volume = {533},
       number = {2},
        pages = {682-695},
          doi = {10.1086/308692},
archivePrefix = {arXiv},
       eprint = {astro-ph/9911459},
 primaryClass = {astro-ph},
       adsurl = {https://ui.adsabs.harvard.edu/abs/2000ApJ...533..682C}
}

@ARTICLE{Camps_2015,
       author = {{Camps}, P. and {Baes}, M.},
        title = "{SKIRT: An advanced dust radiative transfer code with a user-friendly architecture}",
      journal = {Astronomy and Computing},
         year = 2015,
        month = mar,
       volume = {9},
        pages = {20-33},
          doi = {10.1016/j.ascom.2014.10.004},
archivePrefix = {arXiv},
       eprint = {1410.1629},
 primaryClass = {astro-ph.IM},
       adsurl = {https://ui.adsabs.harvard.edu/abs/2015A&C.....9...20C}
}

@ARTICLE{Carnall_2019,
       author = {{Carnall}, Adam C. and {Leja}, Joel and {Johnson}, Benjamin D. and {McLure}, Ross J. and {Dunlop}, James S. and {Conroy}, Charlie},
        title = "{How to Measure Galaxy Star Formation Histories. I. Parametric Models}",
      journal = {\apj},
         year = 2019,
        month = mar,
       volume = {873},
       number = {1},
          eid = {44},
        pages = {44},
          doi = {10.3847/1538-4357/ab04a2},
archivePrefix = {arXiv},
       eprint = {1811.03635},
 primaryClass = {astro-ph.GA},
       adsurl = {https://ui.adsabs.harvard.edu/abs/2019ApJ...873...44C}
}

@ARTICLE{Casey_2023,
       author = {{Casey}, Caitlin M. and {Kartaltepe}, Jeyhan S. and {Drakos}, Nicole E. and {Franco}, Maximilien and {Harish}, Santosh and {Paquereau}, Louise and {Ilbert}, Olivier and {Rose}, Caitlin and {Cox}, Isabella G. and {Nightingale}, James W. and {Robertson}, Brant E. and {Silverman}, John D. and {Koekemoer}, Anton M. and {Massey}, Richard and {McCracken}, Henry Joy and {Rhodes}, Jason and {Akins}, Hollis B. and {Allen}, Natalie and {Amvrosiadis}, Aristeidis and {Arango-Toro}, Rafael C. and {Bagley}, Micaela B. and {Bongiorno}, Angela and {Capak}, Peter L. and {Champagne}, Jaclyn B. and {Chartab}, Nima and {Ch{\'a}vez Ortiz}, {\'O}scar A. and {Chworowsky}, Katherine and {Cooke}, Kevin C. and {Cooper}, Olivia R. and {Darvish}, Behnam and {Ding}, Xuheng and {Faisst}, Andreas L. and {Finkelstein}, Steven L. and {Fujimoto}, Seiji and {Gentile}, Fabrizio and {Gillman}, Steven and {Gould}, Katriona M.~L. and {Gozaliasl}, Ghassem and {Hayward}, Christopher C. and {He}, Qiuhan and {Hemmati}, Shoubaneh and {Hirschmann}, Michaela and {Jahnke}, Knud and {Jin}, Shuowen and {Khostovan}, Ali Ahmad and {Kokorev}, Vasily and {Lambrides}, Erini and {Laigle}, Clotilde and {Larson}, Rebecca L. and {Leung}, Gene C.~K. and {Liu}, Daizhong and {Liaudat}, Tobias and {Long}, Arianna S. and {Magdis}, Georgios and {Mahler}, Guillaume and {Mainieri}, Vincenzo and {Manning}, Sinclaire M. and {Maraston}, Claudia and {Martin}, Crystal L. and {McCleary}, Jacqueline E. and {McKinney}, Jed and {McPartland}, Conor J.~R. and {Mobasher}, Bahram and {Pattnaik}, Rohan and {Renzini}, Alvio and {Rich}, R. Michael and {Sanders}, David B. and {Sattari}, Zahra and {Scognamiglio}, Diana and {Scoville}, Nick and {Sheth}, Kartik and {Shuntov}, Marko and {Sparre}, Martin and {Suzuki}, Tomoko L. and {Talia}, Margherita and {Toft}, Sune and {Trakhtenbrot}, Benny and {Urry}, C. Megan and {Valentino}, Francesco and {Vanderhoof}, Brittany N. and {Vardoulaki}, Eleni and {Weaver}, John R. and {Whitaker}, Katherine E. and {Wilkins}, Stephen M. and {Yang}, Lilan and {Zavala}, Jorge A.},
        title = "{COSMOS-Web: An Overview of the JWST Cosmic Origins Survey}",
      journal = {\apj},
         year = 2023,
        month = sep,
       volume = {954},
       number = {1},
          eid = {31},
        pages = {31},
          doi = {10.3847/1538-4357/acc2bc},
archivePrefix = {arXiv},
       eprint = {2211.07865},
 primaryClass = {astro-ph.GA},
       adsurl = {https://ui.adsabs.harvard.edu/abs/2023ApJ...954...31C}
}

@ARTICLE{Cenci_2025,
       author = {{Cenci}, Elia and {Habouzit}, Melanie},
        title = "{Little Red Dots as direct-collapse black hole nurseries}",
      journal = {\mnras},
         year = 2025,
        month = sep,
       volume = {542},
       number = {3},
        pages = {2597-2609},
          doi = {10.1093/mnras/staf1362},
archivePrefix = {arXiv},
       eprint = {2508.14897},
 primaryClass = {astro-ph.GA},
       adsurl = {https://ui.adsabs.harvard.edu/abs/2025MNRAS.542.2597C}
}

@ARTICLE{Chabrier_2003,
       author = {{Chabrier}, Gilles},
        title = "{Galactic Stellar and Substellar Initial Mass Function}",
      journal = {\pasp},
         year = 2003,
        month = jul,
       volume = {115},
       number = {809},
        pages = {763-795},
          doi = {10.1086/376392},
archivePrefix = {arXiv},
       eprint = {astro-ph/0304382},
 primaryClass = {astro-ph},
       adsurl = {https://ui.adsabs.harvard.edu/abs/2003PASP..115..763C}
}

@ARTICLE{Charlot_2000,
       author = {{Charlot}, St{\'e}phane and {Fall}, S. Michael},
        title = "{A Simple Model for the Absorption of Starlight by Dust in Galaxies}",
      journal = {\apj},
         year = 2000,
        month = aug,
       volume = {539},
       number = {2},
        pages = {718-731},
          doi = {10.1086/309250},
archivePrefix = {arXiv},
       eprint = {astro-ph/0003128},
 primaryClass = {astro-ph},
       adsurl = {https://ui.adsabs.harvard.edu/abs/2000ApJ...539..718C}
}

@ARTICLE{Choi_2016,
       author = {{Choi}, Jieun and {Dotter}, Aaron and {Conroy}, Charlie and {Cantiello}, Matteo and {Paxton}, Bill and {Johnson}, Benjamin D.},
        title = "{Mesa Isochrones and Stellar Tracks (MIST). I. Solar-scaled Models}",
      journal = {\apj},
         year = 2016,
        month = jun,
       volume = {823},
       number = {2},
          eid = {102},
        pages = {102},
          doi = {10.3847/0004-637X/823/2/102},
archivePrefix = {arXiv},
       eprint = {1604.08592},
 primaryClass = {astro-ph.SR},
       adsurl = {https://ui.adsabs.harvard.edu/abs/2016ApJ...823..102C}
}

@ARTICLE{Ciesla_2015,
       author = {{Ciesla}, L. and {Charmandaris}, V. and {Georgakakis}, A. and {Bernhard}, E. and {Mitchell}, P.~D. and {Buat}, V. and {Elbaz}, D. and {LeFloc'h}, E. and {Lacey}, C.~G. and {Magdis}, G.~E. and {Xilouris}, M.},
        title = "{Constraining the properties of AGN host galaxies with spectral energy distribution modelling}",
      journal = {\aap},
         year = 2015,
        month = apr,
       volume = {576},
          eid = {A10},
        pages = {A10},
          doi = {10.1051/0004-6361/201425252},
archivePrefix = {arXiv},
       eprint = {1501.03672},
 primaryClass = {astro-ph.GA},
       adsurl = {https://ui.adsabs.harvard.edu/abs/2015A&A...576A..10C}
}

@ARTICLE{Clausen_2025,
       author = {{Clausen}, Thorbj{\o}rn and {Steinhardt}, Charles L. and {Shao}, Arden and {Senthil Kumar}, Gaurav},
        title = "{Performance of photometric template fitting for ultra-high-redshift galaxies}",
      journal = {\aap},
         year = 2025,
        month = may,
       volume = {697},
          eid = {A160},
        pages = {A160},
          doi = {10.1051/0004-6361/202453247},
archivePrefix = {arXiv},
       eprint = {2412.01893},
 primaryClass = {astro-ph.IM},
       adsurl = {https://ui.adsabs.harvard.edu/abs/2025A&A...697A.160C}
}

@ARTICLE{Conroy_2009,
       author = {{Conroy}, Charlie and {Gunn}, James E. and {White}, Martin},
        title = "{The Propagation of Uncertainties in Stellar Population Synthesis Modeling. I. The Relevance of Uncertain Aspects of Stellar Evolution and the Initial Mass Function to the Derived Physical Properties of Galaxies}",
      journal = {\apj},
         year = 2009,
        month = jul,
       volume = {699},
       number = {1},
        pages = {486-506},
          doi = {10.1088/0004-637X/699/1/486},
archivePrefix = {arXiv},
       eprint = {0809.4261},
 primaryClass = {astro-ph},
       adsurl = {https://ui.adsabs.harvard.edu/abs/2009ApJ...699..486C}
}

@software{Conroy_2010,
       author = {{Conroy}, Charlie and {Gunn}, James E.},
        title = "{FSPS: Flexible Stellar Population Synthesis}",
 howpublished = {Astrophysics Source Code Library, record ascl:1010.043},
         year = 2010,
        month = oct,
          eid = {ascl:1010.043},
       adsurl = {https://ui.adsabs.harvard.edu/abs/2010ascl.soft10043C}
}

@ARTICLE{Cowley_2018,
       author = {{Cowley}, William I. and {Baugh}, Carlton M. and {Cole}, Shaun and {Frenk}, Carlos S. and {Lacey}, Cedric G.},
        title = "{Predictions for deep galaxy surveys with JWST from {\ensuremath{\Lambda}}CDM}",
      journal = {\mnras},
         year = 2018,
        month = feb,
       volume = {474},
       number = {2},
        pages = {2352-2372},
          doi = {10.1093/mnras/stx2897},
archivePrefix = {arXiv},
       eprint = {1702.02146},
 primaryClass = {astro-ph.GA},
       adsurl = {https://ui.adsabs.harvard.edu/abs/2018MNRAS.474.2352C}
}

@article{Crain_2015,
    author = {Crain, Robert A. and Schaye, Joop and Bower, Richard G. and Furlong, Michelle and Schaller, Matthieu and Theuns, Tom and Dalla Vecchia, Claudio and Frenk, Carlos S. and McCarthy, Ian G. and Helly, John C. and Jenkins, Adrian and Rosas-Guevara, Yetli M. and White, Simon D. M. and Trayford, James W.},
    title = {The EAGLE simulations of galaxy formation: calibration of subgrid physics and model variations},
    journal = {\mnras},
    volume = {450},
    number = {2},
    pages = {1937-1961},
    year = {2015},
    month = {04},
    issn = {0035-8711},
    doi = {10.1093/mnras/stv725},
    url = {https://doi.org/10.1093/mnras/stv725},
    eprint = {https://academic.oup.com/mnras/article-pdf/450/2/1937/3067585/stv725.pdf},
}

@ARTICLE{Croton_2006,
       author = {{Croton}, Darren J. and {Springel}, Volker and {White}, Simon D.~M. and {De Lucia}, G. and {Frenk}, C.~S. and {Gao}, L. and {Jenkins}, A. and {Kauffmann}, G. and {Navarro}, J.~F. and {Yoshida}, N.},
        title = "{The many lives of active galactic nuclei: cooling flows, black holes and the luminosities and colours of galaxies}",
      journal = {\mnras},
         year = 2006,
        month = jan,
       volume = {365},
       number = {1},
        pages = {11-28},
          doi = {10.1111/j.1365-2966.2005.09675.x},
archivePrefix = {arXiv},
       eprint = {astro-ph/0508046},
 primaryClass = {astro-ph},
       adsurl = {https://ui.adsabs.harvard.edu/abs/2006MNRAS.365...11C}
}

@ARTICLE{Croton_2016,
       author = {{Croton}, Darren J. and {Stevens}, Adam R.~H. and {Tonini}, Chiara and {Garel}, Thibault and {Bernyk}, Maksym and {Bibiano}, Antonio and {Hodkinson}, Luke and {Mutch}, Simon J. and {Poole}, Gregory B. and {Shattow}, Genevieve M.},
        title = "{Semi-Analytic Galaxy Evolution (SAGE): Model Calibration and Basic Results}",
      journal = {\apjs},
         year = 2016,
        month = feb,
       volume = {222},
       number = {2},
          eid = {22},
        pages = {22},
          doi = {10.3847/0067-0049/222/2/22},
archivePrefix = {arXiv},
       eprint = {1601.04709},
 primaryClass = {astro-ph.GA},
       adsurl = {https://ui.adsabs.harvard.edu/abs/2016ApJS..222...22C}
}

@ARTICLE{Davies_2021,
       author = {{Davies}, L.~J.~M. and {Thorne}, J.~E. and {Robotham}, A.~S.~G. and {Bellstedt}, S. and {Driver}, S.~P. and {Adams}, N.~J. and {Bilicki}, M. and {Bowler}, R.~A.~A. and {Bravo}, M. and {Cortese}, L. and {Foster}, C. and {Grootes}, M.~W. and {H{\"a}u{\ss}ler}, B. and {Hashemizadeh}, A. and {Holwerda}, B.~W. and {Hurley}, P. and {Jarvis}, M.~J. and {Lidman}, C. and {Maddox}, N. and {Meyer}, M. and {Paolillo}, M. and {Phillipps}, S. and {Radovich}, M. and {Siudek}, M. and {Vaccari}, M. and {Windhorst}, R.~A.},
        title = "{Deep Extragalactic VIsible Legacy Survey (DEVILS): consistent multiwavelength photometry for the DEVILS regions (COSMOS, XMMLSS, and ECDFS)}",
      journal = {\mnras},
         year = 2021,
        month = sep,
       volume = {506},
       number = {1},
        pages = {256-287},
          doi = {10.1093/mnras/stab1601},
archivePrefix = {arXiv},
       eprint = {2106.06241},
 primaryClass = {astro-ph.GA},
       adsurl = {https://ui.adsabs.harvard.edu/abs/2021MNRAS.506..256D}
}

@ARTICLE{Desai_2017,
       author = {{Desai}, A. and {Ajello}, M. and {Omodei}, N. and {Hartmann}, D.~H. and {Dom{\'\i}nguez}, A. and {Paliya}, V.~S. and {Helgason}, K. and {Finke}, J. and {Meyer}, M.},
        title = "{Probing the EBL Evolution at High Redshift Using GRBs Detected with the Fermi-LAT}",
      journal = {\apj},
         year = 2017,
        month = nov,
       volume = {850},
       number = {1},
          eid = {73},
        pages = {73},
          doi = {10.3847/1538-4357/aa917c},
archivePrefix = {arXiv},
       eprint = {1710.02535},
 primaryClass = {astro-ph.HE},
       adsurl = {https://ui.adsabs.harvard.edu/abs/2017ApJ...850...73D}
}

@ARTICLE{Devriendt_1999,
       author = {{Devriendt}, J.~E.~G. and {Guiderdoni}, B. and {Sadat}, R.},
        title = "{Galaxy modelling. I. Spectral energy distributions from far-UV to sub-mm wavelengths}",
      journal = {\aap},
         year = 1999,
        month = oct,
       volume = {350},
        pages = {381-398},
          doi = {10.48550/arXiv.astro-ph/9906332},
archivePrefix = {arXiv},
       eprint = {astro-ph/9906332},
 primaryClass = {astro-ph},
       adsurl = {https://ui.adsabs.harvard.edu/abs/1999A&A...350..381D}
}

@ARTICLE{Donnan_2023,
       author = {{Donnan}, C.~T. and {McLeod}, D.~J. and {Dunlop}, J.~S. and {McLure}, R.~J. and {Carnall}, A.~C. and {Begley}, R. and {Cullen}, F. and {Hamadouche}, M.~L. and {Bowler}, R.~A.~A. and {Magee}, D. and {McCracken}, H.~J. and {Milvang-Jensen}, B. and {Moneti}, A. and {Targett}, T.},
        title = "{The evolution of the galaxy UV luminosity function at redshifts z ≃ 8 - 15 from deep JWST and ground-based near-infrared imaging}",
      journal = {\mnras},
         year = 2023,
        month = feb,
       volume = {518},
       number = {4},
        pages = {6011-6040},
          doi = {10.1093/mnras/stac3472},
archivePrefix = {arXiv},
       eprint = {2207.12356},
 primaryClass = {astro-ph.GA},
       adsurl = {https://ui.adsabs.harvard.edu/abs/2023MNRAS.518.6011D}
}

@ARTICLE{Dotter_2016,
       author = {{Dotter}, Aaron},
        title = "{MESA Isochrones and Stellar Tracks (MIST) 0: Methods for the Construction of Stellar Isochrones}",
      journal = {\apjs},
         year = 2016,
        month = jan,
       volume = {222},
       number = {1},
          eid = {8},
        pages = {8},
          doi = {10.3847/0067-0049/222/1/8},
archivePrefix = {arXiv},
       eprint = {1601.05144},
 primaryClass = {astro-ph.SR},
       adsurl = {https://ui.adsabs.harvard.edu/abs/2016ApJS..222....8D}
}

@ARTICLE{Drakos_2022,
       author = {{Drakos}, Nicole E. and {Villasenor}, Bruno and {Robertson}, Brant E. and {Hausen}, Ryan and {Dickinson}, Mark E. and {Ferguson}, Henry C. and {Furlanetto}, Steven R. and {Greene}, Jenny E. and {Madau}, Piero and {Shapley}, Alice E. and {Stark}, Daniel P. and {Wechsler}, Risa H.},
        title = "{Deep Realistic Extragalactic Model (DREaM) Galaxy Catalogs: Predictions for a Roman Ultra-deep Field}",
      journal = {\apj},
         year = 2022,
        month = feb,
       volume = {926},
       number = {2},
          eid = {194},
        pages = {194},
          doi = {10.3847/1538-4357/ac46fb},
archivePrefix = {arXiv},
       eprint = {2110.10703},
 primaryClass = {astro-ph.GA},
       adsurl = {https://ui.adsabs.harvard.edu/abs/2022ApJ...926..194D}
}

@ARTICLE{Driver_2010,
       author = {{Driver}, Simon P. and {Robotham}, Aaron S.~G.},
        title = "{Quantifying cosmic variance}",
      journal = {\mnras},
         year = 2010,
        month = oct,
       volume = {407},
       number = {4},
        pages = {2131-2140},
          doi = {10.1111/j.1365-2966.2010.17028.x},
archivePrefix = {arXiv},
       eprint = {1005.2538},
 primaryClass = {astro-ph.CO},
       adsurl = {https://ui.adsabs.harvard.edu/abs/2010MNRAS.407.2131D}
}

@ARTICLE{Driver_2016,
       author = {{Driver}, Simon P. and {Andrews}, Stephen K. and {Davies}, Luke J. and {Robotham}, Aaron S.~G. and {Wright}, Angus H. and {Windhorst}, Rogier A. and {Cohen}, Seth and {Emig}, Kim and {Jansen}, Rolf A. and {Dunne}, Loretta},
        title = "{Measurements of Extragalactic Background Light from the Far UV to the Far IR from Deep Ground- and Space-based Galaxy Counts}",
      journal = {\apj},
         year = 2016,
        month = aug,
       volume = {827},
       number = {2},
          eid = {108},
        pages = {108},
          doi = {10.3847/0004-637X/827/2/108},
archivePrefix = {arXiv},
       eprint = {1605.01523},
 primaryClass = {astro-ph.GA},
       adsurl = {https://ui.adsabs.harvard.edu/abs/2016ApJ...827..108D}
}

@ARTICLE{Dwek_2013,
       author = {{Dwek}, Eli and {Krennrich}, Frank},
        title = "{The extragalactic background light and the gamma-ray opacity of the universe}",
      journal = {Astroparticle Physics},
         year = 2013,
        month = mar,
       volume = {43},
        pages = {112-133},
          doi = {10.1016/j.astropartphys.2012.09.003},
archivePrefix = {arXiv},
       eprint = {1209.4661},
 primaryClass = {astro-ph.CO},
       adsurl = {https://ui.adsabs.harvard.edu/abs/2013APh....43..112D}
}

@ARTICLE{Eisenstein_2026,
       author = {{Eisenstein}, Daniel J. and {Willott}, Chris and {Alberts}, Stacey and {Arribas}, Santiago and {Bonaventura}, Nina and {Bunker}, Andrew J. and {Cameron}, Alex J. and {Carniani}, Stefano and {Charlot}, Stephane and {Curtis-Lake}, Emma and {D'Eugenio}, Francesco and {Ferruit}, Pierre and {Giardino}, Giovanna and {Hainline}, Kevin and {Hausen}, Ryan and {Jakobsen}, Peter and {Johnson}, Benjamin D. and {Maiolino}, Roberto and {Rauscher}, Bernard J. and {Rieke}, Marcia and {Rieke}, George and {Rix}, Hans-Walter and {Robertson}, Brant and {Stark}, Daniel P. and {Tacchella}, Sandro and {Williams}, Christina C. and {Willmer}, Christopher N.~A. and {Baker}, William M. and {Baum}, Stefi and {Bhatawdekar}, Rachana and {Boyett}, Kristan and {Chen}, Zuyi and {Chevallard}, Jacopo and {Circosta}, Chiara and {Curti}, Mirko and {Danhaive}, A. Lola and {DeCoursey}, Christa and {Endsley}, Ryan and {de Graaff}, Anna and {Dressler}, Alan and {Egami}, Eiichi and {Helton}, Jakob M. and {Hviding}, Raphael E. and {Ji}, Zhiyuan and {Jones}, Gareth C. and {Kumari}, Nimisha and {L{\"u}tzgendorf}, Nora and {Laseter}, Isaac and {Looser}, Tobias J. and {Lyu}, Jianwei and {Maseda}, Michael V. and {Nelson}, Erica and {Parlanti}, Eleonora and {Perna}, Michele and {Pusk{\'a}s}, D{\'a}vid and {Rawle}, Tim and {Rodr{\'\i}guez Del Pino}, Bruno and {Rujopakarn}, Wiphu and {Sandles}, Lester and {Saxena}, Aayush and {Scholtz}, Jan and {Sharpe}, Katherine and {Shivaei}, Irene and {Silcock}, Maddie S. and {Simmonds}, Charlotte and {Skarbinski}, Maya and {Smit}, Renske and {Stone}, Meredith and {Suess}, Katherine A. and {Sun}, Fengwu and {Tang}, Mengtao and {Topping}, Michael W. and {{\"U}bler}, Hannah and {Villanueva}, Natalia C. and {Wallace}, Imaan E.~B. and {Whitler}, Lily and {Witstok}, Joris and {Woodrum}, Charity},
        title = "{Overview of the JWST Advanced Deep Extragalactic Survey (JADES)}",
      journal = {\apjs},
         year = 2026,
        month = mar,
       volume = {283},
       number = {1},
          eid = {6},
        pages = {6},
          doi = {10.3847/1538-4365/ae3163},
archivePrefix = {arXiv},
       eprint = {2306.02465},
 primaryClass = {astro-ph.GA},
       adsurl = {https://ui.adsabs.harvard.edu/abs/2026ApJS..283....6E}
}

@ARTICLE{Ferland_2017,
       author = {{Ferland}, G.~J. and {Chatzikos}, M. and {Guzm{\'a}n}, F. and
         {Lykins}, M.~L. and {van Hoof}, P.~A.~M. and {Williams}, R.~J.~R. and
         {Abel}, N.~P. and {Badnell}, N.~R. and {Keenan}, F.~P. and
         {Porter}, R.~L. and {Stancil}, P.~C.},
        title = "{The 2017 Release Cloudy}",
      journal = {\rmxaa},
         year = 2017,
        month = oct,
       volume = {53},
        pages = {385-438},
archivePrefix = {arXiv},
       eprint = {1705.10877},
 primaryClass = {astro-ph.GA},
       adsurl = {https://ui.adsabs.harvard.edu/abs/2017RMxAA..53..385F}
}

@ARTICLE{Finkelstein_2022,
       author = {{Finkelstein}, Steven L. and {Bagley}, Micaela B. and {Arrabal Haro}, Pablo and {Dickinson}, Mark and {Ferguson}, Henry C. and {Kartaltepe}, Jeyhan S. and {Papovich}, Casey and {Burgarella}, Denis and {Kocevski}, Dale D. and {Huertas-Company}, Marc and {Iyer}, Kartheik G. and {Koekemoer}, Anton M. and {Larson}, Rebecca L. and {P{\'e}rez-Gonz{\'a}lez}, Pablo G. and {Rose}, Caitlin and {Tacchella}, Sandro and {Wilkins}, Stephen M. and {Chworowsky}, Katherine and {Medrano}, Aubrey and {Morales}, Alexa M. and {Somerville}, Rachel S. and {Yung}, L.~Y. Aaron and {Fontana}, Adriano and {Giavalisco}, Mauro and {Grazian}, Andrea and {Grogin}, Norman A. and {Kewley}, Lisa J. and {Kirkpatrick}, Allison and {Kurczynski}, Peter and {Lotz}, Jennifer M. and {Pentericci}, Laura and {Pirzkal}, Nor and {Ravindranath}, Swara and {Ryan}, Russell E. and {Trump}, Jonathan R. and {Yang}, Guang and {Almaini}, Omar and {Amor{\'\i}n}, Ricardo O. and {Annunziatella}, Marianna and {Backhaus}, Bren E. and {Barro}, Guillermo and {Behroozi}, Peter and {Bell}, Eric F. and {Bhatawdekar}, Rachana and {Bisigello}, Laura and {Bromm}, Volker and {Buat}, V{\'e}ronique and {Buitrago}, Fernando and {Calabr{\`o}}, Antonello and {Casey}, Caitlin M. and {Castellano}, Marco and {Ch{\'a}vez Ortiz}, {\'O}scar A. and {Ciesla}, Laure and {Cleri}, Nikko J. and {Cohen}, Seth H. and {Cole}, Justin W. and {Cooke}, Kevin C. and {Cooper}, M.~C. and {Cooray}, Asantha R. and {Costantin}, Luca and {Cox}, Isabella G. and {Croton}, Darren and {Daddi}, Emanuele and {Dav{\'e}}, Romeel and {de La Vega}, Alexander and {Dekel}, Avishai and {Elbaz}, David and {Estrada-Carpenter}, Vicente and {Faber}, Sandra M. and {Fern{\'a}ndez}, Vital and {Finkelstein}, Keely D. and {Freundlich}, Jonathan and {Fujimoto}, Seiji and {Garc{\'\i}a-Argum{\'a}nez}, {\'A}ngela and {Gardner}, Jonathan P. and {Gawiser}, Eric and {G{\'o}mez-Guijarro}, Carlos and {Guo}, Yuchen and {Hamblin}, Kurt and {Hamilton}, Timothy S. and {Hathi}, Nimish P. and {Holwerda}, Benne W. and {Hirschmann}, Michaela and {Hutchison}, Taylor A. and {Jaskot}, Anne E. and {Jha}, Saurabh W. and {Jogee}, Shardha and {Juneau}, St{\'e}phanie and {Jung}, Intae and {Kassin}, Susan A. and {Le Bail}, Aur{\'e}lien and {Leung}, Gene C.~K. and {Lucas}, Ray A. and {Magnelli}, Benjamin and {Mantha}, Kameswara Bharadwaj and {Matharu}, Jasleen and {McGrath}, Elizabeth J. and {McIntosh}, Daniel H. and {Merlin}, Emiliano and {Mobasher}, Bahram and {Newman}, Jeffrey A. and {Nicholls}, David C. and {Pandya}, Viraj and {Rafelski}, Marc and {Ronayne}, Kaila and {Santini}, Paola and {Seill{\'e}}, Lise-Marie and {Shah}, Ekta A. and {Shen}, Lu and {Simons}, Raymond C. and {Snyder}, Gregory F. and {Stanway}, Elizabeth R. and {Straughn}, Amber N. and {Teplitz}, Harry I. and {Vanderhoof}, Brittany N. and {Vega-Ferrero}, Jes{\'u}s and {Wang}, Weichen and {Weiner}, Benjamin J. and {Willmer}, Christopher N.~A. and {Wuyts}, Stijn and {Zavala}, Jorge A. and {Ceers Team}},
        title = "{A Long Time Ago in a Galaxy Far, Far Away: A Candidate z {\ensuremath{\sim}} 12 Galaxy in Early JWST CEERS Imaging}",
      journal = {\apjl},
         year = 2022,
        month = dec,
       volume = {940},
       number = {2},
          eid = {L55},
        pages = {L55},
          doi = {10.3847/2041-8213/ac966e},
archivePrefix = {arXiv},
       eprint = {2207.12474},
 primaryClass = {astro-ph.GA},
       adsurl = {https://ui.adsabs.harvard.edu/abs/2022ApJ...940L..55F}
}

@ARTICLE{Finkelstein_2025,
       author = {{Finkelstein}, Steven L. and {Bagley}, Micaela B. and {Arrabal Haro}, Pablo and {Dickinson}, Mark and {Ferguson}, Henry C. and {Kartaltepe}, Jeyhan S. and {Kocevski}, Dale D. and {Koekemoer}, Anton M. and {Lotz}, Jennifer M. and {Papovich}, Casey and {P{\'e}rez-Gonz{\'a}lez}, Pablo G. and {Pirzkal}, Nor and {Somerville}, Rachel S. and {Trump}, Jonathan R. and {Yang}, Guang and {Yung}, L.~Y. Aaron and {Fontana}, Adriano and {Grazian}, Andrea and {Grogin}, Norman A. and {Kewley}, Lisa J. and {Kirkpatrick}, Allison and {Larson}, Rebecca L. and {Pentericci}, Laura and {Ravindranath}, Swara and {Wilkins}, Stephen M. and {Almaini}, Omar and {Amor{\'\i}n}, Ricardo O. and {Barro}, Guillermo and {Bhatawdekar}, Rachana and {Bisigello}, Laura and {Brooks}, Madisyn and {Buat}, V{\'e}ronique and {Buitrago}, Fernando and {Burgarella}, Denis and {Calabr{\`o}}, Antonello and {Castellano}, Marco and {Cheng}, Yingjie and {Cleri}, Nikko J. and {Cole}, Justin W. and {Cooper}, M.~C. and {Cooper}, Olivia R. and {Costantin}, Luca and {Cox}, Isa G. and {Croton}, Darren and {Daddi}, Emanuele and {Davis}, Kelcey and {Dekel}, Avishai and {Elbaz}, David and {Fern{\'a}ndez}, Vital and {Fujimoto}, Seiji and {Gandolfi}, Giovanni and {Gardner}, Jonathan P. and {Gawiser}, Eric and {Giavalisco}, Mauro and {G{\'o}mez-Guijarro}, Carlos and {Guo}, Yuchen and {Gupta}, Ansh R. and {Hathi}, Nimish P. and {Harish}, Santosh and {Henry}, Aur{\'e}lien and {Hirschmann}, Michaela and {Hu}, Weida and {Hutchison}, Taylor A. and {Iyer}, Kartheik G. and {Jaskot}, Anne E. and {Jha}, Saurabh W. and {Jung}, Intae and {Kassin}, Susan A. and {Kokorev}, Vasily and {Kurczynski}, Peter and {Leung}, Gene C.~K. and {Llerena}, Mario and {Long}, Arianna S. and {Lucas}, Ray A. and {Lu}, Shiying and {McGrath}, Elizabeth J. and {McIntosh}, Daniel H. and {Merlin}, Emiliano and {Mobasher}, Bahram and {Morales}, Alexa M. and {Napolitano}, Lorenzo and {Pacucci}, Fabio and {Pandya}, Viraj and {Rafelski}, Marc and {Rodighiero}, Giulia and {Rose}, Caitlin and {Santini}, Paola and {Seill{\'e}}, Lise-Marie and {Simons}, Raymond C. and {Shen}, Lu and {Straughn}, Amber N. and {Tacchella}, Sandro and {Taylor}, Anthony J. and {Vanderhoof}, Brittany N. and {Vega-Ferrero}, Jes{\'u}s and {Weiner}, Benjamin J. and {Willmer}, Christopher N.~A. and {Zhu}, Peixin and {Bell}, Eric F. and {Wuyts}, Stijn and {Holwerda}, Benne W. and {Wang}, Xin and {Wang}, Weichen and {Zavala}, Jorge A. and {CEERS Collaboration}},
        title = "{The Cosmic Evolution Early Release Science Survey (CEERS)}",
      journal = {\apjl},
         year = 2025,
        month = apr,
       volume = {983},
       number = {1},
          eid = {L4},
        pages = {L4},
          doi = {10.3847/2041-8213/adbbd3},
archivePrefix = {arXiv},
       eprint = {2501.04085},
 primaryClass = {astro-ph.GA},
       adsurl = {https://ui.adsabs.harvard.edu/abs/2025ApJ...983L...4F}
}

@ARTICLE{Fontana_2014,
       author = {{Fontana}, A. and {Dunlop}, J.~S. and {Paris}, D. and {Targett}, T.~A. and {Boutsia}, K. and {Castellano}, M. and {Galametz}, A. and {Grazian}, A. and {McLure}, R. and {Merlin}, E. and {Pentericci}, L. and {Wuyts}, S. and {Almaini}, O. and {Caputi}, K. and {Chary}, R.-R. and {Cirasuolo}, M. and {Conselice}, C.~J. and {Cooray}, A. and {Daddi}, E. and {Dickinson}, M. and {Faber}, S.~M. and {Fazio}, G. and {Ferguson}, H.~C. and {Giallongo}, E. and {Giavalisco}, M. and {Grogin}, N.~A. and {Hathi}, N. and {Koekemoer}, A.~M. and {Koo}, D.~C. and {Lucas}, R.~A. and {Nonino}, M. and {Rix}, H.~W. and {Renzini}, A. and {Rosario}, D. and {Santini}, P. and {Scarlata}, C. and {Sommariva}, V. and {Stark}, D.~P. and {van der Wel}, A. and {Vanzella}, E. and {Wild}, V. and {Yan}, H. and {Zibetti}, S.},
        title = "{The Hawk-I UDS and GOODS Survey (HUGS): Survey design and deep K-band number counts}",
      journal = {\aap},
         year = 2014,
        month = oct,
       volume = {570},
          eid = {A11},
        pages = {A11},
          doi = {10.1051/0004-6361/201423543},
archivePrefix = {arXiv},
       eprint = {1409.7082},
 primaryClass = {astro-ph.GA},
       adsurl = {https://ui.adsabs.harvard.edu/abs/2014A&A...570A..11F}
}

@ARTICLE{Fortuni_2023,
       author = {{Fortuni}, Flaminia and {Merlin}, Emiliano and {Fontana}, Adriano and {Giocoli}, Carlo and {Romelli}, Erik and {Graziani}, Luca and {Santini}, Paola and {Castellano}, Marco and {Charlot}, St{\'e}phane and {Chevallard}, Jacopo},
        title = "{FORECAST: A flexible software to forward model cosmological hydrodynamical simulations mimicking real observations}",
      journal = {\aap},
         year = 2023,
        month = sep,
       volume = {677},
          eid = {A102},
        pages = {A102},
          doi = {10.1051/0004-6361/202346725},
archivePrefix = {arXiv},
       eprint = {2305.19166},
 primaryClass = {astro-ph.IM},
       adsurl = {https://ui.adsabs.harvard.edu/abs/2023A&A...677A.102F}
}

@ARTICLE{Franco_2026,
       author = {{Franco}, Maximilien and {Casey}, Caitlin M. and {Koekemoer}, Anton M. and {Liu}, Daizhong and {Bagley}, Micaela B. and {McCracken}, Henry Joy and {Kartaltepe}, Jeyhan S. and {Akins}, Hollis B. and {Ilbert}, Olivier and {Shuntov}, Marko and {Harish}, Santosh and {Robertson}, Brant E. and {Arango-Toro}, Rafael C. and {Battisti}, Andrew J. and {Chartab}, Nima and {Drakos}, Nicole E. and {Faisst}, Andreas L. and {Flayhart}, Carter and {Gozaliasl}, Ghassem and {Hirschmann}, Michaela and {Massey}, Richard and {Rhodes}, Jason and {Sattari}, Zahra and {Scognamiglio}, Diana and {Weaver}, John R. and {Yang}, Lilan and {Zavala}, Jorge A. and {Berman}, Edward M. and {Gentile}, Fabrizio and {Gillman}, Steven and {Long}, Arianna S. and {Magdis}, Georgios and {McCleary}, Jacqueline E. and {McKinney}, Jed and {Mobasher}, Bahram and {Paquereau}, Louise and {Rest}, Armin and {Sanders}, David B. and {Toft}, Sune and {Yu}, Si-Yue},
        title = "{COSMOS-Web: Comprehensive Data Reduction for Wide-area JWST NIRCam Imaging}",
      journal = {\apj},
         year = 2026,
        month = mar,
       volume = {999},
       number = {2},
          eid = {200},
        pages = {200},
          doi = {10.3847/1538-4357/ae3aa1},
archivePrefix = {arXiv},
       eprint = {2506.03256},
 primaryClass = {astro-ph.IM},
       adsurl = {https://ui.adsabs.harvard.edu/abs/2026ApJ...999..200F}
}

@ARTICLE{GAIA_2016,
       author = {{Gaia Collaboration} and {Prusti}, T. and {de Bruijne}, J.~H.~J. and {Brown}, A.~G.~A. and {Vallenari}, A. and {Babusiaux}, C. and {Bailer-Jones}, C.~A.~L. and {Bastian}, U. and {Biermann}, M. and {Evans}, D.~W. and {Eyer}, L. and {Jansen}, F. and {Jordi}, C. and {Klioner}, S.~A. and {Lammers}, U. and {Lindegren}, L. and {Luri}, X. and {Mignard}, F. and {Milligan}, D.~J. and {Panem}, C. and {Poinsignon}, V. and {Pourbaix}, D. and {Randich}, S. and {Sarri}, G. and {Sartoretti}, P. and {Siddiqui}, H.~I. and {Soubiran}, C. and {Valette}, V. and {van Leeuwen}, F. and {Walton}, N.~A. and {Aerts}, C. and {Arenou}, F. and {Cropper}, M. and {Drimmel}, R. and {H{\o}g}, E. and {Katz}, D. and {Lattanzi}, M.~G. and {O'Mullane}, W. and {Grebel}, E.~K. and {Holland}, A.~D. and {Huc}, C. and {Passot}, X. and {Bramante}, L. and {Cacciari}, C. and {Casta{\~n}eda}, J. and {Chaoul}, L. and {Cheek}, N. and {De Angeli}, F. and {Fabricius}, C. and {Guerra}, R. and {Hern{\'a}ndez}, J. and {Jean-Antoine-Piccolo}, A. and {Masana}, E. and {Messineo}, R. and {Mowlavi}, N. and {Nienartowicz}, K. and {Ord{\'o}{\~n}ez-Blanco}, D. and {Panuzzo}, P. and {Portell}, J. and {Richards}, P.~J. and {Riello}, M. and {Seabroke}, G.~M. and {Tanga}, P. and {Th{\'e}venin}, F. and {Torra}, J. and {Els}, S.~G. and {Gracia-Abril}, G. and {Comoretto}, G. and {Garcia-Reinaldos}, M. and {Lock}, T. and {Mercier}, E. and {Altmann}, M. and {Andrae}, R. and {Astraatmadja}, T.~L. and {Bellas-Velidis}, I. and {Benson}, K. and {Berthier}, J. and {Blomme}, R. and {Busso}, G. and {Carry}, B. and {Cellino}, A. and {Clementini}, G. and {Cowell}, S. and {Creevey}, O. and {Cuypers}, J. and {Davidson}, M. and {De Ridder}, J. and {de Torres}, A. and {Delchambre}, L. and {Dell'Oro}, A. and {Ducourant}, C. and {Fr{\'e}mat}, Y. and {Garc{\'\i}a-Torres}, M. and {Gosset}, E. and {Halbwachs}, J. -L. and {Hambly}, N.~C. and {Harrison}, D.~L. and {Hauser}, M. and {Hestroffer}, D. and {Hodgkin}, S.~T. and {Huckle}, H.~E. and {Hutton}, A. and {Jasniewicz}, G. and {Jordan}, S. and {Kontizas}, M. and {Korn}, A.~J. and {Lanzafame}, A.~C. and {Manteiga}, M. and {Moitinho}, A. and {Muinonen}, K. and {Osinde}, J. and {Pancino}, E. and {Pauwels}, T. and {Petit}, J. -M. and {Recio-Blanco}, A. and {Robin}, A.~C. and {Sarro}, L.~M. and {Siopis}, C. and {Smith}, M. and {Smith}, K.~W. and {Sozzetti}, A. and {Thuillot}, W. and {van Reeven}, W. and {Viala}, Y. and {Abbas}, U. and {Abreu Aramburu}, A. and {Accart}, S. and {Aguado}, J.~J. and {Allan}, P.~M. and {Allasia}, W. and {Altavilla}, G. and {{\'A}lvarez}, M.~A. and {Alves}, J. and {Anderson}, R.~I. and {Andrei}, A.~H. and {Anglada Varela}, E. and {Antiche}, E. and {Antoja}, T. and {Ant{\'o}n}, S. and {Arcay}, B. and {Atzei}, A. and {Ayache}, L. and {Bach}, N. and {Baker}, S.~G. and {Balaguer-N{\'u}{\~n}ez}, L. and {Barache}, C. and {Barata}, C. and {Barbier}, A. and {Barblan}, F. and {Baroni}, M. and {Barrado y Navascu{\'e}s}, D. and {Barros}, M. and {Barstow}, M.~A. and {Becciani}, U. and {Bellazzini}, M. and {Bellei}, G. and {Bello Garc{\'\i}a}, A. and {Belokurov}, V. and {Bendjoya}, P. and {Berihuete}, A. and {Bianchi}, L. and {Bienaym{\'e}}, O. and {Billebaud}, F. and {Blagorodnova}, N. and {Blanco-Cuaresma}, S. and {Boch}, T. and {Bombrun}, A. and {Borrachero}, R. and {Bouquillon}, S. and {Bourda}, G. and {Bouy}, H. and {Bragaglia}, A. and {Breddels}, M.~A. and {Brouillet}, N. and {Br{\"u}semeister}, T. and {Bucciarelli}, B. and {Budnik}, F. and {Burgess}, P. and {Burgon}, R. and {Burlacu}, A. and {Busonero}, D. and {Buzzi}, R. and {Caffau}, E. and {Cambras}, J. and {Campbell}, H. and {Cancelliere}, R. and {Cantat-Gaudin}, T. and {Carlucci}, T. and {Carrasco}, J.~M. and {Castellani}, M. and {Charlot}, P. and {Charnas}, J. and {Charvet}, P. and {Chassat}, F. and {Chiavassa}, A. and {Clotet}, M. and {Cocozza}, G. and {Collins}, R.~S. and {Collins}, P. and {Costigan}, G. and {Crifo}, F. and {Cross}, N.~J.~G. and {Crosta}, M. and {Crowley}, C. and {Dafonte}, C. and {Damerdji}, Y. and {Dapergolas}, A. and {David}, P. and {David}, M. and {De Cat}, P. and {de Felice}, F. and {de Laverny}, P. and {De Luise}, F. and {De March}, R. and {de Martino}, D. and {de Souza}, R. and {Debosscher}, J. and {del Pozo}, E. and {Delbo}, M. and {Delgado}, A. and {Delgado}, H.~E. and {di Marco}, F. and {Di Matteo}, P. and {Diakite}, S. and {Distefano}, E. and {Dolding}, C. and {Dos Anjos}, S. and {Drazinos}, P. and {Dur{\'a}n}, J. and {Dzigan}, Y. and {Ecale}, E. and {Edvardsson}, B. and {Enke}, H. and {Erdmann}, M. and {Escolar}, D. and {Espina}, M. and {Evans}, N.~W. and {Eynard Bontemps}, G. and {Fabre}, C. and {Fabrizio}, M. and {Faigler}, S. and {Falc{\~a}o}, A.~J. and {Farr{\`a}s Casas}, M. and {Faye}, F. and {Federici}, L. and {Fedorets}, G. and {Fern{\'a}ndez-Hern{\'a}ndez}, J. and {Fernique}, P. and {Fienga}, A. and {Figueras}, F. and {Filippi}, F. and {Findeisen}, K. and {Fonti}, A. and {Fouesneau}, M. and {Fraile}, E. and {Fraser}, M. and {Fuchs}, J. and {Furnell}, R. and {Gai}, M. and {Galleti}, S. and {Galluccio}, L. and {Garabato}, D. and {Garc{\'\i}a-Sedano}, F. and {Gar{\'e}}, P. and {Garofalo}, A. and {Garralda}, N. and {Gavras}, P. and {Gerssen}, J. and {Geyer}, R. and {Gilmore}, G. and {Girona}, S. and {Giuffrida}, G. and {Gomes}, M. and {Gonz{\'a}lez-Marcos}, A. and {Gonz{\'a}lez-N{\'u}{\~n}ez}, J. and {Gonz{\'a}lez-Vidal}, J.~J. and {Granvik}, M. and {Guerrier}, A. and {Guillout}, P. and {Guiraud}, J. and {G{\'u}rpide}, A. and {Guti{\'e}rrez-S{\'a}nchez}, R. and {Guy}, L.~P. and {Haigron}, R. and {Hatzidimitriou}, D. and {Haywood}, M. and {Heiter}, U. and {Helmi}, A. and {Hobbs}, D. and {Hofmann}, W. and {Holl}, B. and {Holland}, G. and {Hunt}, J.~A.~S. and {Hypki}, A. and {Icardi}, V. and {Irwin}, M. and {Jevardat de Fombelle}, G. and {Jofr{\'e}}, P. and {Jonker}, P.~G. and {Jorissen}, A. and {Julbe}, F. and {Karampelas}, A. and {Kochoska}, A. and {Kohley}, R. and {Kolenberg}, K. and {Kontizas}, E. and {Koposov}, S.~E. and {Kordopatis}, G. and {Koubsky}, P. and {Kowalczyk}, A. and {Krone-Martins}, A. and {Kudryashova}, M. and {Kull}, I. and {Bachchan}, R.~K. and {Lacoste-Seris}, F. and {Lanza}, A.~F. and {Lavigne}, J. -B. and {Le Poncin-Lafitte}, C. and {Lebreton}, Y. and {Lebzelter}, T. and {Leccia}, S. and {Leclerc}, N. and {Lecoeur-Taibi}, I. and {Lemaitre}, V. and {Lenhardt}, H. and {Leroux}, F. and {Liao}, S. and {Licata}, E. and {Lindstr{\o}m}, H.~E.~P. and {Lister}, T.~A. and {Livanou}, E. and {Lobel}, A. and {L{\"o}ffler}, W. and {L{\'o}pez}, M. and {Lopez-Lozano}, A. and {Lorenz}, D. and {Loureiro}, T. and {MacDonald}, I. and {Magalh{\~a}es Fernandes}, T. and {Managau}, S. and {Mann}, R.~G. and {Mantelet}, G. and {Marchal}, O. and {Marchant}, J.~M. and {Marconi}, M. and {Marie}, J. and {Marinoni}, S. and {Marrese}, P.~M. and {Marschalk{\'o}}, G. and {Marshall}, D.~J. and {Mart{\'\i}n-Fleitas}, J.~M. and {Martino}, M. and {Mary}, N. and {Matijevi{\v{c}}}, G. and {Mazeh}, T. and {McMillan}, P.~J. and {Messina}, S. and {Mestre}, A. and {Michalik}, D. and {Millar}, N.~R. and {Miranda}, B.~M.~H. and {Molina}, D. and {Molinaro}, R. and {Molinaro}, M. and {Moln{\'a}r}, L. and {Moniez}, M. and {Montegriffo}, P. and {Monteiro}, D. and {Mor}, R. and {Mora}, A. and {Morbidelli}, R. and {Morel}, T. and {Morgenthaler}, S. and {Morley}, T. and {Morris}, D. and {Mulone}, A.~F. and {Muraveva}, T. and {Musella}, I. and {Narbonne}, J. and {Nelemans}, G. and {Nicastro}, L. and {Noval}, L. and {Ord{\'e}novic}, C. and {Ordieres-Mer{\'e}}, J. and {Osborne}, P. and {Pagani}, C. and {Pagano}, I. and {Pailler}, F. and {Palacin}, H. and {Palaversa}, L. and {Parsons}, P. and {Paulsen}, T. and {Pecoraro}, M. and {Pedrosa}, R. and {Pentik{\"a}inen}, H. and {Pereira}, J. and {Pichon}, B. and {Piersimoni}, A.~M. and {Pineau}, F. -X. and {Plachy}, E. and {Plum}, G. and {Poujoulet}, E. and {Pr{\v{s}}a}, A. and {Pulone}, L. and {Ragaini}, S. and {Rago}, S. and {Rambaux}, N. and {Ramos-Lerate}, M. and {Ranalli}, P. and {Rauw}, G. and {Read}, A. and {Regibo}, S. and {Renk}, F. and {Reyl{\'e}}, C. and {Ribeiro}, R.~A. and {Rimoldini}, L. and {Ripepi}, V. and {Riva}, A. and {Rixon}, G. and {Roelens}, M. and {Romero-G{\'o}mez}, M. and {Rowell}, N. and {Royer}, F. and {Rudolph}, A. and {Ruiz-Dern}, L. and {Sadowski}, G. and {Sagrist{\`a} Sell{\'e}s}, T. and {Sahlmann}, J. and {Salgado}, J. and {Salguero}, E. and {Sarasso}, M. and {Savietto}, H. and {Schnorhk}, A. and {Schultheis}, M. and {Sciacca}, E. and {Segol}, M. and {Segovia}, J.~C. and {Segransan}, D. and {Serpell}, E. and {Shih}, I. -C. and {Smareglia}, R. and {Smart}, R.~L. and {Smith}, C. and {Solano}, E. and {Solitro}, F. and {Sordo}, R. and {Soria Nieto}, S. and {Souchay}, J. and {Spagna}, A. and {Spoto}, F. and {Stampa}, U. and {Steele}, I.~A. and {Steidelm{\"u}ller}, H. and {Stephenson}, C.~A. and {Stoev}, H. and {Suess}, F.~F. and {S{\"u}veges}, M. and {Surdej}, J. and {Szabados}, L. and {Szegedi-Elek}, E. and {Tapiador}, D. and {Taris}, F. and {Tauran}, G. and {Taylor}, M.~B. and {Teixeira}, R. and {Terrett}, D. and {Tingley}, B. and {Trager}, S.~C. and {Turon}, C. and {Ulla}, A. and {Utrilla}, E. and {Valentini}, G. and {van Elteren}, A. and {Van Hemelryck}, E. and {van Leeuwen}, M. and {Varadi}, M. and {Vecchiato}, A. and {Veljanoski}, J. and {Via}, T. and {Vicente}, D. and {Vogt}, S. and {Voss}, H. and {Votruba}, V. and {Voutsinas}, S. and {Walmsley}, G. and {Weiler}, M. and {Weingrill}, K. and {Werner}, D. and {Wevers}, T. and {Whitehead}, G. and {Wyrzykowski}, {\L}. and {Yoldas}, A. and {{\v{Z}}erjal}, M. and {Zucker}, S. and {Zurbach}, C. and {Zwitter}, T. and {Alecu}, A. and {Allen}, M. and {Allende Prieto}, C. and {Amorim}, A. and {Anglada-Escud{\'e}}, G. and {Arsenijevic}, V. and {Azaz}, S. and {Balm}, P. and {Beck}, M. and {Bernstein}, H. -H. and {Bigot}, L. and {Bijaoui}, A. and {Blasco}, C. and {Bonfigli}, M. and {Bono}, G. and {Boudreault}, S. and {Bressan}, A. and {Brown}, S. and {Brunet}, P. -M. and {Bunclark}, P. and {Buonanno}, R. and {Butkevich}, A.~G. and {Carret}, C. and {Carrion}, C. and {Chemin}, L. and {Ch{\'e}reau}, F. and {Corcione}, L. and {Darmigny}, E. and {de Boer}, K.~S. and {de Teodoro}, P. and {de Zeeuw}, P.~T. and {Delle Luche}, C. and {Domingues}, C.~D. and {Dubath}, P. and {Fodor}, F. and {Fr{\'e}zouls}, B. and {Fries}, A. and {Fustes}, D. and {Fyfe}, D. and {Gallardo}, E. and {Gallegos}, J. and {Gardiol}, D. and {Gebran}, M. and {Gomboc}, A. and {G{\'o}mez}, A. and {Grux}, E. and {Gueguen}, A. and {Heyrovsky}, A. and {Hoar}, J. and {Iannicola}, G. and {Isasi Parache}, Y. and {Janotto}, A. -M. and {Joliet}, E. and {Jonckheere}, A. and {Keil}, R. and {Kim}, D. -W. and {Klagyivik}, P. and {Klar}, J. and {Knude}, J. and {Kochukhov}, O. and {Kolka}, I. and {Kos}, J. and {Kutka}, A. and {Lainey}, V. and {LeBouquin}, D. and {Liu}, C. and {Loreggia}, D. and {Makarov}, V.~V. and {Marseille}, M.~G. and {Martayan}, C. and {Martinez-Rubi}, O. and {Massart}, B. and {Meynadier}, F. and {Mignot}, S. and {Munari}, U. and {Nguyen}, A. -T. and {Nordlander}, T. and {Ocvirk}, P. and {O'Flaherty}, K.~S. and {Olias Sanz}, A. and {Ortiz}, P. and {Osorio}, J. and {Oszkiewicz}, D. and {Ouzounis}, A. and {Palmer}, M. and {Park}, P. and {Pasquato}, E. and {Peltzer}, C. and {Peralta}, J. and {P{\'e}turaud}, F. and {Pieniluoma}, T. and {Pigozzi}, E. and {Poels}, J. and {Prat}, G. and {Prod'homme}, T. and {Raison}, F. and {Rebordao}, J.~M. and {Risquez}, D. and {Rocca-Volmerange}, B. and {Rosen}, S. and {Ruiz-Fuertes}, M.~I. and {Russo}, F. and {Sembay}, S. and {Serraller Vizcaino}, I. and {Short}, A. and {Siebert}, A. and {Silva}, H. and {Sinachopoulos}, D. and {Slezak}, E. and {Soffel}, M. and {Sosnowska}, D. and {Strai{\v{z}}ys}, V. and {ter Linden}, M. and {Terrell}, D. and {Theil}, S. and {Tiede}, C. and {Troisi}, L. and {Tsalmantza}, P. and {Tur}, D. and {Vaccari}, M. and {Vachier}, F. and {Valles}, P. and {Van Hamme}, W. and {Veltz}, L. and {Virtanen}, J. and {Wallut}, J. -M. and {Wichmann}, R. and {Wilkinson}, M.~I. and {Ziaeepour}, H. and {Zschocke}, S.},
        title = "{The Gaia mission}",
      journal = {\aap},
         year = 2016,
        month = nov,
       volume = {595},
          eid = {A1},
        pages = {A1},
          doi = {10.1051/0004-6361/201629272},
archivePrefix = {arXiv},
       eprint = {1609.04153},
 primaryClass = {astro-ph.IM},
       adsurl = {https://ui.adsabs.harvard.edu/abs/2016A&A...595A...1G}
}

@ARTICLE{GAIA_2023,
       author = {{Gaia Collaboration} and {Vallenari}, A. and {Brown}, A.~G.~A. and {Prusti}, T. and {de Bruijne}, J.~H.~J. and {Arenou}, F. and {Babusiaux}, C. and {Biermann}, M. and {Creevey}, O.~L. and {Ducourant}, C. and {Evans}, D.~W. and {Eyer}, L. and {Guerra}, R. and {Hutton}, A. and {Jordi}, C. and {Klioner}, S.~A. and {Lammers}, U.~L. and {Lindegren}, L. and {Luri}, X. and {Mignard}, F. and {Panem}, C. and {Pourbaix}, D. and {Randich}, S. and {Sartoretti}, P. and {Soubiran}, C. and {Tanga}, P. and {Walton}, N.~A. and {Bailer-Jones}, C.~A.~L. and {Bastian}, U. and {Drimmel}, R. and {Jansen}, F. and {Katz}, D. and {Lattanzi}, M.~G. and {van Leeuwen}, F. and {Bakker}, J. and {Cacciari}, C. and {Casta{\~n}eda}, J. and {De Angeli}, F. and {Fabricius}, C. and {Fouesneau}, M. and {Fr{\'e}mat}, Y. and {Galluccio}, L. and {Guerrier}, A. and {Heiter}, U. and {Masana}, E. and {Messineo}, R. and {Mowlavi}, N. and {Nicolas}, C. and {Nienartowicz}, K. and {Pailler}, F. and {Panuzzo}, P. and {Riclet}, F. and {Roux}, W. and {Seabroke}, G.~M. and {Sordo}, R. and {Th{\'e}venin}, F. and {Gracia-Abril}, G. and {Portell}, J. and {Teyssier}, D. and {Altmann}, M. and {Andrae}, R. and {Audard}, M. and {Bellas-Velidis}, I. and {Benson}, K. and {Berthier}, J. and {Blomme}, R. and {Burgess}, P.~W. and {Busonero}, D. and {Busso}, G. and {C{\'a}novas}, H. and {Carry}, B. and {Cellino}, A. and {Cheek}, N. and {Clementini}, G. and {Damerdji}, Y. and {Davidson}, M. and {de Teodoro}, P. and {Nu{\~n}ez Campos}, M. and {Delchambre}, L. and {Dell'Oro}, A. and {Esquej}, P. and {Fern{\'a}ndez-Hern{\'a}ndez}, J. and {Fraile}, E. and {Garabato}, D. and {Garc{\'\i}a-Lario}, P. and {Gosset}, E. and {Haigron}, R. and {Halbwachs}, J. -L. and {Hambly}, N.~C. and {Harrison}, D.~L. and {Hern{\'a}ndez}, J. and {Hestroffer}, D. and {Hodgkin}, S.~T. and {Holl}, B. and {Jan{\ss}en}, K. and {Jevardat de Fombelle}, G. and {Jordan}, S. and {Krone-Martins}, A. and {Lanzafame}, A.~C. and {L{\"o}ffler}, W. and {Marchal}, O. and {Marrese}, P.~M. and {Moitinho}, A. and {Muinonen}, K. and {Osborne}, P. and {Pancino}, E. and {Pauwels}, T. and {Recio-Blanco}, A. and {Reyl{\'e}}, C. and {Riello}, M. and {Rimoldini}, L. and {Roegiers}, T. and {Rybizki}, J. and {Sarro}, L.~M. and {Siopis}, C. and {Smith}, M. and {Sozzetti}, A. and {Utrilla}, E. and {van Leeuwen}, M. and {Abbas}, U. and {{\'A}brah{\'a}m}, P. and {Abreu Aramburu}, A. and {Aerts}, C. and {Aguado}, J.~J. and {Ajaj}, M. and {Aldea-Montero}, F. and {Altavilla}, G. and {{\'A}lvarez}, M.~A. and {Alves}, J. and {Anders}, F. and {Anderson}, R.~I. and {Anglada Varela}, E. and {Antoja}, T. and {Baines}, D. and {Baker}, S.~G. and {Balaguer-N{\'u}{\~n}ez}, L. and {Balbinot}, E. and {Balog}, Z. and {Barache}, C. and {Barbato}, D. and {Barros}, M. and {Barstow}, M.~A. and {Bartolom{\'e}}, S. and {Bassilana}, J. -L. and {Bauchet}, N. and {Becciani}, U. and {Bellazzini}, M. and {Berihuete}, A. and {Bernet}, M. and {Bertone}, S. and {Bianchi}, L. and {Binnenfeld}, A. and {Blanco-Cuaresma}, S. and {Blazere}, A. and {Boch}, T. and {Bombrun}, A. and {Bossini}, D. and {Bouquillon}, S. and {Bragaglia}, A. and {Bramante}, L. and {Breedt}, E. and {Bressan}, A. and {Brouillet}, N. and {Brugaletta}, E. and {Bucciarelli}, B. and {Burlacu}, A. and {Butkevich}, A.~G. and {Buzzi}, R. and {Caffau}, E. and {Cancelliere}, R. and {Cantat-Gaudin}, T. and {Carballo}, R. and {Carlucci}, T. and {Carnerero}, M.~I. and {Carrasco}, J.~M. and {Casamiquela}, L. and {Castellani}, M. and {Castro-Ginard}, A. and {Chaoul}, L. and {Charlot}, P. and {Chemin}, L. and {Chiaramida}, V. and {Chiavassa}, A. and {Chornay}, N. and {Comoretto}, G. and {Contursi}, G. and {Cooper}, W.~J. and {Cornez}, T. and {Cowell}, S. and {Crifo}, F. and {Cropper}, M. and {Crosta}, M. and {Crowley}, C. and {Dafonte}, C. and {Dapergolas}, A. and {David}, M. and {David}, P. and {de Laverny}, P. and {De Luise}, F. and {De March}, R. and {De Ridder}, J. and {de Souza}, R. and {de Torres}, A. and {del Peloso}, E.~F. and {del Pozo}, E. and {Delbo}, M. and {Delgado}, A. and {Delisle}, J. -B. and {Demouchy}, C. and {Dharmawardena}, T.~E. and {Di Matteo}, P. and {Diakite}, S. and {Diener}, C. and {Distefano}, E. and {Dolding}, C. and {Edvardsson}, B. and {Enke}, H. and {Fabre}, C. and {Fabrizio}, M. and {Faigler}, S. and {Fedorets}, G. and {Fernique}, P. and {Fienga}, A. and {Figueras}, F. and {Fournier}, Y. and {Fouron}, C. and {Fragkoudi}, F. and {Gai}, M. and {Garcia-Gutierrez}, A. and {Garcia-Reinaldos}, M. and {Garc{\'\i}a-Torres}, M. and {Garofalo}, A. and {Gavel}, A. and {Gavras}, P. and {Gerlach}, E. and {Geyer}, R. and {Giacobbe}, P. and {Gilmore}, G. and {Girona}, S. and {Giuffrida}, G. and {Gomel}, R. and {Gomez}, A. and {Gonz{\'a}lez-N{\'u}{\~n}ez}, J. and {Gonz{\'a}lez-Santamar{\'\i}a}, I. and {Gonz{\'a}lez-Vidal}, J.~J. and {Granvik}, M. and {Guillout}, P. and {Guiraud}, J. and {Guti{\'e}rrez-S{\'a}nchez}, R. and {Guy}, L.~P. and {Hatzidimitriou}, D. and {Hauser}, M. and {Haywood}, M. and {Helmer}, A. and {Helmi}, A. and {Sarmiento}, M.~H. and {Hidalgo}, S.~L. and {Hilger}, T. and {H{\l}adczuk}, N. and {Hobbs}, D. and {Holland}, G. and {Huckle}, H.~E. and {Jardine}, K. and {Jasniewicz}, G. and {Jean-Antoine Piccolo}, A. and {Jim{\'e}nez-Arranz}, {\'O}. and {Jorissen}, A. and {Juaristi Campillo}, J. and {Julbe}, F. and {Karbevska}, L. and {Kervella}, P. and {Khanna}, S. and {Kontizas}, M. and {Kordopatis}, G. and {Korn}, A.~J. and {K{\'o}sp{\'a}l}, {\'A}. and {Kostrzewa-Rutkowska}, Z. and {Kruszy{\'n}ska}, K. and {Kun}, M. and {Laizeau}, P. and {Lambert}, S. and {Lanza}, A.~F. and {Lasne}, Y. and {Le Campion}, J. -F. and {Lebreton}, Y. and {Lebzelter}, T. and {Leccia}, S. and {Leclerc}, N. and {Lecoeur-Taibi}, I. and {Liao}, S. and {Licata}, E.~L. and {Lindstr{\o}m}, H.~E.~P. and {Lister}, T.~A. and {Livanou}, E. and {Lobel}, A. and {Lorca}, A. and {Loup}, C. and {Madrero Pardo}, P. and {Magdaleno Romeo}, A. and {Managau}, S. and {Mann}, R.~G. and {Manteiga}, M. and {Marchant}, J.~M. and {Marconi}, M. and {Marcos}, J. and {Marcos Santos}, M.~M.~S. and {Mar{\'\i}n Pina}, D. and {Marinoni}, S. and {Marocco}, F. and {Marshall}, D.~J. and {Martin Polo}, L. and {Mart{\'\i}n-Fleitas}, J.~M. and {Marton}, G. and {Mary}, N. and {Masip}, A. and {Massari}, D. and {Mastrobuono-Battisti}, A. and {Mazeh}, T. and {McMillan}, P.~J. and {Messina}, S. and {Michalik}, D. and {Millar}, N.~R. and {Mints}, A. and {Molina}, D. and {Molinaro}, R. and {Moln{\'a}r}, L. and {Monari}, G. and {Mongui{\'o}}, M. and {Montegriffo}, P. and {Montero}, A. and {Mor}, R. and {Mora}, A. and {Morbidelli}, R. and {Morel}, T. and {Morris}, D. and {Muraveva}, T. and {Murphy}, C.~P. and {Musella}, I. and {Nagy}, Z. and {Noval}, L. and {Oca{\~n}a}, F. and {Ogden}, A. and {Ordenovic}, C. and {Osinde}, J.~O. and {Pagani}, C. and {Pagano}, I. and {Palaversa}, L. and {Palicio}, P.~A. and {Pallas-Quintela}, L. and {Panahi}, A. and {Payne-Wardenaar}, S. and {Pe{\~n}alosa Esteller}, X. and {Penttil{\"a}}, A. and {Pichon}, B. and {Piersimoni}, A.~M. and {Pineau}, F. -X. and {Plachy}, E. and {Plum}, G. and {Poggio}, E. and {Pr{\v{s}}a}, A. and {Pulone}, L. and {Racero}, E. and {Ragaini}, S. and {Rainer}, M. and {Raiteri}, C.~M. and {Rambaux}, N. and {Ramos}, P. and {Ramos-Lerate}, M. and {Re Fiorentin}, P. and {Regibo}, S. and {Richards}, P.~J. and {Rios Diaz}, C. and {Ripepi}, V. and {Riva}, A. and {Rix}, H. -W. and {Rixon}, G. and {Robichon}, N. and {Robin}, A.~C. and {Robin}, C. and {Roelens}, M. and {Rogues}, H.~R.~O. and {Rohrbasser}, L. and {Romero-G{\'o}mez}, M. and {Rowell}, N. and {Royer}, F. and {Ruz Mieres}, D. and {Rybicki}, K.~A. and {Sadowski}, G. and {S{\'a}ez N{\'u}{\~n}ez}, A. and {Sagrist{\`a} Sell{\'e}s}, A. and {Sahlmann}, J. and {Salguero}, E. and {Samaras}, N. and {Sanchez Gimenez}, V. and {Sanna}, N. and {Santove{\~n}a}, R. and {Sarasso}, M. and {Schultheis}, M. and {Sciacca}, E. and {Segol}, M. and {Segovia}, J.~C. and {S{\'e}gransan}, D. and {Semeux}, D. and {Shahaf}, S. and {Siddiqui}, H.~I. and {Siebert}, A. and {Siltala}, L. and {Silvelo}, A. and {Slezak}, E. and {Slezak}, I. and {Smart}, R.~L. and {Snaith}, O.~N. and {Solano}, E. and {Solitro}, F. and {Souami}, D. and {Souchay}, J. and {Spagna}, A. and {Spina}, L. and {Spoto}, F. and {Steele}, I.~A. and {Steidelm{\"u}ller}, H. and {Stephenson}, C.~A. and {S{\"u}veges}, M. and {Surdej}, J. and {Szabados}, L. and {Szegedi-Elek}, E. and {Taris}, F. and {Taylor}, M.~B. and {Teixeira}, R. and {Tolomei}, L. and {Tonello}, N. and {Torra}, F. and {Torra}, J. and {Torralba Elipe}, G. and {Trabucchi}, M. and {Tsounis}, A.~T. and {Turon}, C. and {Ulla}, A. and {Unger}, N. and {Vaillant}, M.~V. and {van Dillen}, E. and {van Reeven}, W. and {Vanel}, O. and {Vecchiato}, A. and {Viala}, Y. and {Vicente}, D. and {Voutsinas}, S. and {Weiler}, M. and {Wevers}, T. and {Wyrzykowski}, {\L}. and {Yoldas}, A. and {Yvard}, P. and {Zhao}, H. and {Zorec}, J. and {Zucker}, S. and {Zwitter}, T.},
        title = "{Gaia Data Release 3. Summary of the content and survey properties}",
      journal = {\aap},
         year = 2023,
        month = jun,
       volume = {674},
          eid = {A1},
        pages = {A1},
          doi = {10.1051/0004-6361/202243940},
archivePrefix = {arXiv},
       eprint = {2208.00211},
 primaryClass = {astro-ph.GA},
       adsurl = {https://ui.adsabs.harvard.edu/abs/2023A&A...674A...1G}
}

@article{Garwood_1936,
	author = {Garwood, F.},
	journal = {Biometrika},
	month = {12},
	number = {3-4},
	pages = {437-442},
	title = {(i) Fiducial Limits for the Poisson Distribution},
	volume = {28},
	year = {1936}}

@ARTICLE{Girardi_2000,
       author = {{Girardi}, L. and {Bressan}, A. and {Bertelli}, G. and {Chiosi}, C.},
        title = "{Evolutionary tracks and isochrones for low- and intermediate-mass stars: From 0.15 to 7 M$_{sun}$, and from Z=0.0004 to 0.03}",
      journal = {\aaps},
         year = 2000,
        month = feb,
       volume = {141},
        pages = {371-383},
          doi = {10.1051/aas:2000126},
archivePrefix = {arXiv},
       eprint = {astro-ph/9910164},
 primaryClass = {astro-ph},
       adsurl = {https://ui.adsabs.harvard.edu/abs/2000A&AS..141..371G}
}

@ARTICLE{Gnedin_2000,
       author = {{Gnedin}, Nickolay Y.},
        title = "{Effect of Reionization on Structure Formation in the Universe}",
      journal = {\apj},
         year = 2000,
        month = oct,
       volume = {542},
       number = {2},
        pages = {535-541},
          doi = {10.1086/317042},
archivePrefix = {arXiv},
       eprint = {astro-ph/0002151},
 primaryClass = {astro-ph},
       adsurl = {https://ui.adsabs.harvard.edu/abs/2000ApJ...542..535G}
}

@ARTICLE{Gordon_2003,
       author = {{Gordon}, Karl D. and {Clayton}, Geoffrey C. and {Misselt}, K.~A. and {Landolt}, Arlo U. and {Wolff}, Michael J.},
        title = "{A Quantitative Comparison of the Small Magellanic Cloud, Large Magellanic Cloud, and Milky Way Ultraviolet to Near-Infrared Extinction Curves}",
      journal = {\apj},
         year = 2003,
        month = sep,
       volume = {594},
       number = {1},
        pages = {279-293},
          doi = {10.1086/376774},
archivePrefix = {arXiv},
       eprint = {astro-ph/0305257},
 primaryClass = {astro-ph},
       adsurl = {https://ui.adsabs.harvard.edu/abs/2003ApJ...594..279G}
}

@article{Gould_1997,
  title = {Pair Production in Photon-Photon Collisions},
  author = {Gould, Robert J. and Schr\'eder, G\'erard P.},
  journal = {Phys. Rev.},
  volume = {155},
  issue = {5},
  pages = {1404--1407},
  numpages = {0},
  year = {1967},
  month = {Mar},
  publisher = {American Physical Society},
  doi = {10.1103/PhysRev.155.1404},
  url = {https://link.aps.org/doi/10.1103/PhysRev.155.1404}
}

@ARTICLE{Greene_2024,
       author = {{Greene}, Jenny E. and {Labbe}, Ivo and {Goulding}, Andy D. and {Furtak}, Lukas J. and {Chemerynska}, Iryna and {Kokorev}, Vasily and {Dayal}, Pratika and {Volonteri}, Marta and {Williams}, Christina C. and {Wang}, Bingjie and {Setton}, David J. and {Burgasser}, Adam J. and {Bezanson}, Rachel and {Atek}, Hakim and {Brammer}, Gabriel and {Cutler}, Sam E. and {Feldmann}, Robert and {Fujimoto}, Seiji and {Glazebrook}, Karl and {de Graaff}, Anna and {Khullar}, Gourav and {Leja}, Joel and {Marchesini}, Danilo and {Maseda}, Michael V. and {Matthee}, Jorryt and {Miller}, Tim B. and {Naidu}, Rohan P. and {Nanayakkara}, Themiya and {Oesch}, Pascal A. and {Pan}, Richard and {Papovich}, Casey and {Price}, Sedona H. and {van Dokkum}, Pieter and {Weaver}, John R. and {Whitaker}, Katherine E. and {Zitrin}, Adi},
        title = "{UNCOVER Spectroscopy Confirms the Surprising Ubiquity of Active Galactic Nuclei in Red Sources at z > 5}",
      journal = {\apj},
         year = 2024,
        month = mar,
       volume = {964},
       number = {1},
          eid = {39},
        pages = {39},
          doi = {10.3847/1538-4357/ad1e5f},
archivePrefix = {arXiv},
       eprint = {2309.05714},
 primaryClass = {astro-ph.GA},
       adsurl = {https://ui.adsabs.harvard.edu/abs/2024ApJ...964...39G}
}

@ARTICLE{Graham_2005,
       author = {{Graham}, Alister W. and {Driver}, Simon P.},
        title = "{A Concise Reference to (Projected) S{\'e}rsic R$^{1/n}$ Quantities, Including Concentration, Profile Slopes, Petrosian Indices, and Kron Magnitudes}",
      journal = {\pasa},
         year = 2005,
        month = jan,
       volume = {22},
       number = {2},
        pages = {118-127},
          doi = {10.1071/AS05001},
archivePrefix = {arXiv},
       eprint = {astro-ph/0503176},
 primaryClass = {astro-ph},
       adsurl = {https://ui.adsabs.harvard.edu/abs/2005PASA...22..118G}
}

@ARTICLE{Gruppioni_2013,
       author = {{Gruppioni}, C. and {Pozzi}, F. and {Rodighiero}, G. and {Delvecchio}, I. and {Berta}, S. and {Pozzetti}, L. and {Zamorani}, G. and {Andreani}, P. and {Cimatti}, A. and {Ilbert}, O. and {Le Floc'h}, E. and {Lutz}, D. and {Magnelli}, B. and {Marchetti}, L. and {Monaco}, P. and {Nordon}, R. and {Oliver}, S. and {Popesso}, P. and {Riguccini}, L. and {Roseboom}, I. and {Rosario}, D.~J. and {Sargent}, M. and {Vaccari}, M. and {Altieri}, B. and {Aussel}, H. and {Bongiovanni}, A. and {Cepa}, J. and {Daddi}, E. and {Dom{\'\i}nguez-S{\'a}nchez}, H. and {Elbaz}, D. and {F{\"o}rster Schreiber}, N. and {Genzel}, R. and {Iribarrem}, A. and {Magliocchetti}, M. and {Maiolino}, R. and {Poglitsch}, A. and {P{\'e}rez Garc{\'\i}a}, A. and {Sanchez-Portal}, M. and {Sturm}, E. and {Tacconi}, L. and {Valtchanov}, I. and {Amblard}, A. and {Arumugam}, V. and {Bethermin}, M. and {Bock}, J. and {Boselli}, A. and {Buat}, V. and {Burgarella}, D. and {Castro-Rodr{\'\i}guez}, N. and {Cava}, A. and {Chanial}, P. and {Clements}, D.~L. and {Conley}, A. and {Cooray}, A. and {Dowell}, C.~D. and {Dwek}, E. and {Eales}, S. and {Franceschini}, A. and {Glenn}, J. and {Griffin}, M. and {Hatziminaoglou}, E. and {Ibar}, E. and {Isaak}, K. and {Ivison}, R.~J. and {Lagache}, G. and {Levenson}, L. and {Lu}, N. and {Madden}, S. and {Maffei}, B. and {Mainetti}, G. and {Nguyen}, H.~T. and {O'Halloran}, B. and {Page}, M.~J. and {Panuzzo}, P. and {Papageorgiou}, A. and {Pearson}, C.~P. and {P{\'e}rez-Fournon}, I. and {Pohlen}, M. and {Rigopoulou}, D. and {Rowan-Robinson}, M. and {Schulz}, B. and {Scott}, D. and {Seymour}, N. and {Shupe}, D.~L. and {Smith}, A.~J. and {Stevens}, J.~A. and {Symeonidis}, M. and {Trichas}, M. and {Tugwell}, K.~E. and {Vigroux}, L. and {Wang}, L. and {Wright}, G. and {Xu}, C.~K. and {Zemcov}, M. and {Bardelli}, S. and {Carollo}, M. and {Contini}, T. and {Le F{\'e}vre}, O. and {Lilly}, S. and {Mainieri}, V. and {Renzini}, A. and {Scodeggio}, M. and {Zucca}, E.},
        title = "{The Herschel PEP/HerMES luminosity function - I. Probing the evolution of PACS selected Galaxies to z ≃ 4}",
      journal = {\mnras},
         year = 2013,
        month = jun,
       volume = {432},
       number = {1},
        pages = {23-52},
          doi = {10.1093/mnras/stt308},
archivePrefix = {arXiv},
       eprint = {1302.5209},
 primaryClass = {astro-ph.CO},
       adsurl = {https://ui.adsabs.harvard.edu/abs/2013MNRAS.432...23G}
}

@ARTICLE{Gutkin_2016,
       author = {{Gutkin}, Julia and {Charlot}, St{\'e}phane and {Bruzual}, Gustavo},
        title = "{Modelling the nebular emission from primeval to present-day star-forming galaxies}",
      journal = {\mnras},
         year = 2016,
        month = oct,
       volume = {462},
       number = {2},
        pages = {1757-1774},
          doi = {10.1093/mnras/stw1716},
archivePrefix = {arXiv},
       eprint = {1607.06086},
 primaryClass = {astro-ph.GA},
       adsurl = {https://ui.adsabs.harvard.edu/abs/2016MNRAS.462.1757G}
}

@ARTICLE{Harish_2025,
       author = {{Harish}, Santosh and {Kartaltepe}, Jeyhan S. and {Liu}, Daizhong and {Koekemoer}, Anton M. and {Casey}, Caitlin M. and {Franco}, Maximilien and {Akins}, Hollis B. and {Ilbert}, Olivier and {Shuntov}, Marko and {Drakos}, Nicole E. and {Engesser}, Mike and {Faisst}, Andreas L. and {Gozaliasl}, Ghassem and {Martin}, Crystal L. and {Hirschmann}, Michaela and {Kokorev}, Vasily and {Lambrides}, Erini and {McCracken}, Henry Joy and {McKinney}, Jed and {Paquereau}, Louise and {Rhodes}, Jason and {Robertson}, Brant E.},
        title = "{COSMOS-Web: MIRI Data Reduction and Number Counts at 7.7 {\ensuremath{\mu}}m Using JWST}",
      journal = {\apj},
         year = 2025,
        month = oct,
       volume = {992},
       number = {1},
          eid = {45},
        pages = {45},
          doi = {10.3847/1538-4357/adfa1e},
archivePrefix = {arXiv},
       eprint = {2506.03306},
 primaryClass = {astro-ph.GA},
       adsurl = {https://ui.adsabs.harvard.edu/abs/2025ApJ...992...45H}
}

@Article{Harris_2020,
 title         = {Array programming with {NumPy}},
 author        = {Charles R. Harris and K. Jarrod Millman and St{\'{e}}fan J.
                 van der Walt and Ralf Gommers and Pauli Virtanen and David
                 Cournapeau and Eric Wieser and Julian Taylor and Sebastian
                 Berg and Nathaniel J. Smith and Robert Kern and Matti Picus
                 and Stephan Hoyer and Marten H. van Kerkwijk and Matthew
                 Brett and Allan Haldane and Jaime Fern{\'{a}}ndez del
                 R{\'{i}}o and Mark Wiebe and Pearu Peterson and Pierre
                 G{\'{e}}rard-Marchant and Kevin Sheppard and Tyler Reddy and
                 Warren Weckesser and Hameer Abbasi and Christoph Gohlke and
                 Travis E. Oliphant},
 year          = {2020},
 month         = sep,
 journal       = {Nature},
 volume        = {585},
 number        = {7825},
 pages         = {357--362},
 doi           = {10.1038/s41586-020-2649-2},
 publisher     = {Springer Science and Business Media {LLC}},
 url           = {https://doi.org/10.1038/s41586-020-2649-2}
}

@ARTICLE{Harvey_2025,
       author = {{Harvey}, Thomas and {Conselice}, Christopher J. and {Adams}, Nathan J. and {Austin}, Duncan and {Juod{\v{z}}balis}, Ignas and {Trussler}, James and {Li}, Qiong and {Ormerod}, Katherine and {Ferreira}, Leonardo and {Lovell}, Christopher C. and {Duan}, Qiao and {Westcott}, Lewi and {Harris}, Honor and {Bhatawdekar}, Rachana and {Coe}, Dan and {Cohen}, Seth H. and {Caruana}, Joseph and {Cheng}, Cheng and {Driver}, Simon P. and {Frye}, Brenda and {Furtak}, Lukas J. and {Grogin}, Norman A. and {Hathi}, Nimish P. and {Holwerda}, Benne W. and {Jansen}, Rolf A. and {Koekemoer}, Anton M. and {Marshall}, Madeline A. and {Nonino}, Mario and {Vijayan}, Aswin P. and {Wilkins}, Stephen M. and {Windhorst}, Rogier and {Willmer}, Christopher N.~A. and {Yan}, Haojing and {Zitrin}, Adi},
        title = "{EPOCHS. IV. SED Modeling Assumptions and Their Impact on the Stellar Mass Function at 6.5 {\ensuremath{\leq}} z {\ensuremath{\leq}} 13.5 Using PEARLS and Public JWST Observations}",
      journal = {\apj},
         year = 2025,
        month = jan,
       volume = {978},
       number = {1},
          eid = {89},
        pages = {89},
          doi = {10.3847/1538-4357/ad8c29},
archivePrefix = {arXiv},
       eprint = {2403.03908},
 primaryClass = {astro-ph.GA},
       adsurl = {https://ui.adsabs.harvard.edu/abs/2025ApJ...978...89H}
}

@ARTICLE{Harvey_2025_b,
       author = {{Harvey}, Thomas and {Conselice}, Christopher J. and {Adams}, Nathan J. and {Austin}, Duncan and {Li}, Qiong and {Rusakov}, Vadim and {Westcott}, Lewi and {Goolsby}, Caio M. and {Lovell}, Christopher C. and {Cochrane}, Rachel K. and {Vijayan}, Aswin P. and {Trussler}, James},
        title = "{Behind the spotlight: a systematic assessment of outshining using NIRCam medium bands in the JADES Origins Field}",
      journal = {\mnras},
         year = 2025,
        month = oct,
       volume = {542},
       number = {4},
        pages = {2998-3027},
          doi = {10.1093/mnras/staf1396},
archivePrefix = {arXiv},
       eprint = {2504.05244},
 primaryClass = {astro-ph.GA},
       adsurl = {https://ui.adsabs.harvard.edu/abs/2025MNRAS.542.2998H}
}

@ARTICLE{Herrero_Carrin_2026,
       author = {{Herrero-Carri{\'o}n}, Diego and {Spinoso}, Daniele and {Izquierdo-Villalba}, David and {Su}, Tong and {Bonoli}, Silvia and {Renard}, Pablo},
        title = "{Back to basics: Little Red Dots as galaxies and dust-obscured AGNs in a synthetic NIRCam sky simulated with L-GalaxiesBH}",
      journal = {\mnras},
         year = 2026,
        month = apr,
       volume = {547},
       number = {4},
          eid = {stag478},
        pages = {stag478},
          doi = {10.1093/mnras/stag478},
archivePrefix = {arXiv},
       eprint = {2511.10725},
 primaryClass = {astro-ph.GA},
       adsurl = {https://ui.adsabs.harvard.edu/abs/2026MNRAS.547ag478H}
}

@ARTICLE{Hess_2013,
       author = {{HESS Collaboration} and {Abramowski}, A. and {Acero}, F. and {Aharonian}, F. and {Akhperjanian}, A.~G. and {Anton}, G. and {Balenderan}, S. and {Balzer}, A. and {Barnacka}, A. and {Becherini}, Y. and {Becker Tjus}, J. and {Bernl{\"o}hr}, K. and {Birsin}, E. and {Biteau}, J. and {Bochow}, A. and {Boisson}, C. and {Bolmont}, J. and {Bordas}, P. and {Brucker}, J. and {Brun}, F. and {Brun}, P. and {Bulik}, T. and {Carrigan}, S. and {Casanova}, S. and {Cerruti}, M. and {Chadwick}, P.~M. and {Charbonnier}, A. and {Chaves}, R.~C.~G. and {Cheesebrough}, A. and {Cologna}, G. and {Conrad}, J. and {Couturier}, C. and {Dalton}, M. and {Daniel}, M.~K. and {Davids}, I.~D. and {Degrange}, B. and {Deil}, C. and {deWilt}, P. and {Dickinson}, H.~J. and {Djannati-Ata{\"\i}}, A. and {Domainko}, W. and {O'C. Drury}, L. and {Dubus}, G. and {Dutson}, K. and {Dyks}, J. and {Dyrda}, M. and {Egberts}, K. and {Eger}, P. and {Espigat}, P. and {Fallon}, L. and {Farnier}, C. and {Fegan}, S. and {Feinstein}, F. and {Fernandes}, M.~V. and {Fernandez}, D. and {Fiasson}, A. and {Fontaine}, G. and {F{\"o}rster}, A. and {F{\"u}{\ss}ling}, M. and {Gajdus}, M. and {Gallant}, Y.~A. and {Garrigoux}, T. and {Gast}, H. and {Giebels}, B. and {Glicenstein}, J.~F. and {Gl{\"u}ck}, B. and {G{\"o}ring}, D. and {Grondin}, M.-H. and {H{\"a}ffner}, S. and {Hague}, J.~D. and {Hahn}, J. and {Hampf}, D. and {Harris}, J. and {Heinz}, S. and {Heinzelmann}, G. and {Henri}, G. and {Hermann}, G. and {Hillert}, A. and {Hinton}, J.~A. and {Hofmann}, W. and {Hofverberg}, P. and {Holler}, M. and {Horns}, D. and {Jacholkowska}, A. and {Jahn}, C. and {Jamrozy}, M. and {Jung}, I. and {Kastendieck}, M.~A. and {Katarzy{\'n}ski}, K. and {Katz}, U. and {Kaufmann}, S. and {Kh{\'e}lifi}, B. and {Klochkov}, D. and {Klu{\'z}niak}, W. and {Kneiske}, T. and {Komin}, Nu. and {Kosack}, K. and {Kossakowski}, R. and {Krayzel}, F. and {Laffon}, H. and {Lamanna}, G. and {Lenain}, J.-P. and {Lennarz}, D. and {Lohse}, T. and {Lopatin}, A. and {Lu}, C.-C. and {Marandon}, V. and {Marcowith}, A. and {Masbou}, J. and {Maurin}, G. and {Maxted}, N. and {Mayer}, M. and {McComb}, T.~J.~L. and {Medina}, M.~C. and {M{\'e}hault}, J. and {Menzler}, U. and {Moderski}, R. and {Mohamed}, M. and {Moulin}, E. and {Naumann}, C.~L. and {Naumann-Godo}, M. and {de Naurois}, M. and {Nedbal}, D. and {Nguyen}, N. and {Niemiec}, J. and {Nolan}, S.~J. and {Ohm}, S. and {de O{\~n}a Wilhelmi}, E. and {Opitz}, B. and {Ostrowski}, M. and {Oya}, I. and {Panter}, M. and {Parsons}, D. and {Paz Arribas}, M. and {Pekeur}, N.~W. and {Pelletier}, G. and {Perez}, J. and {Petrucci}, P.-O. and {Peyaud}, B. and {Pita}, S. and {P{\"u}hlhofer}, G. and {Punch}, M. and {Quirrenbach}, A. and {Raue}, M. and {Reimer}, A. and {Reimer}, O. and {Renaud}, M. and {de los Reyes}, R. and {Rieger}, F. and {Ripken}, J. and {Rob}, L. and {Rosier-Lees}, S. and {Rowell}, G. and {Rudak}, B. and {Rulten}, C.~B. and {Sahakian}, V. and {Sanchez}, D.~A. and {Santangelo}, A. and {Schlickeiser}, R. and {Schulz}, A. and {Schwanke}, U. and {Schwarzburg}, S. and {Schwemmer}, S. and {Sheidaei}, F. and {Skilton}, J.~L. and {Sol}, H. and {Spengler}, G. and {Stawarz}, {\L}. and {Steenkamp}, R. and {Stegmann}, C. and {Stinzing}, F. and {Stycz}, K. and {Sushch}, I. and {Szostek}, A. and {Tavernet}, J.-P. and {Terrier}, R. and {Tluczykont}, M. and {Valerius}, K. and {van Eldik}, C. and {Vasileiadis}, G. and {Venter}, C. and {Viana}, A. and {Vincent}, P. and {V{\"o}lk}, H.~J. and {Volpe}, F. and {Vorobiov}, S. and {Vorster}, M. and {Wagner}, S.~J. and {Ward}, M. and {White}, R. and {Wierzcholska}, A. and {Wouters}, D. and {Zacharias}, M. and {Zajczyk}, A. and {Zdziarski}, A.~A. and {Zech}, A. and {Zechlin}, H.-S.},
        title = "{Measurement of the extragalactic background light imprint on the spectra of the brightest blazars observed with H.E.S.S.}",
      journal = {\aap},
         year = 2013,
        month = feb,
       volume = {550},
          eid = {A4},
        pages = {A4},
          doi = {10.1051/0004-6361/201220355},
archivePrefix = {arXiv},
       eprint = {1212.3409},
 primaryClass = {astro-ph.HE},
       adsurl = {https://ui.adsabs.harvard.edu/abs/2013A&A...550A...4H}
}

@ARTICLE{Hirschmann_2019,
       author = {{Hirschmann}, Michaela and {Charlot}, Stephane and {Feltre}, Anna and {Naab}, Thorsten and {Somerville}, Rachel S. and {Choi}, Ena},
        title = "{Synthetic nebular emission from massive galaxies - II. Ultraviolet-line diagnostics of dominant ionizing sources}",
      journal = {\mnras},
         year = 2019,
        month = jul,
       volume = {487},
       number = {1},
        pages = {333-353},
          doi = {10.1093/mnras/stz1256},
archivePrefix = {arXiv},
       eprint = {1811.07909},
 primaryClass = {astro-ph.GA},
       adsurl = {https://ui.adsabs.harvard.edu/abs/2019MNRAS.487..333H}
}

@ARTICLE{Hou_2016,
       author = {{Hou}, Jun and {Frenk}, Carlos. S. and {Lacey}, Cedric G. and {Bose}, Sownak},
        title = "{Constraining SN feedback: a tug of war between reionization and the Milky Way satellites}",
      journal = {\mnras},
         year = 2016,
        month = dec,
       volume = {463},
       number = {2},
        pages = {1224-1239},
          doi = {10.1093/mnras/stw2033},
archivePrefix = {arXiv},
       eprint = {1512.04595},
 primaryClass = {astro-ph.GA},
       adsurl = {https://ui.adsabs.harvard.edu/abs/2016MNRAS.463.1224H}
}

@ARTICLE{Hubble_1926,
       author = {{Hubble}, E.~P.},
        title = "{Extragalactic nebulae.}",
      journal = {\apj},
         year = 1926,
        month = dec,
       volume = {64},
        pages = {321-369},
          doi = {10.1086/143018},
       adsurl = {https://ui.adsabs.harvard.edu/abs/1926ApJ....64..321H}
}

@Article{Hunter_2007,
  Author    = {Hunter, J. D.},
  Title     = {Matplotlib: A 2D graphics environment},
  Journal   = {Computing in Science \& Engineering},
  Volume    = {9},
  Number    = {3},
  Pages     = {90--95},
  publisher = {IEEE COMPUTER SOC},
  doi       = {10.1109/MCSE.2007.55},
  year      = 2007
}

@ARTICLE{Inoue_2014,
       author = {{Inoue}, Akio K. and {Shimizu}, Ikkoh and {Iwata}, Ikuru and {Tanaka}, Masayuki},
        title = "{An updated analytic model for attenuation by the intergalactic medium}",
      journal = {\mnras},
         year = 2014,
        month = aug,
       volume = {442},
       number = {2},
        pages = {1805-1820},
          doi = {10.1093/mnras/stu936},
archivePrefix = {arXiv},
       eprint = {1402.0677},
 primaryClass = {astro-ph.CO},
       adsurl = {https://ui.adsabs.harvard.edu/abs/2014MNRAS.442.1805I}
}

@ARTICLE{Johnson_2026,
       author = {{Johnson}, Benjamin D. and {Robertson}, Brant E. and {Eisenstein}, Daniel J. and {Tacchella}, Sandro and {Pusk{\'a}s}, D{\'a}vid and {Duan}, Qiao and {Wu}, Zihao and {Hainline}, Kevin and {Rieke}, Marcia and {Willott}, Chris and {Willmer}, Christopher N.~A. and {Trussler}, James A.~A. and {Alberts}, Stacey and {Arribas}, Santiago and {Baker}, William M. and {Bunker}, Andrew J. and {Cameron}, Alex J. and {Carniani}, Stefano and {Carreira}, Courtney and {Cargile}, Phillip A. and {Curtis-Lake}, Emma and {Egami}, Eiichi and {Hausen}, Ryan and {Helton}, Jakob M. and {Ji}, Zhiyuan and {Maiolino}, Roberto and {P{\'e}rez-Gonz{\'a}lez}, Pablo G. and {Rinaldi}, Pierluigi and {Sun}, Fengwu and {Sun}, Yang and {Villanueva}, Natalia C. and {Williams}, Christina C. and {Zhu}, Yongda},
        title = "{JWST Advanced Deep Extragalactic Survey (JADES) Data Release 5: NIRCam Imaging in GOODS-S and GOODS-N}",
      journal = {arXiv e-prints},
         year = 2026,
        month = jan,
          eid = {arXiv:2601.15954},
        pages = {arXiv:2601.15954},
          doi = {10.48550/arXiv.2601.15954},
archivePrefix = {arXiv},
       eprint = {2601.15954},
 primaryClass = {astro-ph.GA},
       adsurl = {https://ui.adsabs.harvard.edu/abs/2026arXiv260115954J}
}

@ARTICLE{Kauffmann_1996,
       author = {{Kauffmann}, Guinevere},
        title = "{Disc galaxies at z=0 and at high redshift: an explanation of the observed evolution of damped Lyalpha absorption systems}",
      journal = {\mnras},
         year = 1996,
        month = jul,
       volume = {281},
       number = {2},
        pages = {475-486},
          doi = {10.1093/mnras/281.2.475},
archivePrefix = {arXiv},
       eprint = {astro-ph/9512123},
 primaryClass = {astro-ph},
       adsurl = {https://ui.adsabs.harvard.edu/abs/1996MNRAS.281..475K}
}

@ARTICLE{Keenan_2010,
       author = {{Keenan}, R.~C. and {Barger}, A.~J. and {Cowie}, L.~L. and {Wang}, W.-H.},
        title = "{The Resolved Near-infrared Extragalactic Background}",
      journal = {\apj},
         year = 2010,
        month = nov,
       volume = {723},
       number = {1},
        pages = {40-46},
          doi = {10.1088/0004-637X/723/1/40},
archivePrefix = {arXiv},
       eprint = {1008.4216},
 primaryClass = {astro-ph.CO},
       adsurl = {https://ui.adsabs.harvard.edu/abs/2010ApJ...723...40K}
}

@ARTICLE{Kennicutt_1998,
       author = {{Kennicutt}, Jr., Robert C.},
        title = "{The Global Schmidt Law in Star-forming Galaxies}",
      journal = {\apj},
         year = 1998,
        month = may,
       volume = {498},
       number = {2},
        pages = {541-552},
          doi = {10.1086/305588},
archivePrefix = {arXiv},
       eprint = {astro-ph/9712213},
 primaryClass = {astro-ph},
       adsurl = {https://ui.adsabs.harvard.edu/abs/1998ApJ...498..541K}
}

@ARTICLE{Klypin_2016,
       author = {{Klypin}, Anatoly and {Yepes}, Gustavo and {Gottl{\"o}ber}, Stefan and {Prada}, Francisco and {He{\ss}}, Steffen},
        title = "{MultiDark simulations: the story of dark matter halo concentrations and density profiles}",
      journal = {\mnras},
         year = 2016,
        month = apr,
       volume = {457},
       number = {4},
        pages = {4340-4359},
          doi = {10.1093/mnras/stw248},
archivePrefix = {arXiv},
       eprint = {1411.4001},
 primaryClass = {astro-ph.CO},
       adsurl = {https://ui.adsabs.harvard.edu/abs/2016MNRAS.457.4340K}
}

@ARTICLE{Koekemoer_2013,
       author = {{Koekemoer}, Anton M. and {Ellis}, Richard S. and {McLure}, Ross J. and {Dunlop}, James S. and {Robertson}, Brant E. and {Ono}, Yoshiaki and {Schenker}, Matthew A. and {Ouchi}, Masami and {Bowler}, Rebecca A.~A. and {Rogers}, Alexander B. and {Curtis-Lake}, Emma and {Schneider}, Evan and {Charlot}, Stephane and {Stark}, Daniel P. and {Furlanetto}, Steven R. and {Cirasuolo}, Michele and {Wild}, V. and {Targett}, T.},
        title = "{The 2012 Hubble Ultra Deep Field (UDF12): Observational Overview}",
      journal = {\apjs},
         year = 2013,
        month = nov,
       volume = {209},
       number = {1},
          eid = {3},
        pages = {3},
          doi = {10.1088/0067-0049/209/1/3},
archivePrefix = {arXiv},
       eprint = {1212.1448},
 primaryClass = {astro-ph.CO},
       adsurl = {https://ui.adsabs.harvard.edu/abs/2013ApJS..209....3K}
}

@ARTICLE{Koo_1992,
       author = {{Koo}, David C. and {Kron}, Richard G.},
        title = "{Evidence for evolution in faint field galaxy samples.}",
      journal = {\araa},
         year = 1992,
        month = jan,
       volume = {30},
        pages = {613-652},
          doi = {10.1146/annurev.aa.30.090192.003145},
       adsurl = {https://ui.adsabs.harvard.edu/abs/1992ARA&A..30..613K}
}

@ARTICLE{Koushan_2021,
       author = {{Koushan}, Soheil and {Driver}, Simon P. and {Bellstedt}, Sabine and {Davies}, Luke J. and {Robotham}, Aaron S.~G. and {Lagos}, Claudia del P. and {Hashemizadeh}, Abdolhosein and {Obreschkow}, Danail and {Thorne}, Jessica E. and {Bremer}, Malcolm and {Holwerda}, B.~W. and {Hopkins}, Andrew M. and {Jarvis}, Matt J. and {Siudek}, Malgorzata and {Windhorst}, Rogier A.},
        title = "{GAMA/DEVILS: constraining the cosmic star formation history from improved measurements of the 0.3-2.2 {\ensuremath{\mu}}m extragalactic background light}",
      journal = {\mnras},
         year = 2021,
        month = may,
       volume = {503},
       number = {2},
        pages = {2033-2052},
          doi = {10.1093/mnras/stab540},
archivePrefix = {arXiv},
       eprint = {2102.12323},
 primaryClass = {astro-ph.CO},
       adsurl = {https://ui.adsabs.harvard.edu/abs/2021MNRAS.503.2033K}
}

@ARTICLE{Kron_1980,
       author = {{Kron}, R.~G.},
        title = "{Photometry of a complete sample of faint galaxies.}",
      journal = {\apjs},
         year = 1980,
        month = jun,
       volume = {43},
        pages = {305-325},
          doi = {10.1086/190669},
       adsurl = {https://ui.adsabs.harvard.edu/abs/1980ApJS...43..305K}
}

@ARTICLE{Labbe_2023,
       author = {{Labb{\'e}}, Ivo and {van Dokkum}, Pieter and {Nelson}, Erica and {Bezanson}, Rachel and {Suess}, Katherine A. and {Leja}, Joel and {Brammer}, Gabriel and {Whitaker}, Katherine and {Mathews}, Elijah and {Stefanon}, Mauro and {Wang}, Bingjie},
        title = "{A population of red candidate massive galaxies  600 Myr after the Big Bang}",
      journal = {\nat},
         year = 2023,
        month = apr,
       volume = {616},
       number = {7956},
        pages = {266-269},
          doi = {10.1038/s41586-023-05786-2},
archivePrefix = {arXiv},
       eprint = {2207.12446},
 primaryClass = {astro-ph.GA},
       adsurl = {https://ui.adsabs.harvard.edu/abs/2023Natur.616..266L}
}

@ARTICLE{Lacey_2016,
       author = {{Lacey}, Cedric G. and {Baugh}, Carlton M. and {Frenk}, Carlos S. and {Benson}, Andrew J. and {Bower}, Richard G. and {Cole}, Shaun and {Gonzalez-Perez}, Violeta and {Helly}, John C. and {Lagos}, Claudia D.~P. and {Mitchell}, Peter D.},
        title = "{A unified multiwavelength model of galaxy formation}",
      journal = {\mnras},
         year = 2016,
        month = nov,
       volume = {462},
       number = {4},
        pages = {3854-3911},
          doi = {10.1093/mnras/stw1888},
archivePrefix = {arXiv},
       eprint = {1509.08473},
 primaryClass = {astro-ph.GA},
       adsurl = {https://ui.adsabs.harvard.edu/abs/2016MNRAS.462.3854L}
}

@ARTICLE{Lagos_2019,
       author = {{Lagos}, Claudia del P. and {Robotham}, Aaron S.~G. and {Trayford}, James W. and {Tobar}, Rodrigo and {Bravo}, Mat{\'\i}as and {Bellstedt}, Sabine and {Davies}, Luke J.~M. and {Driver}, Simon P. and {Elahi}, Pascal J. and {Obreschkow}, Danail and {Power}, Chris},
        title = "{From the far-ultraviolet to the far-infrared - galaxy emission at 0 {\ensuremath{\leq}} z {\ensuremath{\leq}} 10 in the SHARK semi-analytic model}",
      journal = {\mnras},
         year = 2019,
        month = nov,
       volume = {489},
       number = {3},
        pages = {4196-4216},
          doi = {10.1093/mnras/stz2427},
archivePrefix = {arXiv},
       eprint = {1908.03423},
 primaryClass = {astro-ph.GA},
       adsurl = {https://ui.adsabs.harvard.edu/abs/2019MNRAS.489.4196L}
}

@ARTICLE{Leja_2019,
       author = {{Leja}, Joel and {Carnall}, Adam C. and {Johnson}, Benjamin D. and {Conroy}, Charlie and {Speagle}, Joshua S.},
        title = "{How to Measure Galaxy Star Formation Histories. II. Nonparametric Models}",
      journal = {\apj},
         year = 2019,
        month = may,
       volume = {876},
       number = {1},
          eid = {3},
        pages = {3},
          doi = {10.3847/1538-4357/ab133c},
archivePrefix = {arXiv},
       eprint = {1811.03637},
 primaryClass = {astro-ph.GA},
       adsurl = {https://ui.adsabs.harvard.edu/abs/2019ApJ...876....3L}
}

@ARTICLE{Leung_2023,
       author = {{Leung}, Gene C.~K. and {Bagley}, Micaela B. and {Finkelstein}, Steven L. and {Ferguson}, Henry C. and {Koekemoer}, Anton M. and {P{\'e}rez-Gonz{\'a}lez}, Pablo G. and {Morales}, Alexa and {Kocevski}, Dale D. and {Yang}, Guang and {Somerville}, Rachel S. and {Wilkins}, Stephen M. and {Yung}, L.~Y. Aaron and {Fujimoto}, Seiji and {Larson}, Rebecca L. and {Papovich}, Casey and {Pirzkal}, Nor and {Berg}, Danielle A. and {Lotz}, Jennifer M. and {Castellano}, Marco and {Ch{\'a}vez Ortiz}, {\'O}scar A. and {Cheng}, Yingjie and {Dickinson}, Mark and {Giavalisco}, Mauro and {Hathi}, Nimish P. and {Hutchison}, Taylor A. and {Jung}, Intae and {Kartaltepe}, Jeyhan S. and {Natarajan}, Priyamvada and {Rothberg}, Barry},
        title = "{NGDEEP Epoch 1: The Faint End of the Luminosity Function at z   9-12 from Ultradeep JWST Imaging}",
      journal = {\apjl},
         year = 2023,
        month = sep,
       volume = {954},
       number = {2},
          eid = {L46},
        pages = {L46},
          doi = {10.3847/2041-8213/acf365},
archivePrefix = {arXiv},
       eprint = {2306.06244},
 primaryClass = {astro-ph.GA},
       adsurl = {https://ui.adsabs.harvard.edu/abs/2023ApJ...954L..46L}
}

@ARTICLE{Levenson_2008,
       author = {{Levenson}, L.~R. and {Wright}, E.~L.},
        title = "{Probing the 3.6 {\ensuremath{\mu}}m CIRB with Spitzer in Three DIRBE Dark Spots}",
      journal = {\apj},
         year = 2008,
        month = aug,
       volume = {683},
       number = {2},
        pages = {585-596},
          doi = {10.1086/589808},
archivePrefix = {arXiv},
       eprint = {0802.1239},
 primaryClass = {astro-ph},
       adsurl = {https://ui.adsabs.harvard.edu/abs/2008ApJ...683..585L}
}

@ARTICLE{Lovell_2023,
       author = {{Lovell}, Christopher C. and {Harrison}, Ian and {Harikane}, Yuichi and {Tacchella}, Sandro and {Wilkins}, Stephen M.},
        title = "{Extreme value statistics of the halo and stellar mass distributions at high redshift: are JWST results in tension with {\ensuremath{\Lambda}}CDM?}",
      journal = {\mnras},
         year = 2023,
        month = jan,
       volume = {518},
       number = {2},
        pages = {2511-2520},
          doi = {10.1093/mnras/stac3224},
archivePrefix = {arXiv},
       eprint = {2208.10479},
 primaryClass = {astro-ph.GA},
       adsurl = {https://ui.adsabs.harvard.edu/abs/2023MNRAS.518.2511L}
}

@ARTICLE{Lovell_2025_syn,
       author = {{Lovell}, Christopher C. and {Roper}, William J. and {Vijayan}, Aswin P. and {Wilkins}, Stephen M. and {Newman}, Sophie and {Seeyave}, Louise},
        title = "{Synthesizer: a Software Package for Synthetic Astronomical Observables}",
      journal = {The Open Journal of Astrophysics},
         year = 2025,
        month = oct,
       volume = {8},
          eid = {152},
        pages = {152},
          doi = {10.33232/001c.145766},
archivePrefix = {arXiv},
       eprint = {2508.03888},
 primaryClass = {astro-ph.IM},
       adsurl = {https://ui.adsabs.harvard.edu/abs/2025OJAp....8E.152L}
}

@ARTICLE{Lu_2025,
       author = {{Lu}, Shengdong and {Frenk}, Carlos S. and {Bose}, Sownak and {Lacey}, Cedric G. and {Cole}, Shaun and {Baugh}, Carlton M. and {Helly}, John C.},
        title = "{A comparison of pre-existing {\ensuremath{\Lambda}}CDM predictions with the abundance of JWST galaxies at high redshift}",
      journal = {\mnras},
         year = 2025,
        month = jan,
       volume = {536},
       number = {1},
        pages = {1018-1034},
          doi = {10.1093/mnras/stae2646},
archivePrefix = {arXiv},
       eprint = {2406.02672},
 primaryClass = {astro-ph.GA},
       adsurl = {https://ui.adsabs.harvard.edu/abs/2025MNRAS.536.1018L}
}

@ARTICLE{Madau_2000,
       author = {{Madau}, Piero and {Pozzetti}, Lucia},
        title = "{Deep galaxy counts, extragalactic background light and the stellar baryon budget}",
      journal = {\mnras},
         year = 2000,
        month = feb,
       volume = {312},
       number = {2},
        pages = {L9-L15},
          doi = {10.1046/j.1365-8711.2000.03268.x},
archivePrefix = {arXiv},
       eprint = {astro-ph/9907315},
 primaryClass = {astro-ph},
       adsurl = {https://ui.adsabs.harvard.edu/abs/2000MNRAS.312L...9M}
}

@ARTICLE{MAGIC-Collaboration_2018,
       author = {{MAGIC Collaboration} and {Albert}, J. and {Aliu}, E. and {Anderhub}, H. and {Antonelli}, L.~A. and {Antoranz}, P. and {Backes}, M. and {Baixeras}, C. and {Barrio}, J.~A. and {Bartko}, H. and {Bastieri}, D. and {Becker}, J.~K. and {Bednarek}, W. and {Berger}, K. and {Bernardini}, E. and {Bigongiari}, C. and {Biland}, A. and {Bock}, R.~K. and {Bonnoli}, G. and {Bordas}, P. and {Bosch-Ramon}, V. and {Bretz}, T. and {Britvitch}, I. and {Camara}, M. and {Carmona}, E. and {Chilingarian}, A. and {Commichau}, S. and {Contreras}, J.~L. and {Cortina}, J. and {Costado}, M.~T. and {Covino}, S. and {Curtef}, V. and {Dazzi}, F. and {De Angelis}, A. and {de Cea del Pozo}, E. and {de los Reyes}, R. and {De Lotto}, B. and {De Maria}, M. and {De Sabata}, F. and {Delgado Mendez}, C. and {Dominguez}, A. and {Dorner}, D. and {Doro}, M. and {Errando}, M. and {Fagiolini}, M. and {Ferenc}, D. and {Fern{\'a}ndez}, E. and {Firpo}, R. and {Fonseca}, M.~V. and {Font}, L. and {Galante}, N. and {Garc{\'\i}a L{\'o}pez}, R.~J. and {Garczarczyk}, M. and {Gaug}, M. and {Goebel}, F. and {Hayashida}, M. and {Herrero}, A. and {H{\"o}hne}, D. and {Hose}, J. and {Hsu}, C.~C. and {Huber}, S. and {Jogler}, T. and {Kneiske}, T.~M. and {Kranich}, D. and {La Barbera}, A. and {Laille}, A. and {Leonardo}, E. and {Lindfors}, E. and {Lombardi}, S. and {Longo}, F. and {L{\'o}pez}, M. and {Lorenz}, E. and {Majumdar}, P. and {Maneva}, G. and {Mankuzhiyil}, N. and {Mannheim}, K. and {Maraschi}, L. and {Mariotti}, M. and {Mart{\'\i}nez}, M. and {Mazin}, D. and {Meucci}, M. and {Meyer}, M. and {Miranda}, J.~M. and {Mirzoyan}, R. and {Mizobuchi}, S. and {Moles}, M. and {Moralejo}, A. and {Nieto}, D. and {Nilsson}, K. and {Ninkovic}, J. and {Otte}, N. and {Oya}, I. and {Panniello}, M. and {Paoletti}, R. and {Paredes}, J.~M. and {Pasanen}, M. and {Pascoli}, D. and {Pauss}, F. and {Pegna}, R.~G. and {Perez-Torres}, M.~A. and {Persic}, M. and {Peruzzo}, L. and {Piccioli}, A. and {Prada}, F. and {Prandini}, E. and {Puchades}, N. and {Raymers}, A. and {Rhode}, W. and {Rib{\'o}}, M. and {Rico}, J. and {Rissi}, M. and {Robert}, A. and {R{\"u}gamer}, S. and {Saggion}, A. and {Saito}, T.~Y. and {Salvati}, M. and {Sanchez-Conde}, M. and {Sartori}, P. and {Satalecka}, K. and {Scalzotto}, V. and {Scapin}, V. and {Schmitt}, R. and {Schweizer}, T. and {Shayduk}, M. and {Shinozaki}, K. and {Shore}, S.~N. and {Sidro}, N. and {Sierpowska-Bartosik}, A. and {Sillanp{\"a}{\"a}}, A. and {Sobczynska}, D. and {Spanier}, F. and {Stamerra}, A. and {Stark}, L.~S. and {Takalo}, L. and {Tavecchio}, F. and {Temnikov}, P. and {Tescaro}, D. and {Teshima}, M. and {Tluczykont}, M. and {Torres}, D.~F. and {Turini}, N. and {Vankov}, H. and {Venturini}, A. and {Vitale}, V. and {Wagner}, R.~M. and {Wittek}, W. and {Zabalza}, V. and {Zandanel}, F. and {Zanin}, R. and {Zapatero}, J.},
        title = "{Very-High-Energy gamma rays from a Distant Quasar: How Transparent Is the Universe?}",
      journal = {Science},
         year = 2008,
        month = jun,
       volume = {320},
       number = {5884},
        pages = {1752},
          doi = {10.1126/science.1157087},
archivePrefix = {arXiv},
       eprint = {0807.2822},
 primaryClass = {astro-ph},
       adsurl = {https://ui.adsabs.harvard.edu/abs/2008Sci...320.1752M}
}

@ARTICLE{Maiolino_2024,
       author = {{Maiolino}, Roberto and {Scholtz}, Jan and {Curtis-Lake}, Emma and {Carniani}, Stefano and {Baker}, William and {de Graaff}, Anna and {Tacchella}, Sandro and {{\"U}bler}, Hannah and {D'Eugenio}, Francesco and {Witstok}, Joris and {Curti}, Mirko and {Arribas}, Santiago and {Bunker}, Andrew J. and {Charlot}, St{\'e}phane and {Chevallard}, Jacopo and {Eisenstein}, Daniel J. and {Egami}, Eiichi and {Ji}, Zhiyuan and {Jones}, Gareth C. and {Lyu}, Jianwei and {Rawle}, Tim and {Robertson}, Brant and {Rujopakarn}, Wiphu and {Perna}, Michele and {Sun}, Fengwu and {Venturi}, Giacomo and {Williams}, Christina C. and {Willott}, Chris},
        title = "{JADES: The diverse population of infant black holes at 4 < z < 11: Merging, tiny, poor, but mighty}",
      journal = {\aap},
         year = 2024,
        month = nov,
       volume = {691},
          eid = {A145},
        pages = {A145},
          doi = {10.1051/0004-6361/202347640},
archivePrefix = {arXiv},
       eprint = {2308.01230},
 primaryClass = {astro-ph.GA},
       adsurl = {https://ui.adsabs.harvard.edu/abs/2024A&A...691A.145M}
}

@ARTICLE{Manzoni_2025,
       author = {{Manzoni}, Giorgio and {Broadhurst}, Tom and {Lim}, Jeremy and {Liu}, Tao and {Smoot}, George and {Baugh}, Carlton M. and {Tompkins}, Scott and {Windhorst}, Rogier and {Driver}, Simon and {Carleton}, Timothy and {Frye}, Brenda and {Fung}, Leo and {Zhang}, Jiashuo and {Cohen}, Seth H. and {Conselice}, Christopher J. and {Grogin}, Norman A. and {Jansen}, Rolf A. and {Koekemoer}, Anton M. and {Ortiz}, III, Rafael and {Pirzkal}, Norbert and {Willmer}, Christopher N.~A.},
        title = "{Explaining JWST Counts with Galaxy Formation Models}",
      journal = {\apj},
         year = 2025,
        month = aug,
       volume = {988},
       number = {2},
          eid = {264},
        pages = {264},
          doi = {10.3847/1538-4357/ade700},
archivePrefix = {arXiv},
       eprint = {2502.04702},
 primaryClass = {astro-ph.GA},
       adsurl = {https://ui.adsabs.harvard.edu/abs/2025ApJ...988..264M}
}
%%%%%%%%%%%%%%%%%%%%%%%%%%%%%%%%%%%%%%%%%%%%%%%%%%

%%%%%%%%%%%%%%%%% APPENDICES %%%%%%%%%%%%%%%%%%%%%
\appendix
\begin{figure*}
    \centering
    \includegraphics[width=\textwidth]{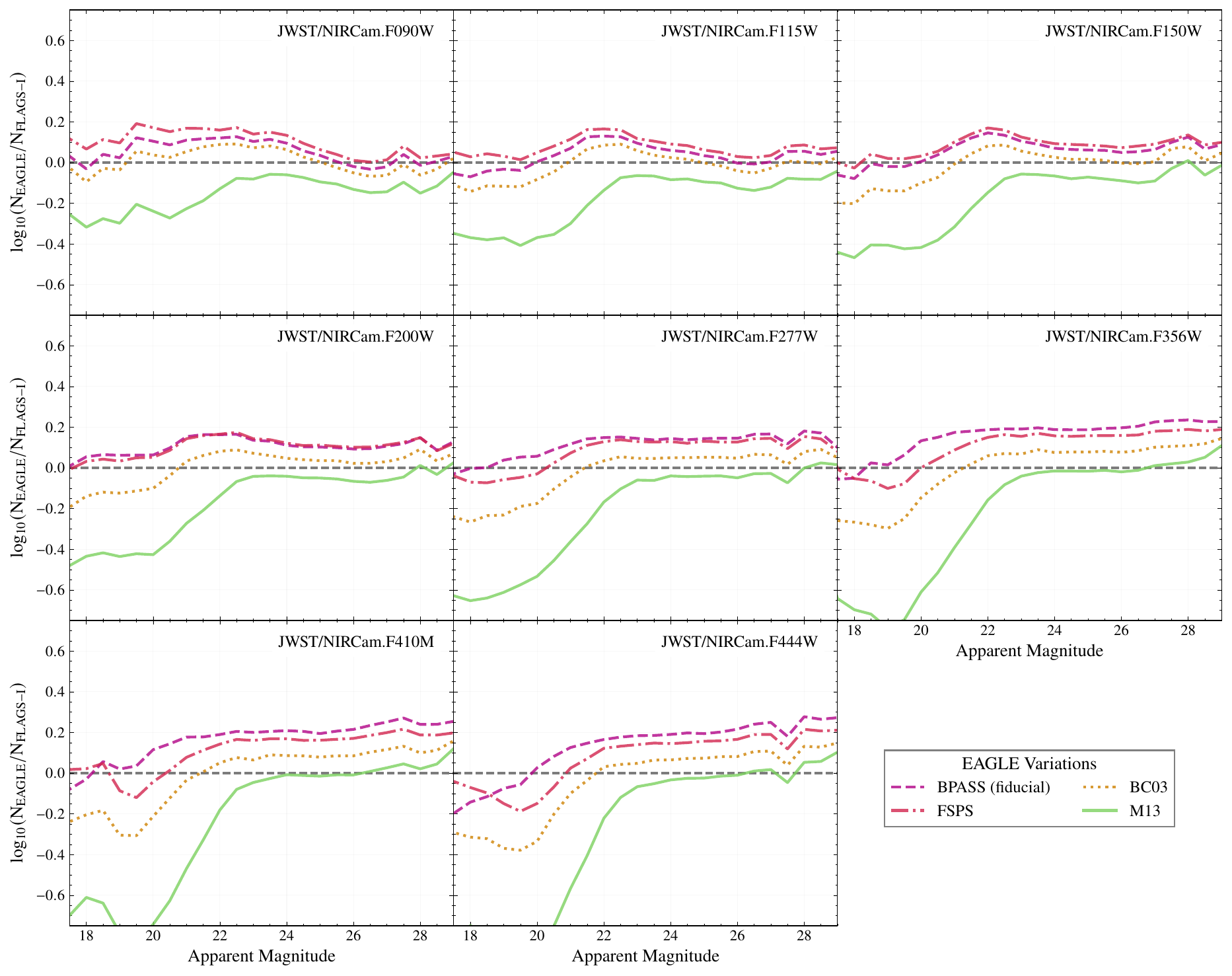}
    \caption{The ratio between the \eagle\ galaxy number counts predicted under different SPS modelling assumptions and the \flags\ observations in eight NIRCam filters. The grey-shaded regions indicate the observational uncertainty bounds.}
    \label{fig:eagle_sps}
\end{figure*}

\section{EAGLE SPS Modelling Variations}\label{app:eagle_sps}

Section \ref{subsec:systematics} demonstrated variations in the galaxy number counts predicted by \eagle\ under different forward modelling assumptions, relative to the fiducial \bpass. Figure \ref{fig:eagle_sps} shows these same variations, but now relative to the galaxy number counts observed in each of the eight NIRCam filters. The same general behaviour is seen for \bpass, \bc\ and \fsps, which reproduce the short-wavelength counts reasonably well, but overpredict the number of faint galaxies at long wavelengths. Overall, the counts predicted when using \bc\ agree with the observations most closely, with the total $\chi^{2}_{\nu}=24.8$ lower than both \fsps\ ($\chi^{2}_{\nu}=68.7$) and \bpass\ ($\chi^{2}_{\nu}=90.8$). While this confirms that the measured physical model performance is highly dependent on the SPS modelling, it is not an indicator that \bc\ is any more representative than the others, unless one assumes that \eagle\ is entirely accurate. It does mean that models such as \jaguar, \sage\ and \spritz, which generally overpredict the counts, would likely appear to perform more poorly if consistently forward modelled with \bpass.

The far poorer performance of \eagle\ when using \ma\ for forward modelling ($\chi^{2}_{\nu}=112.5$) should not be immediately taken as an indicator of poorer SPS modelling. The faint counts at $>3 \ \mathrm{\mu m}$ are recovered the most accurately by far in this configuration, but this comes at the cost of greatly underestimating the $m_{\mathrm{AB}}\lesssim22$ counts in all filters. Given the generally good and consistent performance of the other models at the bright end, this could indicate an issue with how \ma\ handles the evolution of old stars. The switch from using isochrones to following the fuel consumption theorem \citep{Renzini_1981} for post-main-sequence stars is the most obvious difference between \ma\ and the other models, and may need revision.

\end{document}